\documentclass[manuscript]{acmart}

\usepackage{booktabs} % For formal tables
\usepackage[ruled]{algorithm2e} % For algorithms
\renewcommand{\algorithmcfname}{ALGORITHM}

\SetAlFnt{\small}
\SetAlCapFnt{\small}
\SetAlCapNameFnt{\small}
\SetAlCapHSkip{0pt}
\IncMargin{-\parindent}

\usepackage{dsfont}
\usepackage{amsmath}
\usepackage{amsthm}
\usepackage{color,fancybox,graphicx,url}
\usepackage{float}
\usepackage[caption = false]{subfig}

\newtheorem{lemma}{Lemma}
\newtheorem{prop}{Proposition}
\newtheorem{cor}{Corollary}
\newtheorem{assumption}{Assumption}
\newtheorem{definition}{Definition}[section]

\newcommand{\Com}{C}  %community
\newcommand{\setR}{\mathbb{R}}
\newcommand{\setT}{\mathcal{T}}

\newcommand{\Bsbl}{\Big [ }
\newcommand{\Bsbr}{\Big ] }

\newcommand{\Infl}{\mbox{{\tiny infl}}}

\newcommand{\NE}{\mbox{\tiny NE}}
\newcommand{\proxy}{prox}

\newcommand{\comSet}{C}
\newcommand{\leafSet}{\comSet}
\newcommand{\infl}{y_{\Infl}}

\newcommand{\delay}[1]{\theta \left (#1 \right )}
\newcommand{\ddelay}[1]{\theta' \left (#1 \right )}
\newcommand{\delayInfl}[2]{\theta_{\Infl} \left ( #1|#2 \right )}

\newcommand{\possibleStrategy}{\mathcal{S}(\coreBudget,\leafBudget)}

\newcommand{\coreBudget}{M_{\Infl}}
\newcommand{\leafBudget}{M}

\newcommand{\alternative}[1]{r_0 \delay{#1} B_0 }
\newcommand{\leafAllocation}[1]{\comAllocation(#1)}

\newcommand{\allAllocation}{\Omega}
\newcommand{\allAllocationNE}{\Omega_{\NE}(\leafBudget,\coreBudget)}
\newcommand{\allAllocationNEOpt}{\Omega^*_{\NE}(\leafBudget,\coreBudget)}

\newcommand{\allAllocationInfl}{\allAllocation_{\Infl}}
\newcommand{\allAllocationInflNE}{\allAllocation_{\Infl}(\leafBudget,\coreBudget)}

\newcommand{\allAllocationInflOpt}{\allAllocation^*_{\Infl}(\leafBudget,\coreBudget)}

\newcommand{\allProxAllocation}{\allAllocation_{\proxy}}
\newcommand{\allProxAllocationNE}{\allAllocation_{\proxy}(\leafBudget,\coreBudget)}

\newcommand{\allAllocationWelfare}{\allAllocation^*(\coreBudget,\leafBudget)}

\newcommand{\comAllocation}{\Lambda}
\newcommand{\comProxAllocation}{\Lambda}
\newcommand{\contAllocation}{X}
\newcommand{\coreInteract}[1]{\mu_{\Infl}(#1)}

\newcommand{\coreInteractVec}{\mu_{\Infl}}
\newcommand{\coreInteractVecOpt}{\mu^*_{\Infl}}

\newcommand{\setNEInfl}{{\mathcal{S}_{\Infl}}}

\newcommand{\coreUtility}[1]{U_{\Infl}(\coreInteractVec|#1)}
\newcommand{\leafOptUtility}[1]{U_{c,#1}(\leafRateVec{#1}|\coreRateVec,\contAllocation)}
\newcommand{\leafOptProxUtility}[1]{U^{\proxy}_{c,#1}(\leafRateVec{#1}|\coreRateVec,\contAllocation)}

\newcommand{\prodOptUtility}[1]{U_{p,#1}(x({#1})|\coreRateVec,\comAllocation)}
\newcommand{\prodOptUtilityInf}[1]{U^{\Infl}_{p,#1}(x({#1})|\coreRateVec,\comAllocation)}
\newcommand{\prodOptUtilityInfNE}[1]{U^{\Infl}_{p,#1}(x({#1})|\coreRateVec^*,\comAllocation^*)}

\newcommand{\delayCoreUtility}[2]{ \delayInfl{#1}{#2} B(#1|#2) }
\newcommand{\delayUtility}[2]{\delay{\mu(#1|#2)} B(#1|#2) }

\newcommand{\optcore}[1]{OPT_{\Infl}(\coreRateVec|#1)}
\newcommand{\optperiphery}[2]{OPT_{c,#2}(\leafAllocation{#2}|#1)}
\newcommand{\optProxperiphery}[2]{OPT^{\proxy}_{c,#2}(\comAllocation({#2})|#1)}
\newcommand{\optcontperiphery}[2]{OPT_{p,#2}(x(#2)|#1)}

\newcommand{\coreRateVec}{\coreInteractVec}

\newcommand{\leafRateVec}[1]{\comAllocation(#1)}
\newcommand{\Clw}[1]{\leafSet \backslash\{#1\}}
\newcommand{\Cw}[1]{\comSet \backslash\{#1\}}
\DeclareMathOperator*{\argmax}{argmax}

\newcommand{\strategySpace}{S(\leafBudget,\coreBudget)}

\newcommand{\para}[1]{\vspace{0.05in}\indent{{\it #1}:}}

\begin{document}

\title{The Role of Social Support and Influencers in Social Media Communities}

\author{Junwei Su}
%\email{junwei.su@mail.utoronto.ca}
\affiliation{%
  \institution{Department of Computer Science, University of Hong Kong}
}

\author{Peter Marbach}
%\email{marbach@cs.toronto.edu}
\affiliation{%
  \institution{Department of Computer Science, University of Toronto}
}


%\input{content/abstract.tex}

\maketitle

\section{Introduction}

How can individual agents or processes coordinate their actions in a decentralized way to achieve a common goal? This question is crucial across various systems—economic, technical, and sociological.
In economic systems, it explores how buyers and sellers coordinate their actions (buying and selling goods and services) to achieve efficient market outcomes and maximize social welfare.
In technical systems, the focus is on algorithms for distributed systems, like multi-agent AI~\cite{sl_multi1,sl_multi2} and distributed computing~\cite{dist}, which enable agents or processes to work together and optimize system performance.
In sociological systems, it examines how individuals within social groups coordinate their actions to follow social norms and rules, ensuring the proper functioning of the group~\cite{durkheim,giddens,bourdieu}.

All these three systems—economic, technical, and sociological—face common challenges. First, scalability: in large systems, agents cannot interact directly with all others, making it difficult to determine which actions best contribute to the shared objective. Second, heterogeneity: agents often have varying levels of information and resources, complicating coordination in complex environments. Third, conflicting individual and collective goals: agents, while pursuing their own interests, may not naturally align with the broader objectives of the community.

Interestingly, economic systems, in the form of markets, offer an example of how these challenges can be overcome. In economic markets, the three challenges are present in the following form: (a) scalability— individual buyers and sellers do not interact with all actors in the market  directly; (b) heterogeneity—actors have different interests and resources; and (c) conflicting goals—each actor aims to maximize personal gain, such as profit, rather than a collective goal like social welfare. The mechanism that addresses these issues is the use of a common currency—money—which enables buyers and sellers to coordinate their actions while pursuing individual goals. Buyers aim to maximize their utility by purchasing goods that meet their needs, while sellers seek to maximize profits by supplying products that satisfy demand. This dynamic, famously described by Adam Smith’s "invisible hand," shows how self-interested behavior can result in optimal societal outcomes, with goods produced and consumed efficiently, maximizing overall welfare.

This raises an intriguing question: Could similar mechanisms, such as a common "currency" for coordination, be applied to other systems? If so, what could this mechanism be that helps agents in technical and sociological contexts overcome the challenges of scalability, heterogeneity, and conflicting interests?

In this paper, we explore this question in the context of social media. Social media platforms have evolved into dynamic environments where communities form around shared interests. Within these communities, individuals continuously make decisions about what content to post and share, aiming to engage the interests of the broader community. This dynamic mirrors that of broader social systems, where individuals coordinate their actions toward achieving common goals.

Using social media communities provides distinct advantages when exploring how agents coordinate their actions in a distributed environment. First, the actions and interactions within social media communities are well-defined, involving activities like posting and sharing content. Second, user activities on social media platforms can be accessed and analyzed, allowing us to empirically test the theoretical results presented in this paper. Finally, we believe that the insights obtained from analyzing social media communities can be extended and applied to more general sociological systems, as well as technical systems such as multi-agent AI systems, as we discuss further in Section~\ref{sec:conclusion}.

In our analysis of social media communities, we first investigate whether there is an equivalent to money in economic markets—a "currency" that coordinates actions in online communities—and whether it can be formalized. We propose that social support serves this role. On social media platforms, social support manifests in various forms, such as likes, shares, comments, and retweets. These actions reflect approval, engagement, and validation from other community members.

Social support serves two key purposes in these communities. First, it acts as a reward system for content creators, guiding them on what type of content to produce in order to maximize their reach and reputation. Second, like money in traditional markets, social support signals the perceived value of content. Content that is considered more valuable by the community will naturally receive more social support. These observations suggest that social support functions similarly to a common currency in an economic market. In the following sections, we formalize this intuition by examining whether social support can be treated as a currency within a content market and how it impacts market equilibrium.

We also study the role of influencers in helping individuals coordinate their actions within social media communities. Influencers are users with a large number of followers who curate, amplify, and share content, thus playing a significant role in shaping the flow of information within the community. While the role of influencers as content aggregators is widely recognized, their influence on content creation and sharing within these communities is less explored. In this paper, we investigate whether and how influencers shape the types of content that are created within social media communities.

In summary, we analyze the following central questions:
\begin{enumerate}
\item[1)] {Social Support as a Currency:}
  Can social support be accurately characterized as a currency in content markets that shapes the actions of individual agents, such as the content that is produced and shared within the community? If so, how efficient is social support in shaping the actions of individual agents, i.e. what is the resulting social welfare?
\item[2)] {The Role of Influencers:}
  Do influencers impact the actions of individual agents, particularly the content that is produced and shared within the community, and if so, how?
\end{enumerate}
To do that,  we model an online community as a content market where agents post (produce), consume (read), and share content. We distinguish between two types of agents: community members who create and consume content, and influencers who aggregate content for the community. We assume that agents act selfishly and aim to achieve their own  objectives.
We model the objectives and actions of these agents in the content market as follows,
\begin{enumerate}
\item The objective of content consumers is to obtain content that is of interest to them. To achieve this goal, content consumers can follow other community members (content producers) as well as influencers. We assume that consumers have limited resources (limited attention) to follow content producers and influencers, and hence strategically decide who they follow, and how much social support they allocate to each agent they follow. 
\item The objective of content producers is to create content to maximize the social support that they receive, i.e. to maximize  the total number of likes or shares that their content receives.
\item The objective of influencers is to attract as many followers as possible. To achieve this goal, influencers aim at  maximizing the social support that they receive from their followers by  curating (sharing)  content that is of interest to their followers~\cite{media_aggregate,jordan2012lattice,link_aggregator}.
\end{enumerate}
Agents aiming to achieve their own  objectives leads to a game (strategic interaction) between content consumers, producers, and influencers, for which  we  study the existence and efficiency of a Nash equilibrium.

More precisely, we study the existence and efficiency of a Nash equilibrium for two scenarios. First, we assume that all agents have perfect information, meaning they know the actions and preferences (utilities) of all other agents. This scenario allows us to examine and characterize the market outcome, specifically the equilibrium and social welfare, for an idealized market with perfect information.

Next, we consider the more realistic case where agents do not have perfect information. In this scenario, content producers cannot observe the actions of all content consumers and, therefore, cannot directly assess the social support they receive from them. Instead, content producers rely on the social support garnered by influencers to determine which content to create. We study how much the social welfare under imperfect information deviates from that of the idealized market with perfect information.

Through this analysis, we aim to provide insights into the functioning of social support as a currency in content markets and the dual role of influencers as both content aggregators and information proxies.  The main results and insights of our analysis are as follows.

\para{Social Support as a Currency} We formalize how social support operates as a currency. In particular, we derive a parallel between how social support works in a content market, and how money works in a market economy.
Similar to a market economy, where one observes a flow of money in one direction and a flow of goods in the opposite direction, in a content market there is a flow of social support in one direction, and a flow of content in the other direction.
The flow of social support goes from the content consumers to the influencer, and from there to the content producers. The flow of content goes from the content producers to the influencer, and from there to the content consumers. Similar to a market economy where money signals the utility of a good, social support signals the utility of content. The social support given by a content consumer to the content shared by the influencer signals the (personal) utility of this content to the consumer. The social support given by the influencer to content created by a producer signals the (aggregated) utility to the content market.

\para{The Role of Influencers}
Influencers play two important roles in the content market. First, as they aggregate content that is of interest to their followers, influencers help content consumers to easily find content that is of interest to them. This role of influencers is well understood. However, influencers play another important role: they serve as an information proxy for content producers. In the case of a large content market, where it is too costly for content producers to follow all content consumers in order to obtain the necessary information to decide which type of content to create, producers can use influencers as a proxy for this information. That is, they can use the social support received by an influencer as a proxy about the total social support they receive from the content consumers. To the best of our knowledge, this role of influencers has not been formally characterized in literature.

\para{Price of Influence}
For the scenario where all agents in the content market have perfect information, we show there exists an efficient Nash equilibrium that maximizes social welfare. This result suggests that using social support as a currency can lead to an efficient market outcome (equilibrium). 
However, for the scenario where content producers have limited information, this is no longer the case. That is, using an influencer as a proxy to obtain information about the social support of content generally leads to suboptimal decisions by the content producers. These suboptimal decisions lead to a reduction in  the social welfare compared with the scenario where agents have perfect information. 
We refer to the reduction of the social welfare as the price of influence.  A surprising result that we obtain is that (under suitable conditions) the price of influence becomes small in large content markets (with a large number of content producers and consumers). This suggests using an influencer as an information proxy leads to a  (near) optimal social welfare in the case of a large content market.

In summary, our analysis shows that social support functions as an effective mechanism for agents to coordinate their actions, addressing three key challenges:
\begin{enumerate}
\item {\em Scale}, where agents interact with only a limited number of others,
\item {\em Heterogeneity}, where agents differ in their interests and objectives, and
\item {\em Conflicting Interests}, where agents aim to optimize their individual objectives.
\end{enumerate}
More specifically, our analysis demonstrates that social support not only helps overcome these challenges but also results in a coordination mechanism that is: \begin{enumerate}
\item[a)] Simple and easily implementable, as agents can rely solely on social support to guide their actions, and
\item[b)] Highly efficient, leading to actions that (nearly) maximize social welfare, particularly in large communities.
\end{enumerate}
Returning to our initial question of how agents can coordinate their actions to achieve a shared objective in a distributed environment, it is natural to consider whether social support as a coordination mechanism can be applied to more general multi-agent systems. We believe this is indeed the case, and that the results and insights from this analysis can be extended and applied to broader sociological and technical systems, including multi-agent AI systems. This is a research direction we are actively pursuing and will explore further in Section~\ref{sec:conclusion}.

The rest of the paper is organized as follows. In Section~\ref{sec:related_work}, we provide an overview of related literature. In Section~\ref{sec:model}, we provide an overview of the model that we use for our analysis, and in Section~\ref{sec:math_model}~-~\ref{sec:social_support}, we provide the formal definition of the model. In Section~\ref{sec:problem_formulation}~-~\ref{sec:results}, we model and analyze the case of a content market with perfect information. In Section~\ref{sec:limited_information}~-~\ref{sec:results_limited_information}, we extend the analysis to the case where agents have limited information. In Section~\ref{sec:discussion}, we provide a discussion of the results that we obtain, and the insights they provide into the questions that we study. In Section~\ref{sec:conclusion}, we discuss possible directions for future research.

%\newpage
\section{Related Work}\label{sec:related_work}
In this section we provide an overview of existing literature that is relevant to our analysis. This includes   modeling  social media as an attention and influence economy, research on influence in economic and sociological theory, as well as research on attention, influence, and content aggregation in social media,

\subsection{Attention Economy and Influence Economy}
The activities on social media have been modeled as an attention economy~\cite{attention_economy}, where agents aim to maximize and monetize the attention they receive, and as an influence economy, where influencers use their received attention to shape the opinions of their followers, promote products or services, and generate income.

\subsubsection{Attention Economy}

The attention economy refers to the economic model centered around capturing and monetizing users' attention on digital platforms, particularly social media~\cite{attention_economy}. Attention is considered a scarce resource, and various entities compete to gain users' limited attention. Social media platforms, advertisers, and content creators strive to engage users and keep them active on the platform for longer periods to maximize revenue through advertisements, sponsored content, subscriptions, and other monetization methods. 
Platforms and advertisers track metrics (attention-based outcomes) such as views, likes, shares, click-through rates, and time spent on content to assess and optimize engagement levels.

The model and analysis that we carry out reflect this economic model of social media where we assume that a) attention is a limited resource, and b) agents use attention-based outcomes (to which we refer as social support) to maximize the attention they obtain.

\subsubsection{Influence Economy}
The influence economy refers to the system and dynamics surrounding social media platforms, where influencers leverage their online presence and following to shape opinions, promote products or services, and generate income. Influencers have the ability to influence their audience's behavior, choices, and preferences through content creation, endorsements, and collaborations. Advertisers and brands collaborate with influencers to reach a highly targeted and engaged audience, utilizing sponsored content, affiliate marketing, brand partnerships, and influencer marketing campaigns.

While we also study the role of influencers and influence, the focus of our analysis differs from the one of the influence economy.
Instead of studying how influencers use their received attention to shape opinions and generate income, we focus on how the social support of content consumers shapes the content shared by influencers and how the social support of influencers shapes the content created by content producers. Our analysis provides a complementary view of influence activities in social media, focusing on the flow of influence from content consumers to influencers and from influencers to content producers.

\subsubsection{Attention vs Social Support}
Similar to an attention economy, attention and social support are distinct concepts in our model. Attention refers to the attention one agent pays to another, while social support represents observable actions such as likes, shares, and clicks. Attention itself might not be directly observable, whereas social support can be measured. We provide a formal definition of how we model attention and social support in Section~\ref{sec:math_model}~and~\ref{sec:social_support}.

\subsection{Influence}
Our focus on whether social support can be considered a currency and the role of influencers aligns with research questions studied in economic and sociological theory.

\subsubsection{Influence and Information Asymmetry}
Milgrom and Roberts present in~\cite{Milgrom} a formal economic model to analyze influence activities within organizations, including lobbying, campaigning, and persuasion. They use game theory and agency theory to study how individuals strategically use influence tactics to shape decision-making processes and outcomes. Information asymmetry plays a crucial role in the emergence of influence activities in their model.

Our approach shares several similarities with the research presented in Milgrom and Roberts~\cite{Milgrom}. First, they develop a formal model that integrates economic principles, game theory, and agency theory to investigate influence activities. Second, they model influence as a strategic interaction among actors who possess diverse information, preferences, and incentives. Thirdly, they emphasize the significance of information asymmetry as a fundamental factor leading to the emergence of influence and influence activities within a social system.

Nevertheless, there are notable distinctions between our approach and that of Milgrom and Roberts~\cite{Milgrom}. First, while Milgrom and Roberts focus on how individuals endeavor to influence decision-makers, within the context of a content market, this corresponds to how content consumers shape the actions of influencers. In our analysis, we not only consider how content consumers influence the decisions of influencers but also how influencers impact the decisions of content producers.

Second, we employ a different approach regarding how actors are rewarded. Specifically, we assume that the reward system manifests as social support, such as likes and retweets, and examine the outcomes when actors seek to optimize social support. In Milgrom and Roberts~\cite{Milgrom}, the reward system is not predetermined but rather the focus of the analysis. In other words, the aim of Milgrom and Roberts is to design a reward system within an organization that encourages influence activities benefiting the organization while discouraging costly influence activities. Conversely, our analysis demonstrates that social support as a reward yields a (near) optimal social welfare outcome. The primary objective of Milgrom and Roberts is not to identify a mechanism that optimizes social welfare but rather to determine the "best" outcome within a given class of mechanisms, which may still be suboptimal.

\subsubsection{Parsons' Hypothesis}
The sociologist Talcott Parsons proposed in the 1960s that similar to money which signals the utility of goods and services in a market, influence is used to signal the utility of social actions and opinions in a social group~\cite{parsons_influence}.
As such, influence and money are both used to facilitate interactions of agents. Money facilitates interactions between economic agents, whereas influence facilitates interactions between social agents.
If true, this hypothesis would have important implications, providing a formal foundation for studying influence in a social system, creating a link to economic theory, and allowing us to leverage existing results and techniques in economic theory to study influence in a social system. However, the hypothesis was never formally or experimentally verified or disproved.

While we focus in this paper on social media, we believe that our results and insights might extend to more general social systems. In particular, we believe that the model and analysis of this paper can be used to formally study Parsons' hypothesis. A detailed discussion of this is outside the scope of this paper, and is research that we are currently undertaking.

\subsection{Influence, Attention and Content Aggregation in Social Media}
Our model and analysis are closely related to studies on influence, attention, and content aggregation in social media.

\subsubsection{Influence}
The emergence of social media and the need to comprehend its impact on the social structure, as well as the beliefs and actions of individuals, have led to the development of a formal study of influence in social media. Extensive literature exists on this topic, including measuring influence and identifying influential users (see~\cite{survey_influence_measure} for a survey), using models to understand the spread of content (see~\cite{survey_cascade_models} for a survey), and maximizing influence (see~\cite{survey_influence_maxmization} for a survey).

Of particular relevance to our approach and analysis are cascade models, which are employed to model influence in social networks and social media~\cite{survey_cascade_models}. These models capture the propagation of content through a social network by assuming that individuals share a given content from their connections with a certain probability. Typically, these models incorporate factors such as tie strength or network prevalence to determine the probability of content sharing~\cite{survey_cascade_models}.

Cascade models focus on characterizing how a behavior or belief spreads across a fixed social network topology with fixed sharing probabilities. In contrast, our model and analysis allow for variable topology and sharing probabilities, which are outcomes of the strategic interactions between content consumers, content producers, and influencers. However, there is an interesting connection between cascade models and our paper's analysis and results. Specifically, our model and analysis provide insights into the interpretation and generation of sharing probabilities in social media, demonstrating that these probabilities can represent the social support paid by one actor to another. Moreover, the social support exchanged between agents is a result of the strategic interactions among content producers, content consumers, and influencers.  Examples of cascade models include the well-known the independent cascade model (ICM)~\cite{Watts_cascade} and linear threshold model (LTM)~\cite{threshold_models}.

Cascade models focus on characterizing how a behavior or belief spreads across a fixed social network topology with fixed sharing probabilities. In contrast, our model and analysis allow for variable topology and sharing probabilities, which are outcomes of the strategic interactions between content consumers, content producers, and influencers. However, there is an interesting connection between cascade models and our paper's analysis and results. Specifically, our model and analysis provide insights into the interpretation and generation of sharing probabilities in social media, demonstrating that these probabilities can represent the social support paid by one actor to another. Moreover, the social support exchanged between agents is a result of the strategic interactions among content producers, content consumers, and influencers.

\subsubsection{Attention}
Research on attention in social media includes studies of the dynamics of attention~\cite{attention_spread_true_false,attention_structure}, using attention for  marketing~\cite{ROI_marketing}, and the attention economy~\cite{attention_economy,attention_economy_microcelebrity}. Formal studies of attention include the model and results presented by Goel and Ronaghi~\cite{Goel_attention}, Borgs et al.~\cite{Borgs_attention}, and Ghosh and McAfee~\cite{Ghosh_attention}.
These studies adopt a game-theoretic approach to characterize attention in social media, although they differ in how they define the utility derived from attention. For instance~\cite{Goel_attention}, attention may be interpreted as the following rate, the number of followers a user has~\cite{Borgs_attention}, or the publication of content~\cite{Ghosh_attention}.

These attention models share similarities with our approach in terms of adopting a game-theoretic perspective to study attention. However, they differ in two fundamental aspects. First, they assume that content producers do not make strategic decisions regarding the type of content they produce, whereas our analysis considers strategic decision-making by content producers. Second, these models do not assume the existence of an influencer but rather direct following relationships between content consumers and producers.

 \subsubsection{Content Aggregation}

The problem we study bears resemblance to research on content aggregation for news and blog posts, with influencers acting as content aggregators. Existing studies on content aggregation directly relevant to our work include the following.

Jordan et al. study in~\cite{jordan2012lattice} how content aggregators decide on the topics for which they curate content, and users decide on which content aggregator to follow. They formulate this interaction as a "location game" in which the chosen location of an aggregator represents the curated content's topic.

May et al. study in~\cite{social_filter} the interaction between bloggers and users (blog readers), where bloggers decide on which news items to post, and users decide on which bloggers to follow.
The bloggers' objective is to gather as many followers as possible. Their study demonstrates that this "blog-positioning game" can result in "efficient" equilibria, where users generally receive the content of their interest. The model presented in~\cite{social_filter} can be interpreted as a "location game" similar to the one studied in~\cite{jordan2012lattice}.

Dellarocas et al. study in~\cite{media_aggregate} the competition between content creators on the World Wide Web (WWW). Specifically, they analyze how the linking ability of content creators to other web pages affects the strategic interaction among them. The content creators, who curate content from web pages, correspond to the influencer in our model.

Our model and analysis differ from these studies in two key aspects. First, the game we study does not follow the framework of a location game, rendering the analysis conducted in~\cite{jordan2012lattice} inapplicable to our model. Second, we model and analyze the strategic interaction between a single influencer and a large number of content consumers and producers. In contrast, ~\cite{media_aggregate} examines the strategic interaction (competition) among multiple influencers.

\subsubsection{Discussion}

Our model and analysis of the role of social support and influencers includes all aspects of the above literature as it incorporates  cascades (through the connection between the flow of attention and flow of content), attention, social support, and content aggregation and recommendation (through the influencer). This has the following implications. First, our model and analysis shows that these ideas concepts (cascades, attention and social support, and content aggregation) are related; each explaining a  specific aspect of a social media community, and combined providing a comprehensive understanding of the role of social support and influence, in social media. Second, as our model combines all of these concepts, one would expect the resulting model to be more complex compared with the exisiting models in the literature. This is indeed the case (see Section~\ref{sec:model}~-~\ref{sec:problem_formulation}). Third, one would then expect that the model leads to more  complicated results. Surprisingly this is not the case, as the statements of our results in Section~\ref{sec:results} and Section~\ref{sec:results_limited_information} turn out to be  concise, and have intuitive interpretations. We discuss this in more detail in Section~\ref{sec:discussion}.

%\newpage
\section{Social Media Community as a Content Market}\label{sec:model}
In this section, we provide an overview of the model and research questions considered in this paper. The mathematical formulation of the model is detailed in Sections~\ref{sec:math_model}~-~\ref{sec:social_support}.

We model a social media community as a content market consisting of two types of agents: community members and influencers. Community members serve as both content consumers and producers; they post (produce) and read (consume) content. Influencers, by contrast, primarily act as content aggregators, gathering content from the community and making it accessible to all members. While our model can be extended to include influencers as content producers, this does not impact the key results and insights of our analysis.

For instance, on X (formerly Twitter), users join communities to post and engage with content of interest. Influencers, in turn, collect and share content to attract and engage their followers.

The two key concepts that we use in our model are:

\begin{enumerate}
\item[a)] Following Rate: Social media users follow others to receive their content, and the amount of attention they allocate to those they follow reflects their level of interest. This attention influences not only how often a user checks the content but also how carefully they read and engage with it. In our analysis, we use the following rate to model this attention. A higher following rate corresponds to greater attention, indicating more frequent content checks, and shorter delays in receiving new content. We formally define following rates in Section~\ref{sec:math_model} and discuss their impact on the delay in receiving content in Section~\ref{sec:content_consumption_utility}.
\item[b)] Social Support: In addition to following, users engage with content through actions such as sharing (retweeting), liking, or commenting. These actions provide feedback on how interesting or useful the content is, which we refer to as social support. We assume that higher levels of social support—measured by the number of shares, likes, and comments—indicate greater community interest in the content.
\end{enumerate}

Using following rates and social support, we model social media communities as a content market, as explained below.

\subsection{Model of Content Market}

The actions and objectives of the agents in the content market are as follows. Community members, acting as content consumers, aim to obtain content that interests them. They achieve this by following influencers or other community members (content producers). Additionally, they can acquire content from sources outside the community, such as other communities, social media platforms, and websites. Content consumers decide on the following rate (attention) they allocate to influencers, content producers, and external sources. We assume that following rate (attention) is a scarce resource, and each consumer has a limit on the total following rate they can allocate. Furthermore, content consumers provide feedback on content, signaling its utility through actions like 'likes' or sharing (retweeting). This feedback, referred to as social support, indicates the value or interest of the content to the community.

As content producers, community members decide on the type of content to produce. Their goal is to create content that maximizes the social support (likes, retweets, etc.) received within the community.

The actions of influencers are similar to those of content consumers. Influencers decide which content producers to follow and the amount of following rate (attention) to allocate to each. Like content consumers, the following rate (attention) is a limited resource for influencers, and they have a constraint on the total following rate they can allocate. However, while content consumers make these decisions to maximize their personal utility, influencers aim to aggregate content and attract as many followers as possible. They achieve this by maximizing the social support they receive from their followers. Since content consumers can also follow other members and external sources, influencers face competition for their attention. To achieve their goals, influencers can share (retweet) content from the producers they follow. We assume that the probability of an influencer sharing content depends on the following rate allocated to that producer. In other words, if an influencer follows a content producer with a high following rate, they are more likely to share that producer's content. We interpret this sharing as the social support that the influencer provides to the content producer.

In summary, the modeling assumptions that we make for the content market that we study are as follows.
\begin{enumerate}
\item %\subsubsection  
{\it Objective of Content Consumers:}
Content consumers aim to maximize the utility of the content that they receive.  
\item %\subsubsection
  {\it Social Support by Content Consumers:} Content consumers are more likely to provide social support for a content item that provides them a high utility.
\item %\subsubsection
{\it Objective of Content Producers:}
Content producers aim to maximize the social support that they receive in the community.
\item %\subsubsection
{\it Objective of Influencers:}
The objective of the influencer is to obtain as much social support as possible from their followers.
\item %\subsubsection
  {\it Social Support by Influencers:} Influencers are more likely to provide social support to a content item of a community member that they follow with a high rate.
\item %\subsubsection  
{\it Following Rate as a Limited Resource:}
Content consumers and influencers have a limit (bound) on the total rate  (attention)  with which they can follow influencers and other community members. Similarly, influencers have a limit (bound) on the total rate (attention)  with which they can follow community members.
\end{enumerate}

\subsection{Terminology: Content Consumers and Content Producers}
Throughout the paper, we use the following terminology: When discussing the actions of a community member focused on obtaining content, we refer to that member as a "content consumer." Conversely, when focusing on a community member's actions related to creating content, we refer to that member as a "content producer." This terminology helps clearly distinguish between the two primary activities of community members: consuming and producing content.

\subsection{Discussion}

Before presenting the mathematical model for the content market, we will discuss the realism and validity of our modeling assumptions.

We believe that the assumption of a limited following rate (attention) is both valid and realistic. This assumption aligns with observations in the attention economy~\cite{attention_economy}. Similarly, our assumptions about the objectives of content consumers, content producers, and influencers are intuitive and reflect the key motivations of agents within the content market. For instance, social support serves as a measurable metric for content producers and influencers to optimize their decisions and assess performance.

However, it is important to acknowledge that some assumptions regarding social support are idealized and may not perfectly align with reality. For example, while we assume that social support reflects the utility of content, there may be other reasons for providing social support beyond utility. Thus, the model represents an idealized content market where social support directly correlates with utility. Future research could extend this model to scenarios where social support does not necessarily reflect utility, exploring how such variations might affect market equilibrium and social welfare. We discuss these potential extensions in more detail in Section~\ref{sec:conclusion}, including their implications for studying social media algorithms, such as content recommendation systems.

%\newpage
\section{Mathematical Model}\label{sec:math_model}
In this section, we formally define the model outlined in Section~\ref{sec:model}. We begin by specifying the content topics that interest content consumers. While we assume that content consumers share a common set of topics; however, they differ in their primary interest (i.e., their most favored topic). Additionally, we assume that content producers vary in their ability to create content on different topics.

Using the model of this section, we formally define the utilities that content consumers obtain in Section~\ref{sec:content_consumption_utility}. In Section~\ref{sec:social_support} we formally define our model of social support.  In Section~\ref{sec:problem_formulation} we define the objectives of influencers, content consumers, and content producers, as maximization problems.

Our analysis focuses on a content market with a single influencer. However, the model can be extended to include multiple influencers without qualitatively altering the results and insights of the analysis. In the following, we denote the set of community members by $\comSet$ and the influencer by $\infl$. Furthermore, we denote with $N$ the size of the community $\comSet$, i.e. number of community members in $\comSet$.

\subsection{Topics of Interest to Community}
The content market  is defined by a  set of content topics $\setT$ that are of interest to the content consumers. In the following,  we assume that each content item $x$ that is created by a content producer is on one of the topics in $\setT$. In addition, we assume that there exists a "measure" that characterizes how closely two content topics are related.
More precisely, we assume that the set of topics $\setT$ is given by a compact metric space.
\begin{assumption}\label{ass:setT}
The  set $\setT$ of content topics that are of interest to the community is given by a non-empty convex and compact metric space.
\end{assumption}
We denote the distance between two topics $x,x' \in \setT$ as $d(x,x')$.
An example of such is given by a bounded subset $\setT$ of $\setR^N$, i.e. we have that $\setT \subset \setR^N$. In this case, we can think of $x \in \setT$ as a vector of $N$ features that characterize the content topic.

Assumption~\ref{ass:setT} serves both technical and intuitive purposes. Technically, it ensures that there is a solution to the agents' optimization problems within our analysis. Intuitively, it implies that the social media community has a coherent focus on the topics of interest, reflecting a shared thematic orientation among community members.

\subsection{Main Interest of Community Members}
For our analysis, we assume that each community member  (content consumer and producer) has a main interest  $y \in \setT$  which is the content topic of greatest interest to them.  We identify a community member by its main interest $y \in \setT$.

As a content consumer, the probability that a community member  $y$ will be interested in a content item of topic $x$ is given by
\begin{equation}\label{eq:p(x|y)}
	p(x|y) = f(d(x,y)), \qquad x,y \in \setT,
\end{equation}
where $d(x,y)$ is the distance between the main interest  $y$ and the content topic $x$, and  $f:\setR_+  \mapsto [0,1]$, $\setR_+ = \{x \in \setR: x \geq 0\}$, is a decreasing function. This implies that community member $y$ is more interested in content that is close to its main interest $y$.

As a content producer, the ability of community member  $y$ to produce content on topic $x$ is represented by
\begin{equation}\label{eq:q(x|y)}
	q(x|y) = g(d(x,y)), \qquad x,y \in \setT,
\end{equation}
where $g:\setR_+  \mapsto [0,1]$ is again a decreasing function. The interpretation of this function is as follows. If producer  $y$ creates a content item on topic $x$, then this content item will be relevant to topic $x$ with probability $q(x|y)$ given by Eq.~\eqref{eq:q(x|y)}. As the function $g$ is decreasing, content producers are better at producing content that is close to their main interest.

For our analysis, we make the following assumptions for the two function $f$ and $g$.

\begin{assumption}\label{ass:fg}
The functions $f:\setR_+   \mapsto [0,1]$ and $g:\setR_+  \mapsto [0,1]$ are continuous and strictly decreasing.
\end{assumption}
The following result shows that, under Assumption~\ref{ass:setT}~and~\ref{ass:fg}, the topics $x \in \setT$ represent shared interests among community members $y \in \comSet$, where each member has a positive probability of being interested in a topic $x$ and a positive probability of being able to create content relevant to that topic.
\begin{lemma}\label{lemma:pq_min}
There exists a $p_{\min} > 0$ and $q_{\min} > 0$ such that
$$ p(x|y) \geq p_{\min}, \qquad  y \in \Com, x \in \setT$$
and
$$ q(x|y) \geq q_{\min}, \qquad y \in \Com, x \in \setT.$$
\end{lemma}

\begin{proof}
Recall that we have that
$$p(x|y) = f(d(x,y)), \qquad x,y \in \setT,$$
and
$$q(x|y) = g(d(x,y)), \qquad x,y \in \setT.$$

By Assumption~\ref{ass:setT} we have that the metric space $\setT$ is compact, and it follows that there exists a $b_0$ such that
$$ d(z,y) \leq b_0, \qquad  x,y \in \setT.$$

Furthermore, by Assumption~\ref{ass:fg} we have that the functions $f$ and $g$ are decreasing with
$$ f(x) > 0, \qquad x \in \setR_+,$$
and
$$ g(x) > 0, \qquad x \in \setR_+.$$
It then follows that for
$$ p_{\min} = f(b_0)$$
and
$$ q_{\min} = g(b_0),$$
we have that
$$ p_{\min} > 0$$
and
$$ q_{\min} > 0,$$
as well as
$$ p(x|y) \geq p_{\min}, \qquad  y \in \Com, x \in \setT$$
and
$$ q(x|y) \geq q_{\min}, \qquad y \in \Com, x \in \setT.$$
This establishes the result of the lemma.
\end{proof}

\subsection{Content Created by Content Producers}
Each content producer $y \in \comSet$ in the community decides on the topic of content they create, denoted as $x(y) \in \setT$, where $\setT$ is the set of all possible content topics.
The content production decision of each content producer $y \in \comSet$ is strategic, and they aim to select a topic $x(y)$ that will attract the maximum social support from both content consumers and influencers. We define the corresponding (utility) optimization problem in Section~\ref{sec:problem_formulation}.

We then define the overall content production $\contAllocation$ of all producers as 
$$ \contAllocation = (x(y))_{y \in \comSet}$$
where $\comSet$ is the set of all content producers.

\subsection{Following Rates of  Content Consumers}
To obtain content of interest, content consumers can follow content producers, the influencer, and sources outside the community. They can also determine the rate at which they follow these entities. Formally, we define the following rates for a content consumer $y \in \comSet$  as follows:

Let $\mu(z|y)$ be the rate with which content consumer $y$ follows content producer $z$, and let
$$\mu(y) = ( \mu(z|y) )_{z \in \comSet \backslash{\{y\}}}$$
be the rate allocation vector of content consumer $y \in \leafSet$.
Similarly, let  $\mu(\infl|y)$ be the rate with which content consumer $y$ follows the influencer $\infl$.
Finally,  let  $\lambda(y)$ be the rate with which content consumer $y$ follows content sources outside the community.
The overall rate allocation of a content consumer $y$ is then given by
$$\comAllocation(y) = (\mu(y),\mu(\infl|y),\lambda(y)),$$
and the overall rate allocation of all content consumers is given by
$$ \comAllocation = (\comAllocation(y))_{y \in \comSet}.$$

We define in Section~\ref{sec:content_consumption_utility}  the utility obtained by content consumer $y$ under an allocation $\comAllocation(y)$. 
We refer to this utility as the content consumption utility.

We assume that there is a bound  $\leafBudget>0$ on the total rate that a content consumer can allocate, and we have that
$$\mu(\infl|y) + \lambda(y) + \sum_{z \in \Clw{y}} \mu(z|y) \leq \leafBudget, \qquad y \in \comSet.$$
We refer to $\leafBudget$ as the rate budget of the content consumers.

\subsection{Following Rates of Influencer}
Similar to content consumers, the influencer $\infl$ can also decide on the rate $\coreInteract{y}$ with which it follows a content producer $z \in \comSet$ in order to obtain (aggregate) content.
Let 
$$ \coreRateVec = ( \coreInteract{z} )_{z \in \comSet}$$
be the corresponding rate allocation vector of the influencer $\infl$. 

Again, we assume that there is a bound  $\coreBudget>0$ on the total rate that the influence can allocate, and we have that
$$\sum_{z \in \leafSet} \coreInteract{z} \leq \coreBudget.$$ 
We refer to $\coreBudget$ as the rate budget of the influencer.

We define in Section~\ref{sec:problem_formulation} the optimization problem that the influencer uses to choose the rate allocation $\coreRateVec$.

%\newpage
\section{Content Consumption Utility}\label{sec:content_consumption_utility}
In this section, we formally define the utility that content consumers obtain from the content they receive. To model the utility that content consumers obtain from content, we assume they are delay-sensitive: the lower the delay with which a content item is obtained, the higher (less discounted) the utility from this content item. The assumption of delay-sensitivity has been broadly used in modeling social network formation, dating back to the seminal work by Jackson and Wolinsky~\cite{Jackson}.

Recall that content consumers can receive content from the following sources: directly from content producers they follow, indirectly through the influencer, and from content sources outside the community. How content consumers allocate these following rates determines the delay with which they obtain content from these three sources: a higher rate of following a source results in a lower delay in receiving content from that source. Below, we define the utility that content consumers obtain from each of these content sources as a function of their following rate allocation.

\subsection{Utility from Directly Following another Agent}
First, we define the utility that a content consumer $y$ obtains by following a content producer $z$ with rate $\mu(z|y)$. Recall that $x(z)$ is the topic on which content producer $z$ creates content.

Suppose that content consumer $y$ receives a reward of value 1 for each content item that is of interest to them. If content consumer $y$ receives content from content producer $z$ without delay, the (expected) utility of content consumer $y$ for receiving content directly from content producer $z$ is given by
$$ r_p B(z|y),$$
where
$$B(z|y) = q(x(z)|z)  p(x(z)|y)$$
is the probability that a content item from content producer $z$ is of interest to content consumer $y$, and $r_p$ is the rate at which content producer $z$ creates new content.

We assume that the overall utility that a content consumer $y$ receives from content produced by $z$ depends not only on the utility $B(z|y)$ but also on the rate $\mu(z|y)$ with which content consumer $y$ follows producer $z$. The larger the rate $\mu(z|y)$, the higher the overall utility that $y$ receives from $z$. More precisely, if content consumer $y$ follows content from content producer $z$ with rate $\mu(z|y)$, then the (expected) utility rate of content consumer $y$ is given by
\begin{equation}\label{eq:UCd_delay2}
    \begin{split}
        \delayUtility{z}{y},\\
    \end{split}
\end{equation}
where the function $\delay{\mu(z|y)}$ models the sensitivity of the overall utility with respect to the following rate $\mu(z|y)$, and hence the delay with which consumer $y$ receives content from producer $z$. 

For our analysis, we make the following assumption for the function $\delay{\mu}$.
\begin{assumption}\label{ass:delay}
  The function $\delay{\mu}: \setR_+ \mapsto \setR_+$ is strictly increasing and strictly concave with
$$ \delay{0} = 0$$
and
$$ \lim_{\mu \to \infty} \delay{\mu} = 1.$$
Furthermore, the function $\delay{\mu}$ is continuously differentiable. 
\end{assumption}  
Assumption~\ref{ass:delay} implies that a low following rate $\mu$ leads to a higher discounting of the utility $B(z|y)$. 

\subsection{Utility from Following the Influencer}
Next, we define the utility of content consumer $y$ for receiving content from content producer $z$ through the influencer $\infl$. For this, suppose that the influencer $\infl$ follows content producer $z$ with rate $\mu_{\Infl}(z)$, and content consumer $y$ follows the influencer $\infl$ with rate $\mu(\infl|y)$. The utility of content consumer $y$ for receiving content from content producer $z$ via the influencer $\infl$ is then given by
\begin{equation}\label{eq:utilit_influencer}
    \delayCoreUtility{z}{y},
\end{equation}  
where
$$\delayInfl{z}{y} = \delay{\coreInteractVec(z)}\delay{\mu(\infl|y)}.$$

\subsection{Utility from Following Sources outside the Community}
Finally, we define the utility of content consumer $y$ for receiving content from content sources outside the community. For this, let $r_0 > 0$ be the total rate (over all content sources) at which new content items are generated by sources outside the community, and let $B_0$ be the probability that a content item produced from these sources is of interest to content consumer $y$. The utility of content consumer $y$ for receiving content from sources outside the community is given by
\begin{equation}\label{eq:utility_other}
    \alternative{\lambda(y)},
\end{equation}  
where $\lambda(y)$ is the rate with which content consumer $y$ follows sources outside the community. 

Given the rate allocation $\coreInteractVec$ of the influencer $\infl$ and the content production allocation $X$ of the content producers, and rate allocation $\comAllocation(y)$ of content consumer $y$, the total content consumption utility $\leafOptUtility{y}$ of content consumer $y$ is given
\begin{equation}\label{eq:consumer_utility}
	\begin{aligned}
		\leafOptUtility{y}
		= \; &  r_p \sum_{z \in \Clw{y}} \delayCoreUtility{z}{y}  \\
		& +  r_p \sum_{z \in \Clw{y}} \delayUtility{z}{y} \\
		& + \alternative{\lambda(y)}.
	\end{aligned}
\end{equation}

\subsection{Discussion}

We make the following two simplifying assumptions in modeling the consumption utility.

First, we assume that the rate $r_p$ at which content producers create new content is uniform across all producers. This simplification aids both the notation and the analysis. While it is possible to extend the analysis to account for different content creation rates among producers, doing so would introduce additional complexity to the notation and analysis without fundamentally altering the results and insights.

Second, Eq.~\eqref{eq:consumer_utility} implies that consumer $y$ can receive content from producer $z$ both directly and indirectly through the influencer $\infl$. In other words, the utility derived from content produced by producer $z$ is double-counted. To simplify the analysis, we assume that this effect can be disregarded. Specifically, we assume that consumer $y$ primarily obtains content through the influencer $\infl$ and only directly follows a small fraction of content producers. Empirical evidence supports this assumption, showing that community members typically obtain content directly from only a minority of other community members. Moreover, it can be formally demonstrated that this assumption holds when the influencer's rate budget $\coreBudget$ is sufficiently large. Although not making this assumption would complicate the analysis, it would not significantly alter the qualitative results and insights obtained.

Overall, these simplifying assumptions enable a more tractable analysis without compromising the core findings and insights.

\section{Social Support}\label{sec:social_support}
We define social support as the probability with which an agent (either a content consumer or an influencer) provides positive feedback indicating that a content item is of interest to them. In social media, this positive feedback can take the form of a like, share, click, etc.

We assume that the probability that a content consumer $y$ provides positive feedback to a content item from content producer $z$ is proportional to the utility that the content item provides to $y$. More precisely, for a content item obtained by directly following producer $z$, the social support is given by
$$\delayUtility{z}{y},$$
where
$$B(z|y) = q(x(z)|z)  p(x(z)|y).$$
For a content item obtained indirectly through the influencer, the social support is given by
$$\delayCoreUtility{z}{y}.$$ 

Note that both expressions for social support take on a value in $[0,1]$ and can therefore be interpreted as a probability. 

Furthermore, we assume that the probability that the influencer $\infl$ provides positive feedback to a content item from content producer $z$ is proportional to the rate at which the influencer follows content producer $z$, and is given by
$$\delay{\coreInteract{z}}.$$
Note that this expression for social support again takes on a value in $[0,1]$ and can hence be interpreted as a probability. The analysis can be extended to incorporate more general expressions for the social support provided by the influencer. Importantly, the analysis and results remain valid as long as the probability of the influencer providing social support to a content item from producer $z$ is an increasing function of the following rate $\coreInteract{z}$.

%\newpage
\section{Game with Perfect Information}\label{sec:problem_formulation}
In this section, we model the objectives of the influencer, content consumers, and content producers as optimization problems, demonstrating that this results in a strategic interaction (game) among these agents in the content market. We assume that all agents have perfect information, meaning they are aware of the actions and preferences (utilities) of all other agents. In Section~\ref{sec:limited_information}, we relax this assumption and consider the case of imperfect information.

\subsection{Optimal Rate Allocation by Influencer}\label{ssec:core_agent_model}
The objective of the influencer is to maximize the social support received from content consumers who follow the influencer. Using the definition of social support provided in Section~\ref{sec:social_support}, the resulting optimization problem for the influencer is as follows.

Recall that $\comAllocation$ represents the overall rate allocation of all content consumers, $\contAllocation$ represents the content production of all content producers, and $\coreInteractVec$ is the rate allocation vector of the influencer $\infl$. Additionally, recall Eq.~\eqref{eq:utilit_influencer}, which defines the utility that content consumers derive from receiving content through the influencer $\infl$, and the definition of social support provided by content consumers given in Section~\ref{sec:social_support}.

Given the overall rate allocation  $\comAllocation$ and content production $\contAllocation$, the objective of the influencer is then to choose a rate allocation $\coreInteractVec$ that solves the following optimization problem,
\begin{equation}\label{eq:core_opt}
	\begin{aligned}
		& \underset{\coreRateVec \geq 0}{\text{maximize}} && \coreUtility{\comAllocation,\contAllocation} =
          r_p \sum_{z \in \leafSet} \sum_{y \in \Clw{z}}   \delayCoreUtility{z}{y} \\
		& \text{subject to} && \sum_{z \in \leafSet} \coreInteract{z} \leq \coreBudget, 
	\end{aligned}
\end{equation}
where 
$$B(z|y) = q(x(z)|z) p(x(z)|y)$$
and
$$\delayInfl{z}{y} = \delay{\coreInteractVec(z)}\delay{\mu(\infl|y)},$$
and $\coreBudget>0$ is the rate budget available to the influencer for following content producers.
We refer to this maximization problem as $\optcore{\comAllocation,\contAllocation}$.

Note that in the case where
$$\mu(\infl|y) = 0, \qquad y \in \comSet,$$
the maximization problem $\optcore{\comAllocation,\contAllocation}$ is not well defined as any allocation $\coreInteractVec$ that satisfies the rate constraints is an optimal solution. To avoid this ambiguity, we make the following assumption for our analysis. 
\begin{assumption}\label{ass:infl_alloc}
  If for the optimization problem  $\optcore{\comAllocation,\contAllocation}$ given by Eq.~\eqref{eq:core_opt} we have that
$$\sum_{y \in \comSet} \mu(\infl|y) = 0,$$
  then for optimal allocation
  $$\coreInteractVecOpt = \arg\max_{\coreInteractVec \geq 0} \optcore{\comAllocation,\contAllocation}$$
we set
$$  \coreInteractVecOpt(z) = \coreBudget/N, \qquad z \in \comSet,$$
where $N$ is the size of the community $\comSet$. 
\end{assumption}

\subsection{Optimal Rate Allocation by  Content Consumers}\label{ssec:consumer_model}
The objective of content consumers $y \in \comSet$  is to obtain content that maximizes the content consumption utility given by Eq.~\eqref{eq:consumer_utility}.

Recall that $\leafOptUtility{y}$ given by Eq.~\eqref{eq:consumer_utility} is the content consumption utility of content consumer $y$ under the rate allocation $\coreInteractVec$ of the influencer $\infl$, the content production $X$ of all content producers.
The objective of content consumer $y$ is then to choose a rate allocation $\comAllocation^*$ that solves the following optimization problem,
\begin{equation}\label{eq:consumer_opt}
	\begin{aligned}
		& \underset{\comAllocation(y) \geq 0}{\text{maximize}}
		& & \leafOptUtility{y}\\
		& \text{subject to} &&\mu(\infl|y) + \lambda(y) + \sum_{z \in \Clw{y}} \mu(z|y) \leq \leafBudget,
	\end{aligned}
\end{equation}
where $\leafBudget>0$ is the rate budget available to content consumer $y$.  We refer to this maximization problem as the consumption utility maximization problem  $\optperiphery{\coreRateVec,\contAllocation}{y}$.

\subsection{Optimal Content Production by Content Producers}\label{ssec:producer_model}
The objective of content producer $z \in \comSet$ is to create content on a topic $x(y)$  that maximizes the social support that $z$ receives from the content consumers. Using the definition of the social support as given in Section~\ref{sec:social_support}, the resulting optimization  problem is given as follows.

Using the definitions given  in Section~\ref{sec:social_support}, the social support $\prodOptUtility{z}$ that producer $z$ receives based on the rate allocation $\coreInteractVec$ of the influencer $\infl$ and the overall rate allocation $\comAllocation$ of all content consumers, is given by
\begin{equation}\label{eq:producer_utility}
	\begin{aligned}
	  \prodOptUtility{z}
		= \; &   r_p \sum_{y \in \Clw{z}} \delayCoreUtility{z}{y}  \\
		& + r_p \sum_{y \in \Clw{z}} \delayUtility{z}{y}.
	\end{aligned}
\end{equation}
The objective of content producer $z$ is then to create content on topic $x^*(z)$  that solves the following optimization problem,
\begin{equation}\label{eq:producer_opt}
	\begin{aligned}
		& \underset{x(z) \in \setT}{\text{maximize}}
		& & \prodOptUtility{z}.
	\end{aligned}
\end{equation}
We refer to this maximization problem as $\optcontperiphery{\coreRateVec,\comAllocation}{z}$.

\subsection{Nash Equilibrium}\label{ssec:stable}
Note that the optimal solution of the maximization problem $\optcore{\comAllocation,\contAllocation}$ of the influencer depends on the overall rate allocation  $\comAllocation$ of the content consumers and the content production $\contAllocation$ of all content producers. Similarly, the optimal solution of the maximization problem $\optperiphery{\coreRateVec,\contAllocation}{y}$ of content consumer $y$ depends on the rate allocation  $\coreInteractVec$ of the influencer and the content production $\contAllocation$ of all content producers.
Finally, the optimal solution of the maximization problem $\optcontperiphery{\coreRateVec,\comAllocation}{z}$ of content producer $z$ depends on the rate allocation  $\coreInteractVec$ of the influencer and the overall rate allocation $\contAllocation$ of all content consumers.
This coupling creates a strategic interaction between the influencer and community members. A Nash equilibrium for the resulting game is given as follows.

Let $\allAllocation = (\coreInteractVec,\comAllocation,\contAllocation) $ be a given allocation that characterizes the rate allocation $\coreInteractVec$ of the influencer, and the overall rate allocation  $\comAllocation$ of all content consumers, and the content allocation $X$ of all content producers. 
\begin{definition}\label{def:stable_allocation}
	An allocation $\allAllocationNE  = (\coreInteractVecOpt,\comAllocation^*,\contAllocation^*)$ is a Nash equilibrium if we have that
	\begin{enumerate}
		\item $\coreInteractVecOpt$ is a solution for the maximization problem $\optcore{\comAllocation^*,\contAllocation^*}$,
		\item $\comAllocation^*(y)$, $y \in \comSet$,  is a solution to the maximization problem $\optperiphery{\coreInteractVecOpt,\contAllocation^*}{y}$,
		and
		\item $x^*(z)$, $z \in \comSet$,  is a solution to the maximization problem  $\optcontperiphery{\coreInteractVecOpt,\comAllocation^*}{z}$.
	\end{enumerate}  
\end{definition}
Definition~\ref{def:stable_allocation} states that under a Nash equilibrium $\allAllocationNE = (\coreInteractVecOpt,\comAllocation^*,\contAllocation^*)$ no agent in the content market   is able to increase the value of their objective function by unilaterally changing their allocation.
In the next two sections, we analyze the existence and properties of a Nash equilibrium.

%\newpage
\section{Results on  Nash Equilibrium for Perfect Information}\label{sec:results}
We first analyze the Nash equilibrium. Note that this equilibrium assumes that all agents have perfect information when deciding on their allocations. For this case, we analyze the existence and efficiency of a Nash equilibrium  $\allAllocationNE$.  We measure the efficiency of a Nash equilibrium by the social welfare (sum of the utilities) over all content consumers.

\subsection{Existence and Efficiency of a  Nash Equilibrium  $\allAllocationNE$}

Our first result states that there always exists a Nash equilibrium.
\begin{prop}\label{prop:existence_NE}
	There exists  a Nash equilibrium  $\allAllocationNE  =  (\coreInteractVecOpt,\comAllocation^*,\contAllocation^*)$.
\end{prop}
We provide a proof for Proposition~\ref{prop:existence_NE} in Appendix~\ref{proof:prop1}.
To prove the proposition, we show that the game defined in Section~\ref{sec:problem_formulation} is a potential game for the  potential function 
\begin{equation}\label{eq:social_welfare}
	\Phi(\allAllocation) = \sum_{y \in \leafSet}  \leafOptUtility{y},
\end{equation}
where $ \leafOptUtility{y}$ is the consumption utility of community member $y$ as given by Eq.~\eqref{eq:consumer_utility}. Note that the potential function $\Phi(\allAllocation)$ is the total consumption utility over all content consumers which contains both the utility obtained from content producers and the influencer, as well as content from other sources. In the following we refer to $\Phi(\allAllocation)$ as the social welfare.

The next result states that there always exists a Nash equilibrium that maximizes the social welfare. To state the result, we use the following notation.
Let $\possibleStrategy$ be the set of all admissible allocations  $\allAllocation = (\coreInteractVec,\comAllocation,\contAllocation)$, i.e. the set of all allocations such  the rate allocation  $\coreInteractVec$ is given by a non-negative vector that satisfy the rate budget constraints $\coreBudget$ of the influencer, $\comAllocation(y)$, $y \in \comSet$ is a non-negative vector that satisfies the rate budget constraint $\leafBudget$ for the content consumers, and we have that $x(z) \in \setT$, $z \in \comSet$.

Using the fact that the  social welfare $\Phi(\allAllocation)$ is a potential function, we obtain the following result.
\begin{prop}\label{prop:social_welfare}
	There exists a Nash equilibrium $\allAllocationWelfare  =  (\coreInteractVecOpt,\comAllocation^*,\contAllocation^*)$  that maximizes the social welfare, i.e. we have that
	$$\allAllocationWelfare = \argmax_{\allAllocation \in \possibleStrategy} \Phi(\allAllocation).$$
\end{prop}
We provide a proof for Proposition~\ref{prop:social_welfare} in Appendix~\ref{proof:prop2}.

\subsection{Efficiency of Nash Equilibria  $\allAllocationNE$}

Proposition~\ref{prop:social_welfare} states that there is a Nash equilibrium $\allAllocationWelfare$ that maximizes the social welfare given by Eq.~\eqref{eq:social_welfare}. However, it does not state that all Nash equilibria $\allAllocationNE$ maximize the social welfare, nor does it provide conditions under which the community interactions converge to a Nash equilibrium $\allAllocationWelfare$. It is possible to provide sufficient conditions for all Nash equilibria to be efficient, and for the community to converge to an efficient Nash equilibrium. Roughly, these conditions impose additional assumptions on the distribution of the community members within the community such that the distribution is unimodal. While this analysis is possible, it is outside the scope of this paper. It is more challenging to identify necessary conditions for all Nash equilibria to be efficient and for an efficient Nash equilibrium to emerge. This question is an open problem.

%\newpage
\section{Game with Imperfect Information}\label{sec:limited_information}
In Sections~\ref{sec:problem_formulation}~and~\ref{sec:results}, we analyzed the existence and efficiency of a Nash equilibrium for the case where content producers have perfect information, i.e. they know the actions and preferences (utilities) of all other agents and have access to the social support that they receive from all content consumers $y \in \comSet$. 
In this section, we consider the more realistic case where agents do not have perfect information. In particular, we assume that content producers can not observe the actions of all content consumers, and hence can not directly observe the social support that they obtain from the content consumers. Instead, content producers rely on the social support obtained from the influencer to decide which content to create. In this section, we define the resulting game based on this interaction. In Section~\ref{sec:results_limited_information} we analyze the existence of a Nash equilibrium for this game, and in Section~\ref{sec:efficiency_NE_Infl} we characterize whether, and by how much, the outcome of the case with limited information deviates from market equilibrium from the idealized of perfect information.

In the following we define the allocation decisions by the influencer, content consumers, and content producers, for the game with imperfect information.

\subsection{Optimal Rate Allocation by Influencer}
We assume that the influencer $\infl$ is able to observe the social support received from the content consumer who follow the influencer.
The influencer $\infl$  then use the maximization  problem $\optcore{\comAllocation,\contAllocation}$  given by Eq.~\eqref{eq:core_opt} to decide on its allocation $\coreRateVec$ by setting
\begin{equation*}
	\begin{aligned}
		& \underset{\coreRateVec \geq 0}{\text{maximize}} && \coreUtility{\comAllocation,\contAllocation} \\
		& \text{subject to} && \sum_{z \in \leafSet} \coreInteract{z} \leq \coreBudget, 
	\end{aligned}
\end{equation*}
where the objective function $\coreUtility{\comAllocation,\contAllocation}$ is as given by Eq.~\eqref{eq:core_opt}.

\subsection{Optimal Rate Allocation by Content Consumers}
We assume that content consumers can observe the content that content producers create, i.e. they can observe the content allocation $\contAllocation$. One way to achieve this is by following the influencer, and observing the content $x(z)$ of producer $z$ that the influencer shares. Using this assumption, content consumer $y \in \comSet$  chooses  an allocation $\comAllocation(y)$
that maximizes $\optperiphery{\coreRateVec,\contAllocation}{y}$ given by
\begin{equation*}
	\begin{aligned}
		& \underset{\comAllocation(y) \geq 0}{\text{maximize}}
		& & \leafOptUtility{y}\\
		& \text{subject to} &&\mu(\infl|y) + \lambda(y) + \sum_{z \in \Clw{y}} \mu(z|y) \leq \leafBudget.
	\end{aligned}
\end{equation*}

\subsection{Optimal Content Production by Content Producers}\label{ssec:inf_opt_content}
We assume that content producers can not directly observe the social support that they receive, and instead use the social support obtained by the influencer as a proxy for the social support their content receives from the content consumers. To formulate the resulting optimization problem, we use the following notation. 

For a given content production vector $\contAllocation$, we use  $X_{\{z,x(z)'\}}$ to denote that the content created by content producer $z \in \comSet$ changed from $x(z)$ to $x(z)')$, while the content created by all other content producers is the same.

Furthermore, for a given rate allocation $\comAllocation$ of the content consumers and a given content allocation $\contAllocation$ of the content producers, let  $\coreInteractVecOpt(\comAllocation,\contAllocation)$ denote  the optimal rate allocation of the influencer $\infl$ as given by Eq.~\eqref{eq:core_opt}. Furthermore, let $\coreInteractVecOpt(z|\comAllocation,\contAllocation)$ be the following rate that the influencer $\infl$ allocates to content producer  $z$ under $\coreInteractVec(\comAllocation,\contAllocation)$.

Using these definitions, the objective of content producer $z$ is to create content on the topic $x^*(z)$  which maximize the social support received from the influencer, i.e. we have that
\begin{equation}\label{eq:max_attention_community}
	x^*(z) =
	\argmax_{x(z) \in \setT} \delay{\coreInteractVecOpt(z|\comAllocation,\contAllocation_{\{z,x(z)\}})}.
\end{equation}

\subsection{Nash Equilibrium}
We define a Nash equilibrium for the case of limited information as follows. 

\begin{definition}\label{def:att_NE}
	An allocation $\allAllocationInflNE  = (\coreInteractVecOpt,\comAllocation^*,\contAllocation^*)$ is a Nash equilibrium  if we have that
	\begin{enumerate}
	\item $\coreInteractVecOpt$ is a solution
          for the maximization problem $\optcore{\comAllocation^*,\contAllocation^*}$,
		\item $\comAllocation^*(y)$, $y \in \comSet$,  is a solution to the maximization problem $\optperiphery{\coreInteractVecOpt,\contAllocation^*}{y}$,
		and
	      \item $x^*(z) =
	\argmax_{x(z) \in \setT} \delay{\coreInteract{z|\comAllocation^*,\contAllocation^*_{\{z,x(z)\}}}}$, $z \in \comSet$.
	\end{enumerate}  
\end{definition}
In the following, we refer to a Nash equilibrium $\allAllocationInflNE  = (\coreInteractVecOpt,\comAllocation^*,\contAllocation^*)$  as given by Definition~\ref{def:att_NE} as a Nash equilibrium for the game with imperfect information, and to a Nash equilibrium $\allAllocationNE  = (\coreInteractVecOpt,\comAllocation^*,\contAllocation^*)$  as given by Definition~\ref{def:stable_allocation} as a Nash equilibrium for the game with perfect information.

\section{Results on Nash Equilibrium for Limited Information}\label{sec:results_limited_information}
In this section we analyze the existence of a Nash equilibrium for the game with imperfect information as given by Definition~\ref{def:att_NE}, and characterize the optimal allocation $x^*(z|\comAllocation,\contAllocation)$ of  content producer $z$ (see Eq.~\eqref{eq:max_attention_community}) under a Nash equilibrium.

\subsection{Existence of a Nash Equilibrium $\allAllocationInflNE$}
Our first result states that there always exists a Nash equilibrium given by Definition~\ref{def:att_NE}.
\begin{prop}\label{prop:att_existence_NE}
	There exists  a Nash equilibrium  $\allAllocationInflNE  =  (\coreInteractVecOpt,\comAllocation^*,\contAllocation^*)$ as given by Definition~\ref{def:att_NE}.
\end{prop}
We provide a proof for Proposition~\ref{prop:att_existence_NE} in Appendix~\ref{proof:prop4}.
%Proposition~\ref{prop:att_existence_NE} is obtained by using Brouwer's fixed-point theorem as the underlying
We note that the game with imperfect information as given by Definition~\ref{def:att_NE} is no longer a potential game with the social welfare  $\Phi(\allAllocation)$ as a potential function. As a result, the proof for Proposition~\ref{prop:att_existence_NE}  differs from the one given for Proposition~\ref{prop:existence_NE} for the game with perfect information.

\subsection{Optimal Decision by Content Producers}
We next characterize the optimal allocation $\coreInteractVecOpt$ of the influencer under a Nash equilibrium as given by Definition~\ref{def:att_NE}. 

Let 
\begin{equation}\label{eq:inf_prodUtility}
	\begin{aligned}
		\prodOptUtilityInf{z}
		= & r_p \sum_{y \in \Clw{z}}   \delayCoreUtility{z}{y},
	\end{aligned}
\end{equation}
be the production utility that producer $y$ receives through the  influencer $\infl$ under the allocation $(\comAllocation,\contAllocation)$ (see Eq.~\eqref{eq:producer_utility}). 
We then obtain the following result.
\begin{prop}\label{prop:att_contProducer}
If $\allAllocationInflNE = (\coreInteractVecOpt,\comAllocation^*,\contAllocation^*)$ is a Nash equilibrium as given by Definition~\ref{def:att_NE} such that for content producer $z \in \comSet$ we have
  $$\coreInteractVecOpt(z) > 0$$
 and
 $$\sum_{y \in \Cw{z}} \mu^*(\infl|y) > 0,$$
then we have 
	$$x^*(z)  =  \argmax_{x(z) \in \setT} \prodOptUtilityInfNE{z}.$$
\end{prop}
We provide a proof for Proposition~\ref{prop:att_contProducer} in Appendix~\ref{proof:prop3}.

Proposition~\ref{prop:att_contProducer} states that a content topic $x^*(z)$ of content producer $z \in \comSet$ under a Nash equilibrium  $\allAllocationInflNE = (\coreInteractVecOpt,\comAllocation^*,\contAllocation^*)$ maximizes the social support $\prodOptUtilityInfNE{z}$ that producer $z$ receives from content shared by the influencer $\infl$, as given by Eq.~\eqref{eq:inf_prodUtility}.

It is important to note that $\prodOptUtilityInf{z}$ differs from the overall social support $\prodOptUtility{z}$ (given by Eq.\eqref{eq:producer_utility}) that content producer $z$ receives in the community, since $\prodOptUtilityInf{z}$ excludes the social support that content producer obtains  from content consumers $y \in \comSet$ who follow producer $z$, and we have $\mu(z|y) > 0$.

\section{Efficiency of Nash Equilibrium $\allAllocationInflOpt$}\label{sec:efficiency_NE_Infl}
Let $\setNEInfl$ be the set of all Nash equilibria $\allAllocationInfl$ as given by Definition~\ref{def:att_NE}, and let
\begin{equation}\label{eq:opt_NE_infl}
	\allAllocationInflOpt = \argmax_{\allAllocationInfl \in \setNEInfl} \Phi(\allAllocationInfl)
\end{equation}
be a Nash equilibrium that maximizes the social welfare over all Nash equilibria as given by Definition~\ref{def:att_NE}.

Proposition~\ref{prop:att_contProducer} states  that by maximizing the social support from the influencer $\infl$, a content producer  $z \in \comSet$  maximizes at a Nash equilibrium the production utility  $\prodOptUtilityInfNE{z}$ that content producer  $z$ obtains from its content that is shared by the influencer $\infl$. An optimal allocation $x^*(z)$ for the production utility  $\prodOptUtilityInfNE{z}$ generally does not maximize the consumption utility over all content consumers (as it ignores the utility of consumers who follow the producer $z$ directly), and hence generally does not maximize the social welfare.

As a result, a Nash equilibrium $\allAllocationInflOpt$ as given by Eq.~\eqref{eq:opt_NE_infl} generally does not maximize the socially optimal allocation as given in Proposition~\ref{prop:social_welfare}, and the social welfare $\Phi(\allAllocationInflOpt)$ obtained under a Nash equilibrium  $\allAllocationInflOpt$  as given by Eq.~\eqref{eq:opt_NE_infl} is generally strictly smaller than the social welfare $\Phi(\allAllocationWelfare)$ obtained under a socially optimal Nash equilibrium  $\allAllocationWelfare$ as given by Proposition.~\ref{prop:social_welfare}. In the following, we analyze the efficiency of an optimal Nash equilibrium  $\allAllocationInflOpt$  as given by Eq.~\eqref{eq:opt_NE_infl}.

To characterize the efficiency of a Nash equilibrium as given by Definition~\ref{def:att_NE}, we study the case where the size $N$ of the community and the rate budget $\coreBudget$ of the influencer are large, and show that for this case that a Nash equilibrium  $\allAllocationInflOpt$ as given by Eq.~\eqref{eq:opt_NE_infl} is socially optimal.
We obtain the following result.
\begin{prop}\label{prop:allocation_inf_rate_Infl}
Let $\leafBudget$ be a given rate budget for the content consumers. 
Then there exists a integer $N_0$ and a constant $k_{\Infl}$ such that if $N > N_0$,  and $\coreBudget > N k_{\Infl}$, the following is true.
For an optimal  Nash equilibrium
$$\allAllocationInflOpt =  (\coreInteractVecOpt,\comAllocation^*,\contAllocation^*)$$
for the game with imperfect information as  given by Eq.~\eqref{eq:opt_NE_infl} we have that
	$$\Phi \Big (\allAllocationInflOpt \Big ) = \Phi \Big (\allAllocationWelfare \Big ).$$
\end{prop}

Proposition~\ref{prop:allocation_inf_rate_Infl} is the key result of our analysis.
It provides the critical insight that, for a sufficiently large community size $N$ and a large enough influencer rate budget $\coreBudget$, the following holds: 
\begin{enumerate}
\item[a)] If the influencer allocates its following rate based on the social support received from content consumers, and
\item[b)]  If content producers use  the social support from  the influencer to decide which type of content to produce,
\end{enumerate}
then this dynamic leads to an efficient outcome, where a  Nash equilibria $\allAllocationInflOpt$ as given by Eq.~\eqref{eq:opt_NE_infl} maximizes the social welfare.
This highlights the fundamental role that social support (as a form of currency) and the influencer (as an information proxy) play in coordinating the actions of individual agents in the content market.
We discuss this result in more detail in Section~\ref{sec:discussion}.

Proposition~\ref{prop:allocation_inf_rate_Infl} is also technically the most challenging to prove. The proof is provided in Appendix~\ref{app:prove_allocation_inf_rate_Infl}.

%\newpage
\section{Discussion of Results}\label{sec:discussion}
In this section, we discuss the results obtained in Sections~\ref{sec:results}, ~\ref{sec:results_limited_information}, and ~\ref{sec:efficiency_NE_Infl}. We explore the insights these results provide into the fundamental roles of social support (as a form of currency) and the influencer (as an information proxy) in coordinating the actions of individual agents within the content market.

\subsection{Social Support as a  Currency}
Proposition~\ref{prop:social_welfare} establishes the existence of a Nash equilibrium that maximizes social welfare, demonstrating that social support leads to an efficient outcome. This result provides a formal understanding of social support as a currency within the content market under consideration. Specifically, it shows that achieving an efficient market equilibrium requires the following three conditions: First, the social support provided by content consumers must effectively signal the utility they derive from content. Second, content producers must aim to maximize the social support they receive for their content. Finally, influencers should strive to maximize the social support obtained from their followers. Deviations from these three conditions are likely to result in suboptimal market outcomes. We discuss how these insights can be applied to study algorithms on social media platforms, such as content recommendation algorithms, in Section~\ref{sec:conclusion}.

\subsection{Role of Influencer}\label{ssec:influencers}
In the content market, influencers serve two crucial roles. First, they act as aggregators of content tailored to the interests of their followers, facilitating the discovery of relevant content by consumers. This role is widely recognized and aligns with the established understanding of influencers in social media and content markets. However, influencers also assume another significant role. In large content markets, where content producers face substantial costs in tracking all content consumers to gather relevant information for content creation decisions, influencers can act as proxies for this information. Specifically, content producers can use the social support received by an influencer as an approximation of the overall social support they would receive from the entire consumer base. The higher the social support garnered by a content item from an influencer, the higher the anticipated level of social support from all content consumers. By leveraging influencers as information proxies, content producers can make nearly optimal decisions, resulting in an efficient market outcome. This particular role of influencers has not been formally characterized in existing literature.

\subsection{Flow of Social Support and Flow of Content}
Our analysis reveals two distinct flows within the content market: the flow of social support and the flow of content. The flow of social support originates from content consumers, moves towards the influencer, and then passes to the content producers. Conversely, the flow of content starts with the content producers, moves through the influencer, and reaches the content consumers.

This structure resembles a market economy, where there are two flows: money flowing in one direction and goods in the opposite direction. Similarly, in the content market, the flow of social support from consumers to the influencer, and then from the influencer to the content producers, is coupled with the flow of content. This coupling operates in both directions: (a) the flow of social support determines the flow of content, and (b) the flow of content determines the flow of social support.

To elaborate on this coupling, we specify that the flow of social support influences the flow of content in the following ways:
\begin{itemize}
\item The social support from content consumers determines the content that the influencer shares.
\item The social support received by the influencer affects the content created by the producers.
\end{itemize}

Similarly, the flow of content influences social support as follows:
\begin{itemize}
\item The utility of the content created by a producer determines the social support that the producer receives from the influencer.
\item The utility of the content shared by the influencer determines the social support the influencer receives from content consumers.
\end{itemize}

A Nash equilibrium in this context corresponds to a fixed point where these two flows are balanced, similar to the equilibrium of money and goods flow in a market economy.

Just as money signals the utility of goods in a market economy, social support signals the utility of content. The social support provided by a content consumer reflects the (personal) utility of a content item to that consumer, while the social support received by the influencer signals the aggregated utility of content to the influencer's followers (see Proposition~\ref{prop:att_contProducer}).

It is worth noting that existing cascade models in the literature primarily focus on how the flow of social support determines the flow of content~\cite{survey_cascade_models,Watts_cascade,threshold_models}. In contrast, our analysis identifies and studies the second flow, namely how the flow of content shapes the flow of social support. Moreover, we demonstrate that these two flows are interlinked as outcomes of the strategic interactions among content consumers, content producers, and influencers.

\subsection{Price of Influence}
Proposition~\ref{prop:social_welfare} demonstrates that an efficient market equilibrium is achieved when agents have perfect information. However, in the scenario discussed in Section~\ref{sec:limited_information}, where content producers have limited information, this is no longer the case. Specifically, using the influencer as a proxy to gauge social support leads to suboptimal decisions by content producers. Consequently, the social welfare $\Phi(\allAllocationInflOpt)$ achieved under a Nash equilibrium, as given by Eq.~\eqref{eq:opt_NE_infl}, is generally strictly lower than the social welfare $\Phi(\allAllocationWelfare)$ achieved under a socially optimal allocation, as given by Proposition~\ref{prop:social_welfare}.

We refer to the reduction $|\Phi(\allAllocationWelfare) - \Phi(\allAllocationInflOpt)|$ in social welfare as the ``price of influence.''

Proposition~\ref{prop:allocation_inf_rate_Infl} helps identify situations where the price of influence is low, meaning the influencer's role as a proxy results in an equilibrium that is close to maximizing social welfare. This proposition suggests that the price of influence decreases when the influencer's rate budget $\coreBudget$ is significantly larger than the rate budget $\leafBudget$ of the community members. Thus, in large communities where the influencer's rate budget greatly exceeds that of the community members, using the influencer as an information proxy is effective. This approach allows content producers to gather necessary information for decision-making by following a single agent (i.e., the influencer) rather than all consumers, leading to a more efficient market outcome.

%\newpage
\section{Conclusion}\label{sec:conclusion}
This paper presents a formal study of the role played by social support and influencers within a content market. The content market being considered here models a community on a social media platform where users both create and consume content of interest to them, while influencers curate and distribute content to attract followers. Our analysis provides a characterization of social support as a currency within this market, and highlights the influencers' dual role as content aggregators and information proxies for social support.

The paper's most significant impact lies in its potential to transform how social media platforms and multi-agent systems are understood and designed. By modeling social support as a currency, the paper provides a novel and powerful framework to analyze the impact of social media content recommendation algorithms on the flow of content within social media. Additionally, the insights gained from this work have the potential to transform the understanding and design of decentralized AI systems and other multi-agent environments where coordination and strategic interaction are critical. We discuss this in more detail below, together with additional important directions for future research.

\subsection{Emergence of Influencers}
Our analysis assumes the existence of an influencer whose objective is to maximize the social support received from their followers. This leads to two crucial questions:
\begin{enumerate}
\item Whether and how such an influencer emerges in a content market.
\item Whether and why the influencer chooses to maximize the social support received from their followers as their objective.
\end{enumerate}

We are currently modeling and analyzing the formation of a content market. Preliminary results demonstrate that an influencer indeed emerges once the content market reaches a certain size. Moreover, under suitable assumptions, the influencer derives significantly higher utility from sharing content rather than creating it. Thus, it is rational for an influencer to focus on maximizing social support through content sharing. The influencer achieves this by aggregating content from producers in a way that maximizes the social support received from their followers. These results provide a formal justification for the assumptions regarding the role and objectives of influencers in our model.

\subsection{Empirical Validation}
Using social media communities as a concrete scenario allows us to test how agents coordinate their actions in a distributed manner. The activities of users on online social media platforms can be accessed and analyzed, providing a way to empirically test and validate the theoretical results presented in this paper. We are currently conducting this empirical study, and preliminary results indicate that the findings from our theoretical analysis are observable in real-life social media communities. This ongoing research will be detailed in a forthcoming paper.

\subsection{Analyzing Social Media Algorithms}
The finding that social support acts as a currency within a content market offers a fresh perspective on social media algorithms, including content recommendation algorithms. Specifically, it suggests that social media algorithms can be viewed as actors or mechanisms that influence the flow of social support and content. Similar to a market economy, one can investigate social media algorithms by characterizing their impact on a) the flow of social support, b) the flow of content, and c) the market equilibrium and the social welfare achieved at the equilibrium.
Furthermore, it suggests that social media platforms, equipped with complete market information, can utilize this insider information to "manipulate" the content market in their favor, resulting in high utility for the platform at the expense of reduced social welfare and user utilities. Exploring these issues constitutes an important direction for future research.

\subsection{Application to General Multi-Agent Systems}
A critical direction for future research is to explore whether and how social support, as a mechanism for coordinating actions, can be applied to more general multi-agent systems, such as multi-agent AI systems or distributed computing systems. These systems often face similar challenges related to coordination and strategic interaction among agents.

The central idea is to investigate whether a form of "social support" can emerge in these environments, where agents provide feedback or rewards based on each other’s contributions to shared goals or system-wide efficiency. This concept could be analogous to how likes or shares signal the value of content in social media.

In multi-agent AI systems, this mechanism could enhance coordination in scenarios where agents must balance individual and collective objectives. For instance, in collaborative tasks or competitive environments like resource allocation or complex strategy games, social support could serve as a feedback-based incentive that helps agents adapt their strategies based on positive reinforcement from others in the system. This could lead to more effective decision-making in settings such as multi-agent reinforcement learning, where agents learn to cooperate or compete within shared environments.

This direction is ongoing research that we are currently undertaking, focusing on the application of social support mechanisms to multi-agent systems and exploring their impact on coordination and strategic interaction among agents.

\subsection{Extensions of the Model and Analysis}
Finally, there are several possible extensions and generalizations of the model and analysis presented in this paper. For instance, while Proposition~\ref{prop:social_welfare} states the existence of a Nash equilibrium $\allAllocationWelfare$ that maximizes social welfare (as defined by Eq.~\eqref{eq:social_welfare}),
it does not guarantee that all Nash equilibria $\allAllocationNE$ maximize social welfare, nor does it provide conditions under which community interactions converge to the Nash equilibrium $\allAllocationWelfare$. It is possible to establish sufficient conditions for all Nash equilibria to be efficient and for the community to converge to an efficient Nash equilibrium. However, identifying the necessary conditions for an efficient Nash equilibrium to emerge remains an open problem.

Proposition~\ref{prop:allocation_inf_rate_Infl} states that in the limit of an infinite rate budget, we have that all Nash equilibria $\allAllocationInflOpt$ as given by Eq.~\ref{eq:opt_NE_infl} are efficient in the sense that they maximize the social welfare.
However, it does not provide a characterization of how quickly the allocation $\allAllocationInflOpt$ converges to the socially optimal allocation $\allAllocationWelfare$, nor does it provide a bound on the price of influence $|\Phi(\allAllocationWelfare) - \Phi(\allAllocationInflOpt)|$ in communities where the price of influence is non-zero. This is another important direction for future research.

%\newpage
\bibliographystyle{unsrt}
\bibliography{content/main}

\begin{thebibliography}{10}

\bibitem{sl_multi1}
Jason Noble and Daniel~W Franks.
\newblock Social learning in a multi-agent system.
\newblock {\em Computing and Informatics}, 22(6):561--574, 2003.

\bibitem{sl_multi2}
Natasha Jaques, Angeliki Lazaridou, Edward Hughes, Caglar Gulcehre, Pedro
  Ortega, DJ~Strouse, Joel~Z Leibo, and Nando De~Freitas.
\newblock Social influence as intrinsic motivation for multi-agent deep
  reinforcement learning.
\newblock In {\em International Conference on Machine Learning}, pages
  3040--3049. PMLR, 2019.

\bibitem{dist}
S~Kartik and C~Siva~Ram Murthy.
\newblock Task allocation algorithms for maximizing reliability of distributed
  computing systems.
\newblock {\em IEEE Transactions on Computers}, 46(6):719--724, 1997.

\bibitem{durkheim}
Emilie Durkheim.
\newblock {\em The Division of Labor in Society}.
\newblock 1893.

\bibitem{giddens}
A.~Giddens.
\newblock {\em The Constitution of Society: Outline of the Theory of
  Structuration}.
\newblock Outline of the Theory of Structuration. University of California
  Press, 1984.

\bibitem{bourdieu}
P.~Bourdieu and L.J.D. Wacquant.
\newblock {\em An Invitation to Reflexive Sociology}.
\newblock University of Chicago Press, 1992.

\bibitem{media_aggregate}
Chrysanthos Dellarocas, Zsolt Katona, and William Rand.
\newblock Media, aggregators, and the link economy: Strategic hyperlink
  formation in content networks.
\newblock {\em Management Science}, 59(10):2360--2379, 2013.

\bibitem{jordan2012lattice}
Patrick~R Jordan, Uri Nadav, Kunal Punera, Andrzej Skrzypacz, and George
  Varghese.
\newblock Lattice games and the economics of aggregators.
\newblock In {\em Proceedings of the 21st international conference on World
  Wide Web}, pages 549--558. ACM, 2012.

\bibitem{link_aggregator}
Chrysanthos Dellarocas, Zsolt Katona, and William Rand.
\newblock Media, aggregators and the link economy: Strategic hyperlink
  formation in content networks.
\newblock {\em School of Management Research Paper}, 30:06--131, 2010.

\bibitem{attention_economy}
Thomas~H Davenport and John~C Beck.
\newblock The attention economy.
\newblock {\em Ubiquity}, 2001(May):1--es, 2001.

\bibitem{Milgrom}
Paul Milgrom and John Roberts.
\newblock An economic approach to influence activities in organizations.
\newblock {\em American Journal of Sociology}, 94:S154--S179, 1988.

\bibitem{parsons_influence}
Talcott Parsons.
\newblock {On the concept of influence}.
\newblock {\em Public Opinion Quarterly}, 27(1):37--62, 01 1963.

\bibitem{survey_influence_measure}
Fabi{\'a}n Riquelme and Pablo Gonz{\'a}lez-Cantergiani.
\newblock Measuring user influence on twitter: A survey.
\newblock {\em Information processing \& management}, 52(5):949--975, 2016.

\bibitem{survey_cascade_models}
Mahdi Jalili and Matja{\v{z}} Perc.
\newblock Information cascades in complex networks.
\newblock {\em Journal of Complex Networks}, 5(5):665--693, 2017.

\bibitem{survey_influence_maxmization}
Yuchen Li, Ju~Fan, Yanhao Wang, and Kian-Lee Tan.
\newblock Influence maximization on social graphs: A survey.
\newblock {\em IEEE Transactions on Knowledge and Data Engineering},
  30(10):1852--1872, 2018.

\bibitem{Watts_cascade}
Duncan~J. Watts.
\newblock A simple model of global cascades on random networks.
\newblock {\em Proceedings of the National Academy of Sciences},
  99(9):5766--5771, 2002.

\bibitem{threshold_models}
Peter Dodds and Duncan~J. Watts.
\newblock {Threshold Models of Social Influence}.
\newblock In {\em {The Oxford Handbook of Analytical Sociology}}. Oxford
  University Press, 2011.

\bibitem{attention_spread_true_false}
Soroush Vosoughi, Deb Roy, and Sinan Aral.
\newblock The spread of true and false news online.
\newblock {\em Science}, 359(6380):1146--1151, 2018.

\bibitem{attention_structure}
Sharad Goel, Duncan~J. Watts, and Daniel~G. Goldstein.
\newblock The structure of online diffusion networks.
\newblock In {\em Proceedings of the 13th ACM Conference on Electronic
  Commerce}, EC '12, page 623–638, New York, NY, USA, 2012. Association for
  Computing Machinery.

\bibitem{ROI_marketing}
Donna~L Hoffman and Marek Fodor.
\newblock Can you measure the roi of your social media marketing?
\newblock {\em MIT Sloan management review}, 2010.

\bibitem{attention_economy_microcelebrity}
Zeynep Tufekci.
\newblock Not this one social movements, the attention economy, and
  microcelebrity networked activism.
\newblock {\em American Behavioral Scientist}, 57(7):848--870, 2013.

\bibitem{Goel_attention}
Ashish Goel and Farnaz Ronaghi.
\newblock A game-theoretic model of attention in social networks.
\newblock In Anthony Bonato and Jeannette Janssen, editors, {\em Algorithms and
  Models for the Web Graph}, pages 78--92, Berlin, Heidelberg, 2012. Springer
  Berlin Heidelberg.

\bibitem{Borgs_attention}
Christian Borgs, Jennifer Chayes, Brian Karrer, Brendan Meeder, R.~Ravi, Ray
  Reagans, and Amin Sayedi.
\newblock Game-theoretic models of information overload in social networks.
\newblock In Ravi Kumar and Dandapani Sivakumar, editors, {\em Algorithms and
  Models for the Web-Graph}, pages 146--161, Berlin, Heidelberg, 2010. Springer
  Berlin Heidelberg.

\bibitem{Ghosh_attention}
Arpita Ghosh and Preston McAfee.
\newblock Incentivizing high-quality user-generated content.
\newblock In {\em Proceedings of the 20th International Conference on World
  Wide Web}, WWW '11, page 137–146, New York, NY, USA, 2011. Association for
  Computing Machinery.

\bibitem{social_filter}
Avner May, Augustin Chaintreau, Nitish Korula, and Silvio Lattanzi.
\newblock Filter \& follow: How social media foster content curation.
\newblock In {\em ACM SIGMETRICS Performance Evaluation Review}, volume~42,
  pages 43--55. ACM, 2014.

\bibitem{Jackson}
Matthew~O. Jackson and Asher Wolinsky.
\newblock A strategic model of social and economic networks.
\newblock {\em Journal of Economic Theory}, 71(1):44--74, 1996.

\bibitem{berge_max}
Claude Berge.
\newblock {\em Topological spaces: Including a treatment of multi-valued
  functions, vector spaces and convexity}.
\newblock Oliver \& Boyd, 1877.

\bibitem{brouwer_fixpoint}
Viktor~Vasilevich Prasolov.
\newblock {\em Elements of combinatorial and differential topology}, volume~74.
\newblock American Mathematical Society, 2022.

\end{thebibliography}
%\newpage
\appendix
\newcommand{\leafAlloc}[0]{\Lambda}
\newcommand{\production}[0]{X}
\newcommand{\Pinf}[0]{P_{\mathrm{infl}}}
\newcommand{\Minf}[0]{\mu_{\mathrm{infl}}}
\newcommand{\Uinf}[0]{U^{\mathrm{infl}}}
\newcommand{\Uinfp}[0]{U^{\mathrm{infl}}_p}

\newcommand{\delayCoreUtilityRate}[3]{r_p \Bsbl B(#1|#2) e^{ - \alpha \left ( \frac{1}{#3} + \frac{1}{\mu(\infl|#2)} \right)} -  c  \Bsbr I(#3)I(\mu(\infl|#2))}
\newcommand{\leafOptUtilityRate}[2]{U_c(#2|\coreRateVec,\contAllocation,#1)}
\newcommand{\prodOptUtilityRate}[2]{U_p(#2|\coreRateVec,\comAllocation,#1)}
\newcommand{\delayCoreUtilityNodelay}[2]{r_p \Bsbl B(#1|#2)  -  c  \Bsbr }

\newcommand{\coreUtilityFull}[2]{U_{\Infl}(#1|#2)}
\newcommand{\leafOptUtilityFull}[2]{U_{c,#1}(#2|\coreRateVec,\contAllocation)}
\newcommand{\prodOptUtilityFull}[2]{U_{p,#1}(#2|\coreRateVec,\comAllocation)}
\newcommand{\prodOptUtilityInfFull}[2]{U^{\Infl}_{p,#1}(#2|\coreRateVec,\comAllocation)}

%\newpage
\section{Proof of Proposition~\ref{prop:existence_NE}}\label{proof:prop1}
In this appendix we prove Proposition~\ref{prop:existence_NE} which states that there exists  a Nash equilibrium  $\allAllocationNE  =  (\coreInteractVecOpt,\comAllocation^*,\contAllocation^*)$ for the game with perfect information as given by Definition~\ref{def:stable_allocation}.

\subsection{Preliminary Results}
We first show that the game with perfect information given by Definition~\ref{def:stable_allocation} is a potential game. We proceed as follows.

Recall that the objective function $\coreUtility{\comAllocation,\contAllocation}$ of the influencer is given by (see Eq.~\ref{eq:core_opt}) 
\begin{equation*}
\begin{split}
  \coreUtility{\comAllocation,\contAllocation} =   r_p \sum_{z \in \leafSet} \sum_{y \in \Clw{z}}   \delayCoreUtility{z}{y}
\end{split}
\end{equation*}
where
$$\delayInfl{z}{y} = \delay{\coreInteractVec(z)}\delay{\mu(\infl|y)};$$
the objective function of a content consumer $y \in \comSet$ is given by (see Eq.~\ref{eq:consumer_opt})
\begin{equation*}
\begin{split}
   \leafOptUtility{y}
  = & \; r_p   \sum_{z \in \Clw{y}} \delayCoreUtility{z}{y}  \\
    & + r_p \sum_{z \in \Clw{y}} \delayUtility{z}{y} \\
    & + \alternative{\lambda(y)},
\end{split}
\end{equation*}
the objective function of a content producer  $z \in \comSet$ is given by (see Eq.~\ref{eq:producer_opt})
\begin{equation*}
\begin{aligned}
   \prodOptUtility{z}
  = &  \; r_p \sum_{y \in \Clw{z}} \delayCoreUtility{z}{y}  \\
  & + r_p \sum_{y \in \Clw{z}} \delayUtility{z}{y},
\end{aligned}
\end{equation*}

The next result states that the game with perfect information as given by Definition~\ref{def:stable_allocation} is a potential game.

\begin{lemma}\label{lemma:potential_game}
	The game with perfect information as given by Definition~\ref{def:stable_allocation} is a potential game with  potential function 
	\begin{equation*}
	\begin{aligned}
	\Phi(\allAllocation) &= \sum_{y \in \leafSet}  \leafOptUtility{y} \\
        &= r_p \sum_{y \in \leafSet} \left [  \sum_{z \in \Clw{y}} \delayCoreUtility{z}{y}  
     + \sum_{z \in \Clw{y}} \delayUtility{z}{y} \right ] \\
    & + \sum_{y \in \leafSet} \alternative{\lambda(y)}.
	\end{aligned}
	\end{equation*}
\end{lemma}

\begin{proof}
Recall that $\allAllocation  = (\coreRateVec,\comAllocation,\contAllocation)$ denotes the allocation decision of all agents in the community, where $\coreRateVec$ is the rate allocation of the influence $\infl$,  $\comAllocation$ is the rate  allocation of all content consumers $y \in \comSet$, and $\contAllocation$ indicates the content production of all content producers $z \in \comSet$. To simplify the notation, for a given rate allocation of the content consumers $\comAllocation$, we use  $\comAllocation_{\{y,\leafAlloc(y)'\}}$ to denote that the  rate allocation of content consumer $y \in \comSet$ changed from $\leafAlloc(y)$ to $\leafAlloc(y)'$, while the rate allocations of all other content consumers is the same. Similarly,  for a given content production vector $\contAllocation$, we use  $X_{\{z,x(z)'\}}$ to denote that the content created by content producer $z \in \comSet$ changed from $x(z)$ to $x(z)')$, while the content created by all other content producers is the same.

To prove the result of the lemma, we then have  to establish the following three properties.

 % (\coreRateVec,\comAllocation,\contAllocation) 
 
First, for a given allocation $\allAllocation  = (\coreRateVec,\comAllocation,\contAllocation)$ we have show to that if the influencer  $\infl$ changes its rate allocation from $\coreRateVec$ to $\coreRateVec'$, then we have that
	\begin{equation}\label{eq:core_change}
		\begin{aligned}
			&\Phi(\coreRateVec,\comAllocation,\contAllocation)  -  \Phi(\coreRateVec',\comAllocation,\contAllocation)
		   = \coreUtilityFull{\coreRateVec}{\comAllocation,\contAllocation} - \coreUtilityFull{\coreRateVec'}{\comAllocation,\contAllocation}.
		\end{aligned}
	\end{equation}
        
        Second, we have to show that if content consumer $y \in \comSet$ changes its rate allocation from $\comAllocation(y)$ to $\comAllocation(y)'$, then we have that
	\begin{equation}\label{eq:consumption_change}
	\begin{aligned}
		&\Phi(\coreRateVec,\comAllocation,\contAllocation)  -  \Phi(\coreRateVec,\comAllocation_{\{y,\comAllocation(y)'\}},\contAllocation)
		   =\leafOptUtility{y} - \leafOptUtilityFull{y}{\comAllocation(y)'}.
	\end{aligned}
	\end{equation}

Finally, we have to show that if content producer $z \in \comSet$ changes its content from  $x(z)$ to $x(z)'$, then we have that
	\begin{equation}\label{eq:production_change}
	\begin{aligned}
		&\Phi(\coreRateVec,\comAllocation,\contAllocation)  -  \Phi(\coreRateVec,\comAllocation,\contAllocation_{\{z,x(z)'\}})
		   =\prodOptUtility{z}- \prodOptUtilityFull{z}{x'(z)}.
	\end{aligned}
	\end{equation}

The result that these three properties hold follows directly from the definitions of the objective functions $\coreUtility{\comAllocation,\contAllocation}$, $\leafOptUtility{y}$, and $\prodOptUtility{z}$, and we omit here a detailed derivation.
\end{proof}

Recall from Section~\ref{sec:results} that $\possibleStrategy$ is the set of all admissible allocations  $\allAllocation = (\coreInteractVec,\comAllocation,\contAllocation)$, i.e. the set of all allocations such  the rate allocation  $\coreInteractVec$ is given by a non-negative vector that satisfy the rate budget constraints $\coreBudget$ of the influencer, $\comAllocation(y)$, $y \in \comSet$ is a non-negative vector that satisfies the rate budget constraint $\leafBudget$ for the content consumers, and we have that $x(z) \in \setT$, $z \in \comSet$.

We then have the following result.
\begin{lemma}\label{lemma:convex_decision_space}
 The set of all admissible allocations $\possibleStrategy$ is a convex and compact set.
\end{lemma}

\begin{proof}
By Assumption~\ref{ass:setT} the set of all possible content allocations $x(z)$ of a content producer are given by a convex and compact set. Furthermore, by definition we have that the set of all non-negative rate allocations $\coreInteractVec$ of the influencer that satisfy the budget constraint $\coreBudget$ is given by a convex and bounded set, and hence a convex and compact set. Finally,  by definition we have that the set of all non-negative rate allocations $\comAllocation(y)$ of a content consumer $y \in \comSet$  that satisfy the budget constraint $\leafBudget$ is given by a convex and bounded set, and hence a convex and compact set.
The result of the lemma then follows. 
\end{proof}

\subsection{Proof of Proposition~\ref{prop:existence_NE}}
Using  Lemma~\ref{lemma:potential_game}, we proof  Proposition~\ref{prop:existence_NE} as follows.

By Lemma~\ref{lemma:potential_game}, the game with perfect information as given by Definition~\ref{def:stable_allocation} is a potential game with potential function
\begin{equation*}
	\begin{aligned}
	\Phi(\allAllocation) &= \sum_{y \in \leafSet}  \leafOptUtility{y}.
	\end{aligned}
\end{equation*}
By Lemma~\ref{lemma:convex_decision_space}, the domain $\possibleStrategy$ of the potential function is convex and compact, and hence there exists an allocation $\allAllocation^*$ that maximizes the potential function $\Phi(\allAllocation)$, i.e. there exists an allocation $\allAllocation^* \in \possibleStrategy$ such that 
$$\allAllocation^* = \arg\max_{\allAllocation \in \possibleStrategy} \Phi(\allAllocation).$$
Note that by the definition of the potential function  $\Phi(\allAllocation)$, and by Eq.~\eqref{eq:core_change}~to~\eqref{eq:production_change}, the allocation $\allAllocation^*$ is a Nash equilibrium as given by Definition~\ref{def:stable_allocation}. The result of  Proposition~\ref{prop:existence_NE} follows.

%\end{proof}

%\newpage
\section{Proof of Proposition~\ref{prop:social_welfare}}\label{proof:prop2}
Proposition~\ref{prop:social_welfare} follows immediately from the proof of Proposition~\ref{prop:existence_NE} which shows that there exists a Nash equilibrium $\allAllocation^*$ that maximizes the potential function $\Phi(\allAllocation)$, i.e. there exists a Nash equilibrium  $\allAllocation^*$ such that 
$$\allAllocation^* = \arg\max_{\allAllocation \in \possibleStrategy} \Phi(\allAllocation),$$
where $\Phi(\allAllocation)$ is the social welfare under allocation $\allAllocation$, i.e. we have that
$$\Phi(\allAllocation) = \sum_{y \in \leafSet}  \leafOptUtility{y}.$$

%\newpage
\section{Proof of Proposition~\ref{prop:att_existence_NE}}\label{proof:prop4}
In this appendix, we provide proof for Proposition~\ref{prop:att_existence_NE} which states that there exists  a Nash equilibrium  $\allAllocationInflNE  =  (\coreInteractVecOpt,\comAllocation^*,\contAllocation^*)$ for the game with imperfect information as given by Definition~\ref{def:att_NE}.

For our proof, we use Berge's maximum theorem~\cite{berge_max} and Brouwer's fixed-point theorem~\cite{brouwer_fixpoint}, which are given as follows.

\begin{theorem}[Maximum Theorem]\label{thm:berge_max_thm}
Let $A$, $B$, and $C$, be metrics spaces. Furthermore, let  $F: A \times B \mapsto C$ be a continuous function, and let $G:B \mapsto A$ be a compact valued and continuous correspondence such that $G(b) \neq \emptyset$, $b \in B$. Then, the maximum value function
$$V(b) = \max_{a \in G(b)} F(a,b), \qquad b \in B,$$
is continuous.
\end{theorem}
We then obtain the following corollary.
\begin{cor}\label{cor:berge_max_thm}
Let $A$ and  $B$ be non-empty compact and convex metrics spaces. Furthermore, let  $F: A \times B \mapsto C$ be a continuous function. Then, the maximum value function
$$V(b) = \max_{a \in A} F(a,b), \qquad b \in B,$$
is continuous.
\end{cor}

\begin{theorem}[Brouwer's Fixed-point Theorem]\label{thm:brouwer_max_thm}
  Let $A$ be a convex and compact metric space. Then every continuous function $f: A \mapsto A$ has a fixed point.
\end{theorem}

Using these two results, the outline of proof for  Proposition~\ref{prop:att_existence_NE} is as follows.

\begin{enumerate}
\item We first construct a mapping  $\mathcal{F}: \possibleStrategy \mapsto \possibleStrategy$ where
  $$\mathcal{F} = f^{(1)} \circ f^{(2)} \circ f^{(2)} $$
is a composition of the the optimal solution to the optimization problem that define a Nash equilibrium for the game with imperfect information as given by Definition~\ref{def:att_NE}. 
    \item We next we use Berge's Maximum theorem  to show that each function $f^{(1)}$, $f^{(2)}$, and $f^{(3)}$,  is continuous, and hence the mapping $\mathcal{F}$ is continuous.
    \item Finally, we use  Brouwer’s fixed-point theorem, we show that mapping $\mathcal{F} = f^{(1)} \circ f^{(2)} \circ f^{(3)} $ has a fixed point, which corresponds to a Nash equilibrium for the game with imperfect information as given by Definition~\ref{def:att_NE}.
\end{enumerate}

%\newpage
\subsection{Mapping $\mathcal{F} = f^{(1)} \circ f^{(2)} \circ f^{(3)}$}
In this appendix we defined and analyze the mapping  $\mathcal{F}: \possibleStrategy \mapsto \possibleStrategy$ given
$$\mathcal{F} = f^{(1)} \circ f^{(2)} \circ f^{(3)},$$
where each function maps an allocation $\allAllocation \in \possibleStrategy$  to an optimal solution to the optimization problem that define a Nash equilibrium for the game with imperfect information as given by Definition~\ref{def:att_NE}. 

Given an allocation $\allAllocation = (\coreInteractVec,\comAllocation,\contAllocation)  \in \possibleStrategy$, the first function $f^{(1)}:  \possibleStrategy \mapsto \possibleStrategy$ is given by
$$\allAllocation^{(1)} = \left (\coreInteractVec^{(1)},\comAllocation^{(1)},\contAllocation^{(1)} \right ) = f^{(1)}(\allAllocation)$$
where
\begin{equation*}
	\begin{aligned}
	\coreInteractVec^{(1)} =	& \argmax_{\coreRateVec \geq 0} && \coreUtility{\comAllocation,\contAllocation} =
          r_p \sum_{z \in \leafSet} \sum_{y \in \Clw{z}}   \delayCoreUtility{z}{y} \\
		& \text{subject to} && \sum_{z \in \leafSet} \coreInteract{z} \leq \coreBudget, 
	\end{aligned}
\end{equation*}
and
$$\comAllocation^{(1)} = \comAllocation \quad \mbox{ and } \quad X^{(1)} =  X.$$
We then have the following result.

\begin{lemma}\label{lemma:continuity_production1}
The function $f^{(1)}:  \possibleStrategy \mapsto  \possibleStrategy$ is continuous.
\end{lemma}

\begin{proof}
  Using Assumption~\ref{ass:setT} and Corollary~\ref{cor:berge_max_thm}, in order to prove the result of the lemma, it suffices to show that the function
$$\coreUtility{\comAllocation,\contAllocation} =  r_p \sum_{z \in \leafSet} \sum_{y \in \Clw{z}}   \delayCoreUtility{z}{y},$$
where 
$$B(z|y) = q(x(z)|z) p(x(z)|y)$$
and
$$\delayInfl{z}{y} = \delay{\coreInteractVec(z)}\delay{\mu(\infl|y)},$$  
is continuous on $\possibleStrategy$. 

Note that the distance function of a metric space is by definition continuous.
By Assumption~\ref{ass:fg}, the functions $f$ and $g$ are continuous. Combining these two results, we obtain that the functions $p(x|y)$ and $q(x|y)$, $x,y \in \setT$, given by
$$p(x|y) = f(d(x,y))$$
and
$$q(x|y) = g(d(x,y)),$$
are continuous.
Furthermore by Assumption~\ref{ass:delay}, the function $\delay{\mu}$, $\mu \geq 0$, is continuous.
It then follows that the function  $f^{(1)}:  \possibleStrategy \mapsto  \possibleStrategy$ is a composition of continuous functions, and hence continuous. The result of the lemma then follows.
\end{proof}

%\newpage
Recall that the objective function of the optimization problem $\optperiphery{\coreRateVec,\contAllocation}{y}$ of content consumer $y \in \comSet$, is given by (see Eq.~\eqref{eq:consumer_opt})
$$	\begin{aligned}
		\leafOptUtility{y}
		= \; &  r_p \sum_{z \in \Clw{y}} \delayCoreUtility{z}{y}  \\
		& +  r_p \sum_{z \in \Clw{y}} \delayUtility{z}{y} \\
		& + \alternative{\lambda(y)}.
	\end{aligned}$$

Given an allocation $\allAllocation = (\coreInteractVec,\comAllocation,\contAllocation)  \in \possibleStrategy$, the second function $f^{(2)}:  \possibleStrategy \mapsto \possibleStrategy$ is given by
$$\allAllocation^{(2)} = \left (\coreInteractVec^{(2)},\comAllocation^{(2)},\contAllocation^{(2)} \right ) = f^{(2)}(\allAllocation)$$
where
\begin{equation*}
	\begin{aligned}
	\comAllocation^{(2)}(y) =	& \argmax_{\comAllocation(y) \geq 0}{\text{maximize}}
		& & \leafOptUtility{y}\\
		& \text{subject to} &&\mu(\infl|y) + \lambda(y) + \sum_{z \in \Clw{y}} \mu(z|y) \leq \leafBudget,
	\end{aligned}
\end{equation*}
and
$$\coreInteractVec^{(2)} = \coreInteractVec \quad \mbox{ and } \quad X^{(2)} =  X.$$
We then have the following result.

\begin{lemma}\label{lemma:continuity_production2}
The function $f^{(2)}:  \possibleStrategy \mapsto  \possibleStrategy$ is continuous.
\end{lemma}
Lemma~\ref{lemma:continuity_production2} can be proved using the same argument as given in the proof for Lemma~\ref{lemma:continuity_production1}. We omit here a detailed derivation.

%\newpage
Recall from Section~\ref{sec:limited_information}  that $\coreInteractVecOpt(\comAllocation,\contAllocation)$ is  the optimal rate allocation of the influencer $\infl$ as given by Eq.~\eqref{eq:core_opt}  for a given rate allocation $\comAllocation$ of the content consumers and a given content allocation $\contAllocation$ of the content producers, i.e. we have that
\begin{equation*}
	\begin{aligned}
	\coreInteractVecOpt(\comAllocation,\contAllocation)  =	& \argmax_{\coreRateVec \geq 0} && \coreUtility{\comAllocation,\contAllocation} =
          r_p \sum_{z \in \leafSet} \sum_{y \in \Clw{z}}   \delayCoreUtility{z}{y} \\
		& \text{subject to} && \sum_{z \in \leafSet} \coreInteract{z} \leq \coreBudget, 
	\end{aligned}
\end{equation*}

Furthermore, recall from Section~\ref{sec:limited_information}  that $\coreInteractVecOpt(z|\comAllocation,\contAllocation)$ is the following rate that the influencer $\infl$ allocates to content producer  $z$ under $\coreInteractVec(\comAllocation,\contAllocation)$.

Given an allocation $\allAllocation = (\coreInteractVec,\comAllocation,\contAllocation)$, the third function $f^{(3)}(\allAllocation)$ is then given by
$$\allAllocation^{(3)} = \left (\coreInteractVec^{(3)},\comAllocation^{(3)},\contAllocation^{(3)} \right ) = f^{(3)}(\allAllocation)$$
where
$$x^{(3)}(z) = 	\argmax_{x(z) \in \setT} \delay{\coreInteractVecOpt(z|\comAllocation,\contAllocation_{\{z,x(z)\}})}, \qquad z \in \comSet,$$
and
$$\coreInteractVec^{(3)} = \coreInteractVec \quad \mbox{ and } \comAllocation^{(3)} = \comAllocation.$$
We then have the following result.
\begin{lemma}\label{lemma:continuity_production3}
The function $f^{(3)}:  \possibleStrategy \mapsto  \possibleStrategy$ is continuous.
\end{lemma}

\begin{proof}
  Using Assumption~\ref{ass:setT} and Corollary~\ref{cor:berge_max_thm}, in order to prove the result of the lemma, it suffices to show that the function $\delay{\coreInteractVecOpt(z|\comAllocation,\contAllocation)}$ is continuous on $\possibleStrategy$. By Assumption~\ref{ass:delay}, the function $\delay{\mu}$, $\mu \geq 0$, is continuous and by Lemma~\ref{lemma:continuity_production1} the function $\coreInteractVecOpt(z|\comAllocation,\contAllocation)$ is continuous on $\possibleStrategy$. It then follows that the function  $\delay{\coreInteractVecOpt(z|\comAllocation,\contAllocation)}$ is continuous on $\possibleStrategy$. This establishes the result of the lemma.
\end{proof}

%\newpage
\subsection{Proof of Proposition~\ref{prop:att_existence_NE}}
Let the three functions $ f^{(1)}$, $f^{(2)}$,  and $f^{(3)}$, be as defined above.
By the definition of  the three functions $ f^{(1)}$, $f^{(2)}$,  and $f^{(3)}$, 
we have that a fixed point of the function   $\mathcal{F}: \possibleStrategy \mapsto \possibleStrategy$ given by $\mathcal{F} = f^{(1)} \circ f^{(2)} \circ f^{(3)}$ is a Nash equilibrium of the game with imperfect information as given by Definition~\ref{def:att_NE}.

By Lemma~\ref{lemma:convex_decision_space}, we have that the set $\possibleStrategy$ is convex and compact. By Lemma~\ref{lemma:continuity_production1}~to~\ref{lemma:continuity_production3}, we have that each of the three functions   $ f^{(1)}$, $f^{(2)}$,  and $f^{(3)}$, is continuous.
This implies that the function $\mathcal{F} = f^{(3)} \circ f^{(2)} \circ f^{(1)} (\Omega)$ is a composition of the continuous functions, and hence is itself continuous. 

By Lemma~\ref{lemma:convex_decision_space} we have that the set $\possibleStrategy$ is convex and compact. As the function $\mathcal{F}$ it then follows from Brouwer's fixed-point theorem that the function $\mathcal{F}$ has a fixed point. 
As a fixed point of the function   $\mathcal{F}: \possibleStrategy \mapsto \possibleStrategy$ given by $\mathcal{F} = f^{(1)} \circ f^{(2)} \circ f^{(3)}$ is a Nash equilibrium of the game with imperfect information as given by Definition~\ref{def:att_NE}, we obtain the result of  Proposition~\ref{prop:att_existence_NE}.
 
%\newpage
% \newcommand{\leafAlloc}[0]{\Lambda}
% \newcommand{\production}[0]{X}
% \newcommand{\Pinf}[0]{P_{\mathrm{infl}}}
% \newcommand{\Minf}[0]{\mu_{\mathrm{infl}}}
% \newcommand{\Uinf}[0]{U^{\mathrm{infl}}}
% \newcommand{\Uinfp}[0]{U^{\mathrm{infl}}_p}

\section{Proof of Proposition~\ref{prop:att_contProducer}}\label{proof:prop3}
In this section we prove  Proposition~\ref{prop:att_contProducer} which states that if $\allAllocationInflNE = (\coreInteractVecOpt,\comAllocation^*,\contAllocation^*)$ is a Nash equilibrium as given by Definition~\ref{def:att_NE} such that for content producer $z \in \comSet$ we have
  $$\coreInteractVecOpt(z) > 0,$$
 and
  $$\sum_{z' \in \Cw{z}} \coreInteractVecOpt(z') > 0,$$
then we have 
$$x^*(z)  =  \argmax_{x(z) \in \setT} \prodOptUtilityInfNE{z},$$
where
$$\prodOptUtilityInf{z} =  r_p \sum_{y \in \Clw{z}}   \delayCoreUtility{z}{y}$$
and
$$\delayInfl{z}{y} = \delay{\coreInteractVec(z)}\delay{\mu(\infl|y)}.$$

\subsection{Preliminary Results}
To prove Proposition~\ref{prop:att_contProducer}, we first derive the following two lemmas.

\begin{lemma}\label{lemma:opt_condition_1}
If  $\allAllocationInflNE = (\coreInteractVecOpt,\comAllocation^*,\contAllocation^*)$ is a Nash equilibrium for the game with imperfect information as given by Definition~\ref{def:att_NE} such that for content producer $z_0 \in \comSet$ we have
$$\coreInteractVecOpt(z_0) > 0$$
and
 $$\sum_{y \in \Cw{z_0}} \mu^*(\infl|y) > 0,$$
then we have that
$$x^*(z_0) = \argmax_{x(z_0) \in \setT} \delay{\coreInteractVecOpt(z_0|\comAllocation^*,\contAllocation^*_{\{z_0,x(z_0)\}})},$$
if and only if
$$x^*(z_0) = \argmax_{x(z_0) \in \setT}  \sum_{y \in \Clw{z_0}}   \delay{\mu^*(\infl|y)} B(z_0|y),$$
where
$$B(z_0|y) = q(x(z_0)|z_0) p(x(z_0)|y).$$
\end{lemma}  

\begin{proof}
  By Assumption~\ref{ass:delay}, the function $\delay{\mu}$, $\mu \geq 0$, is strictly increasing. It then follows that if there exists a content type $x$ such that
$$  \delay{\coreInteractVecOpt(z_0|\comAllocation^*,\contAllocation^*_{\{z_0,x(z_0)\}})} > 0,$$
then content producer $z_0$ maximizes  $\delay{\coreInteractVecOpt(z_0|\comAllocation^*,\contAllocation^*_{\{z_0,x(z_0)\}})}$, if and only if producer $z_0$ maximizes the rate  $\coreInteractVecOpt(z_0|\comAllocation^*,\contAllocation^*_{\{z_0,x(z_0)\}})$.

  By definition, the rate $\coreInteractVecOpt(z_0|\comAllocation^*,\contAllocation^*_{\{z_0,x(z_0)\}})$ is given by the rate that the influencer $\infl$ allocates to content producer $z_0$ under an optimal solution $\coreInteractVecOpt$ to the maximization problem  $\optcore{\comAllocation^*,\contAllocation^*_{\{z_0,x(z_0)\}}}$ given by (see  Eq.~\eqref{eq:core_opt})
\begin{equation*}
	\begin{aligned}
		& \underset{\coreRateVec \geq 0}{\text{maximize}} &&
          r_p \sum_{z \in \leafSet} \sum_{y \in \Clw{z}}   \delayCoreUtility{z}{y} \\
		& \text{subject to} && \sum_{z \in \leafSet} \coreInteract{z} \leq \coreBudget.
	\end{aligned}
\end{equation*}
Using Assumption~\ref{ass:infl_alloc}, without loss of generality we can assume that  an optimal solution $\coreInteractVecOpt$ to the maximization problem  $\optcore{\comAllocation^*,\contAllocation^*_{\{z_0,x(z_0)\}}}$ we have
$$\sum_{z \in \comSet} \coreInteractVecOpt(z) = \coreBudget.$$

This implies that for  an optimal solution $\coreInteractVecOpt$ to the maximization problem  $\optcore{\comAllocation^*,\contAllocation^*_{\{z_0,x(z_0)\}}}$, there exists a Lagrange multiplier $\alpha^*(x(z_0))$ such that
\begin{equation}\label{eq:lagrange_1}
  \coreInteractVecOpt(z) = G \left ( \frac{\alpha^*(x(z_0))}{r_p \gamma(z)} \right ), \qquad z \in \comSet,
\end{equation}  
and
\begin{equation}\label{eq:lagrange_2}
  \sum_{z \in \comSet}   G \left ( \frac{\alpha^*(x(z_0))}{r_p \gamma(z)} \right ) = \coreBudget,
\end{equation}  
where
$$\gamma(z) =  \sum_{y \in \Clw{z}} \delay{\mu^*(\infl|y)}   q(x(z)|z) p(x(z)|y) \qquad z \in \comSet,$$
and $G(b)$, $b \geq 0$, is the inverse function of $\ddelay{\mu}$, $\mu \geq 0$. Note that by Assumption~\ref{ass:delay}, the derivative  $\ddelay{\mu}$, $\mu \geq 0$ is strictly decreasing, and hence the function $G(b)$, $b \geq 0$, is well-defined.

As by Assumption~\ref{ass:delay} the function $\delay{\mu}$, $\mu \geq 0$, is concave, it follows that  $\coreInteractVecOpt$ is an  optimal solution to the maximization problem  $\optcore{\comAllocation,\contAllocation_{\{z,x(z_0)\}}}$ if and only if the optimality conditions given by Eq.~\eqref{eq:lagrange_1}~and~\eqref{eq:lagrange_2} hold.

Note that by Assumption~\ref{ass:delay}, the function $G(b)$, $b \geq 0$,  is is strictly decreasing. As a result we have that  the lower the value of $\alpha^*(x(z_0))/\gamma(z)$, the higher the value of the function  $G(\alpha^*(x(z_0))/r_p \gamma(z))$. 
As by assumption we have that
$$\sum_{y \in \Cw{z_0}} \mu^*(\infl|y) > 0,$$
and hence by Assumption~\ref{ass:delay} we have that
$$\sum_{y \in \Cw{z_0}} \delay{\mu^*(\infl|y)} > 0,$$
it then follows that if the content production $x^*(z)$, $z \in \Cw{z_0}$, of all content producers other than producer $z_0$ are fixed, then the following is true.
Content producer $z_0$ maximizes $G(\alpha(x(z_0))/\gamma(z)$, and hence  maximizes $ \coreInteractVecOpt(z)$, if and only if $x(z_0)$ maximizes $\gamma(z_0)$ and we have that
$$x^*(z_0) = \argmax_{x(z_0) \in \setT}  r_p \sum_{y \in \Clw{z_0}}    \delay{\mu^*(\infl|y)} B(z_0|y),$$
where
$$B(z_0|y) = q(x(z_0)|z) p(x(z_0)|y).$$
The result of the lemma then follows. 
\end{proof}

%\newpage
\begin{lemma}\label{lemma:opt_condition_2}
  If $\allAllocation = (\coreInteractVec,\comAllocation,\contAllocation)$ is an allocation such that for content producer $z \in \comSet$ we have
  $$\coreInteractVec(z) > 0$$
and
$$\sum_{y \in \Cw{z}} \mu(\infl|y) > 0,$$
then we have that
$$x^*(z) = \argmax_{x(z) \in \setT} \prodOptUtilityInf{z},$$
if and only if
$$x^*(z) = \argmax_{x(z) \in \setT}  \sum_{y \in \Clw{z}}   \delay{\mu(\infl|y)} B(z|y),$$
where
$$B(z|y) = q(x(z)|z) p(x(z)|y).$$
\end{lemma}

The result of Lemma~\ref{lemma:opt_condition_2} follows immediately from the  definition of the function $\prodOptUtilityInf{z}$ (see Eq.~\eqref{eq:inf_prodUtility}), i.e. we have that
$$\prodOptUtilityInf{z} =  r_p \sum_{y \in \Clw{z}}   \delayCoreUtility{z}{y},$$
where
$$\delayInfl{z}{y} = \delay{\coreInteractVec(z)}\delay{\mu(\infl|y)}.$$

%\newpage
\subsection{Proof of Proposition~\ref{prop:att_contProducer}}
The result of  Proposition~\ref{prop:att_contProducer} follows immediately from Lemma~\ref{lemma:opt_condition_1} and Lemma~\ref{lemma:opt_condition_2}. 

%\newpage
\section{Proof for Proposition~\ref{prop:allocation_inf_rate_Infl}}~\label{app:prop_limit}
In this appendix we prove Proposition~\ref{prop:allocation_inf_rate_Infl}, which states that in a large community where the influencer allocates a large rate budget $\coreBudget$, there exists a Nash equilibrium $\allAllocationInflOpt$ for the game with imperfect information that maximizes the social welfare, i.e. we have that
$$\Phi \Big (\allAllocationInflOpt \Big ) = \Phi \Big (\allAllocationWelfare \Big ),$$
where
$$\allAllocationWelfare = \argmax_{\allAllocation \in \possibleStrategy} \Phi(\allAllocation).$$

In order to prove this result, we introduce another interaction between content producer and consumers, and the influencer, to which we refer as the proxy game. Under this game, content consumers are constrained to only follow the influencer, i.e. consumers can obtain content only through the influencer. In the following, we show that  in a large community where the influencer allocates a large rate budget $\coreBudget$ we have that $\allAllocationNE$ is Nash equilibrium for the game with perfect information, and  $\allAllocationInflNE$ is Nash equilibrium for the game with imperfect information, if an only if they are a Nash equilibrium for the proxy game. We then use this result to prove Proposition~\ref{prop:allocation_inf_rate_Infl}.

To do that, we proceed as follows. In Subsection~\ref{app:def_proxy_game}, we formally define the proxy game. In Subsections~\ref{app:NE_proxy_perfect}~and~\ref{app:NE_proxy_imperfect}, we provide  the result that  $\allAllocationNE$ is Nash equilibrium for the game with perfect information, and  $\allAllocationInflNE$ is Nash equilibrium for the game with imperfect information, if an only if they are a Nash equilibrium for the proxy game. Finally, we Subsection~\ref{app:prove_allocation_inf_rate_Infl}, we prove Proposition~\ref{prop:allocation_inf_rate_Infl}.

\subsection{Proxy Game}\label{app:def_proxy_game}

Under the proxy  game, content consumers are constrained to only follow the influencer, i.e. content consumers can obtain content only through the influencer.
The optimization problem of content consumer $y \in \Com$ is now to choose a rate allocation $\comProxAllocation^*(y)$ that solves the following optimization problem,
\begin{equation}\label{eq:proxy_consumer_opt}
	\begin{aligned}
		& \underset{\comProxAllocation(y) \geq 0}{\text{maximize}}
		& & \alternative{\lambda(y)} + r_p \sum_{z \in \Clw{y}} & \delayCoreUtility{z}{y} \\
                	& \text{subject to} && \lambda(y) + \mu(\infl|y) \leq \leafBudget,
	\end{aligned}
\end{equation}
where $\leafBudget>0$ is the rate budget available to content consumer $y$.  We refer to this maximization problem as   $\optProxperiphery{\coreRateVec,\contAllocation}{y}$.

The optimization problem of the influencer and content producer stay the same as given by Eq.~\eqref{eq:core_opt}, and Eq.~\eqref{eq:producer_opt}, respectively.

A Nash equilibrium for the proxy game is then defined as follows.

\begin{definition}\label{def:proxy_NE}
	An allocation $\allProxAllocationNE  = (\coreInteractVecOpt,\comProxAllocation^*,\contAllocation^*)$ is a Nash equilibrium for the proxy  game if we have that
	\begin{enumerate}
		\item $\coreInteractVecOpt$ is a solution for the maximization problem $\optcore{\comAllocation^*,\contAllocation^*}$,
		\item $\comProxAllocation^*(y)$, $y \in \comSet$,  is a solution to the maximization problem $\optProxperiphery{\coreInteractVecOpt,\contAllocation^*}{y}$  with $$\mu(z|y) = 0, \qquad z \in \Cw{y},$$
		\item $x^*(y)$, $y \in \comSet$,  is a solution to the maximization problem  $\optcontperiphery{\coreInteractVecOpt,\comAllocation^*}{y}$.
	\end{enumerate}  
\end{definition}

The following result states that the proxy game is a potential game.

\begin{lemma}\label{lemma:potential_proxy_game}
The proxy game is a potential game with  potential function 
$$	\Phi^{\proxy}(\allProxAllocation) = \sum_{y \in \leafSet}  \leafOptProxUtility{y},$$
where $ \leafOptProxUtility{y}$ is the consumption utility of community member $y$ under the proxy game given by 
$$ \leafOptProxUtility{y} = \alternative{\lambda(y)} + r_p \sum_{z \in \Clw{y}} \delayCoreUtility{z}{y}.$$
\end{lemma}

Using the result of Lemma~\ref{lemma:potential_proxy_game}, we obtain the following result.
\begin{prop}\label{prop:existence_NE_proxy_game}
	There exists  a Nash equilibrium  $\allProxAllocationNE  =  (\coreInteractVecOpt,\comAllocation^*,\contAllocation^*)$ for the proxy game as given by Definition~\ref{def:proxy_NE}.
\end{prop}  

We can prove Lemma~\ref{lemma:potential_proxy_game} and Proposition~\ref{prop:existence_NE_proxy_game} using the same argument as given to prove Lemma~\ref{lemma:potential_game} and Proposition~\ref{prop:existence_NE} for the game with perfect information as defined in Subsection~\ref{ssec:stable}. We omit here a detailed proof.

We provide an analysis of the properties of a Nash equilibrium $\allProxAllocation$ for the proxy game in Appendix~\ref{app:analysis_proxy}.

%\newpage
\subsection{Properties of Nash equilibrium $\allAllocationNE$}\label{app:NE_proxy_perfect}

The following result connects  a Nash equilibrium $\allAllocationNE$ for the game with perfect information with a Nash equilibrium for the proxy game.

\begin{prop}\label{prop:NE_proxy_perfect}
Let $\leafBudget$ be a given rate budget for the content consumers. 
Then there exists a integer $N_0$ and a constant $k_{\Infl}$ such that if $N > N_0$,  and $\coreBudget > N k_{\Infl}$, the following is true. 
An $\allAllocationNE  = (\coreInteractVec,\comAllocation^*,\contAllocation^*)$ is a  Nash equilibrium of the game with perfect information as given by Definition~\ref{def:stable_allocation}, if and only if  $\allAllocationNE$ is a  Nash equilibrium for the proxy game as given by Definition~\ref{def:proxy_NE}. 
\end{prop}
We provide a proof for Proposition~\ref{prop:NE_proxy_perfect} in Appendix~\ref{app:analysis_perfect}. While the result of Proposition~\ref{prop:NE_proxy_perfect} is intuitive, its proof is surprisingly involved.

%\newpage
\subsection{Properties of Nash equilibrium $\allAllocationInflOpt$}\label{app:NE_proxy_imperfect}
In addition, we have the following result that connects  a Nash equilibrium $\allAllocationInflNE$ for the game with imperfect information with a Nash equilibrium for the proxy game.

\begin{prop}\label{prop:NE_proxy_imperfect}
Let $\leafBudget$ be a given rate budget for the content consumers. 
Then there exists a integer $N_0$ and a constant $k_{\Infl}$ such that if $N > N_0$,  and $\coreBudget > N k_{\Infl}$, the following is true. 
An allocation $\allAllocationInfl  = (\coreInteractVec,\comAllocation^*,\contAllocation^*)$ is a  Nash equilibrium of the game with imperfect information as given by Definition~\ref{def:att_NE}, if and only if $\allAllocationInflOpt$ is a  Nash equilibrium for the proxy game  as given by Definition~\ref{def:proxy_NE}. 
\end{prop}
We provide a proof for Proposition~\ref{prop:NE_proxy_imperfect} in Appendix~\ref{app:analysis_imperfect}.

\subsection{Proof of Proposition~\ref{prop:allocation_inf_rate_Infl}}\label{app:prove_allocation_inf_rate_Infl}

Using Proposition~\ref{prop:NE_proxy_perfect} and  Proposition~\ref{prop:NE_proxy_imperfect}, we can prove Proposition~\ref{prop:allocation_inf_rate_Infl} as follows.

Combining  Proposition~\ref{prop:NE_proxy_perfect} and  Proposition~\ref{prop:NE_proxy_imperfect}, we obtain the following result. For a given rate budget  $\leafBudget$ for the content consumers, then there exists a integer $N_0$ and a constant $k_{\Infl}$ such that if $N > N_0$,  and $\coreBudget > N k_{\Infl}$, the following is true. An $\allAllocationNE  = (\coreInteractVec,\comAllocation^*,\contAllocation^*)$ is a  Nash equilibrium of the game with perfect information as given by Definition~\ref{def:stable_allocation}, if and only if  $\allAllocationInflNE$ is a  Nash equilibrium for the game with imperfect information as given by Definition~\ref{def:att_NE}.

Furthermore, by Proposition~\ref{prop:social_welfare} there always exists a Nash equilibrium $\allAllocationNEOpt$ that maximizes the social welfare, i.e. we have
$$\Phi \Big (\allAllocationNEOpt \Big ) = \Phi \Big (\allAllocationWelfare \Big ),$$
where
$$\allAllocationWelfare = \argmax_{\allAllocation \in \possibleStrategy} \Phi(\allAllocation).$$

Combining the above two results, it obtain the following result. For a given rate budget  $\leafBudget$ for the content consumers, there exists a integer $N_0$ and a constant $k_{\Infl}$ such that if $N > N_0$,  and $\coreBudget > N k_{\Infl}$, then the following is true.
Let $\allAllocationNEOpt$ is a Nash equilibrium for the game with perfect information that maximizes the social welfare. Then
$$\allAllocationInflOpt =  \allAllocationNEOpt,$$
 is a Nash equilibrium  $\allAllocationInflOpt$ with imperfect information that maximizes the social welfare. The result of Proposition~\ref{prop:allocation_inf_rate_Infl} then follows.

%\newpage
\section{Analysis of  of Nash Equilibrium $\allProxAllocationNE$ for Proxy Game}\label{app:analysis_proxy}

In this section, we analyze the properties of a Nash equilibrium of the proxy game as given by Definition~\ref{def:proxy_NE}. Recall that under this game, content consumers are constrained to only follow the influencer, i.e. consumers can obtain content only through the influencer.

Recall that for the proxy game, the objective of content consumer $y \in \Com$ is now to choose a rate allocation $\comAllocation^*(y)$ that solves the following optimization problem  $\optProxperiphery{\coreRateVec,\contAllocation}{y}$ given by (see Eq.~\eqref{eq:proxy_consumer_opt})
$$	\begin{aligned}
		& \underset{\comAllocation(y) \geq 0}{\text{maximize}}
		& & \alternative{\lambda(y)} + r_p \sum_{z \in \Clw{y}} \delayCoreUtility{z}{y} \\
          & \text{subject to} && \lambda(y) + \mu(\infl|y) \leq \leafBudget,\\
          &&& \sum_{z \in \Cw{y}} \mu(z|y) = 0.
	\end{aligned}
$$
where $\leafBudget>0$ is the rate budget available to content consumer $y$.

The optimization problem of the influencer and content producer stay the same and are  given by Eq.~\eqref{eq:core_opt}, and Eq.~\eqref{eq:consumer_opt}, respectively.

In the following we analyze the properties of a Nash equilibrium $\allProxAllocationNE$ of the proxy game.

%\newpage

\subsection{Optimality Conditions for  Nash Equilibrium  $\allProxAllocationNE$}

We first derive necessary and sufficient conditions for a  $\allProxAllocationNE  = (\coreInteractVecOpt,\comAllocation^*,\contAllocation^*)$  to be a Nash equilibrium for  the proxy game as given by Definition~\ref{def:proxy_NE} 

\begin{lemma}\label{lemma:conditions_proxy_NE}
Under Assumption~\ref{ass:infl_alloc},  an $\allProxAllocationNE  = (\coreInteractVecOpt,\comAllocation^*,\contAllocation^*)$  is a  Nash equilibrium as given by Definition~\ref{def:proxy_NE}, if an only if the following conditions hold true,
\begin{enumerate}
\item[a)]   For all content producers $z \in \comSet$, the content allocation $x^*(z)$ is a solution for the maximization problem  $\optcontperiphery{\coreRateVec,\comAllocation}{y}$.

\item[b)] For all content consumers $y \in \comSet$ we have that
  $$\sum_{z \in \Cw{y}} \mu^*(z|y) = 0.$$

\item[c)] For all content consumers $y \in \comSet$ we have that
  $$ \lambda^*(y) + \mu^*(\infl|y)  = \leafBudget.$$

\item[d)] If for content consumer $y \in \comSet$ we have that $\lambda^*(y) > 0$, then we have
$$\ddelay{\lambda^*(y)} r_0 B_0 \geq  \ddelay{\mu^*(\infl|y)} r_p \sum_{z \in \Cw{y}} \delay{\coreInteractVecOpt(z)} B(z|y).$$

\item[e)] If  for content consumer $y \in \comSet$ we have that $\mu^*(\infl|y) > 0$, then we have

$$\ddelay{\mu^*(\infl|y)}r_p  \sum_{z \in \Cw{y}} \delay{\coreInteractVecOpt(z)} B(z|y) \geq \ddelay{\lambda^*(y)} r_0 B_0.$$

\item[f)] We have that
  $$\sum_{z \in \comSet} \coreInteractVecOpt(z) =  \coreBudget.$$  
  
\item[g)] If for content producer $z \in \comSet$ we have that $\coreInteractVecOpt(z) > 0$, then we have for $z' \in \Cw{z}$ that
  $$ \ddelay{\coreInteractVecOpt(z)} \sum_{y \in \Cw{z}} \delay{\mu^*(\infl|y)} B(z|y)  \geq \ddelay{\coreInteractVecOpt(z')} \sum_{y \in \Cw{z'}} \delay{\mu^*(\infl|y)}B(z|y).$$
\end{enumerate}

\end{lemma}

\begin{proof}
Recall that  $\allProxAllocationNE  = (\coreInteractVecOpt,\comAllocation^*,\contAllocation^*)$  is a  Nash equilibrium of the proxy game as given by Definition~\ref{def:proxy_NE}, if an only if the following conditions hold true,
	\begin{enumerate}
		\item[1)] $\coreInteractVecOpt$ is a solution for the maximization problem $\optcore{\comAllocation^*,\contAllocation^*}$,
		\item[2)] $\comProxAllocation^*(y)$, $y \in \comSet$,  is a solution to the maximization problem $\optProxperiphery{\coreInteractVecOpt,\contAllocation^*}{y}$  with $$\mu(z|y) = 0, \qquad z \in \Cw{y},$$
		\item[3)] $x^*(y)$, $y \in \comSet$,  is a solution to the maximization problem  $\optcontperiphery{\coreInteractVecOpt,\comAllocation^*}{y}$.
	\end{enumerate}

To show that the result of the lemma holds, we have to show that the above Properties 1)-3) hold if and only if the Properties a)-g) in the statement of the lemma hold.

As Property a) is identical to Property 3), in order to show that the above Properties 1)-3) hold if and only if the Properties a)-f) in the statement of the lemma hold, it suffices to prove the following two statements.

First, we have that  $\comProxAllocation^*(y)$, $y \in \comSet$,  is a solution to the maximization problem $\optProxperiphery{\coreInteractVecOpt,\contAllocation^*}{y}$ (Property 1)), if and only if Properties b)-e) hold.

Second, we have that  $\coreInteractVecOpt$ is a solution for the maximization problem $\optcore{\comAllocation^*,\contAllocation^*}$ (Property 2)), if and only if Properties f) and g) hold.

We first prove that that  $\comProxAllocation^*(y)$, $y \in \comSet$,  is a solution to the maximization problem $\optProxperiphery{\coreInteractVecOpt,\contAllocation^*}{y}$ (Property 1)), if and only if Properties b)-e) hold.

Note that Property b) follows from the definition of the proxy game which states that content consumers can get content only through the influencer, and hence we have that if  $\comProxAllocation^*(y)$, $y \in \comSet$,  is a solution to the maximization problem $\optProxperiphery{\coreInteractVecOpt,\contAllocation^*}{y}$ with
$$\sum_{z \in \Cw{y}} \mu^*(z|y) = 0.$$

Note that by definition we have that $B_0 > 0$. As a result, if Property c) does not hold and we have that
$$ \lambda^*(y) + \mu^*(\infl|y)  < \leafBudget,$$
then we increase the objective function in  the maximization problem $\optProxperiphery{\coreInteractVecOpt,\contAllocation^*}{y}$ by setting
$$\lambda(y) =  \lambda^*(y) + \epsilon_0,$$
where
$$ \epsilon_0 =  \leafBudget - \left [  \lambda^*(y) + \mu^*(\infl|y) \right ] > 0.$$
As a result, we have that if  $\comProxAllocation^*(y)$, $y \in \comSet$,  is a solution to the maximization problem $\optProxperiphery{\coreInteractVecOpt,\contAllocation^*}{y}$, then we have that, then we have that
$$ \lambda^*(y) + \mu^*(\infl|y)  = \leafBudget.$$

Using the two constraints given by Properties b) and c), Properties d) and e) correspond to the first-order optimality conditions for the optimization problem   $\optProxperiphery{\coreInteractVecOpt,\contAllocation^*}{y}$. As by Assumption~\ref{ass:delay} the function $\delay{\mu}$, $\mu \geq 0$, is concave, it follows that  $\comProxAllocation^*(y)$, $y \in \comSet$,  is a solution to the maximization problem $\optProxperiphery{\coreInteractVecOpt,\contAllocation^*}{y}$, if and only if it satisfies the first-order optimality conditions.

Combining the above results, it follows  $\comProxAllocation^*(y)$, $y \in \comSet$,  is a solution to the maximization problem $\optProxperiphery{\coreInteractVecOpt,\contAllocation^*}{y}$ (Property 1)), if and only if Properties b)-e) hold.

Using the same line of argument as given for Property 1), one can show that  under Assumption~\ref{ass:infl_alloc} we have that $\coreInteractVecOpt$ is a solution for the maximization problem $\optcore{\comAllocation^*,\contAllocation^*}$ (Property 2)), if and only if Properties f) and g) hold. We omit here a detailed derivation. 

The result of the lemma then follows. 
\end{proof}

%\newpage

\subsection{Properties of $\allProxAllocationNE$ for  $ \sum_{y \in \Com} \mu^*(\infl|y) > 0$}

We next characterize the properties of a Nash equilibrium $\allProxAllocation= (\coreInteractVecOpt,\comAllocation^*,\contAllocation^*)$ such that
\begin{equation}\label{eq:prox_infl_pos_rate}
  \sum_{y \in \comSet} \mu^*(\infl|y) > 0.
\end{equation}  
In Appendix~\ref{app:prop_NE_proxy} we provide conditions under which the inequality given by Eq.~\eqref{eq:prox_infl_pos_rate} holds.

\begin{lemma}\label{lemma:proxy_infl_pos_utility}
If $\allProxAllocationNE  = (\coreInteractVecOpt,\comAllocation^*,\contAllocation^*)$  is a  Nash equilibrium for the proxy game given by Definition~\ref{def:proxy_NE} such that 
$$\sum_{y \in \comSet} \mu^*(\infl|y) > 0, $$
then we have that
$$\sum_{y \in \comSet} \delay{\mu^*(\infl|y)} > 0.$$
\end{lemma}

\begin{proof}
By Assumption~\ref{ass:delay} we have that the function $\delay{\mu}$, $\mu \geq 0$, is concave. As a result, we have that
$$\sum_{y \in \comSet} \delay{\mu^*(\infl|y)} \geq \delay{ \sum_{y \in \comSet} \mu^*(\infl|y)}.$$
Furthermore,  by Assumption~\ref{ass:delay} we have that the function $\delay{\mu}$, $\mu \geq 0$, is positive. As a result, if we have that
$$\sum_{y \in \comSet} \mu^*(\infl|y) > 0, $$
then we obtain that
$$\sum_{y \in \comSet} \delay{\mu^*(\infl|y)} \geq \delay{ \sum_{y \in \comSet} \mu^*(\infl|y)} > 0.$$
This establishes the result of the lemma.
\end{proof}

%\newpage

\begin{lemma}\label{lemma:proxy_alpha}
Let $\alpha$, $0 < \alpha < 0.5$, be a given constant, then the following is true.
If $\allProxAllocationNE  = (\coreInteractVecOpt,\comAllocation^*,\contAllocation^*)$  is a  Nash equilibrium for the proxy game given by Definition~\ref{def:proxy_NE} such that 
$$\sum_{y \in \comSet} \delay{\mu^*(\infl|y)} > 0,$$
then there exists a content producer $z_0 \in \comSet$ such that
$$\sum_{y \in \Cw{z}} \delay{\mu^*(\infl|y)} > \alpha \sum_{y \in \comSet} \delay{\mu^*(\infl|y)}, \qquad z \in \Cw{z_0}.$$
\end{lemma}

\begin{proof}
Let $\epsilon_0$ be given by
$$\epsilon_0 = \sum_{y \in \comSet} \delay{\mu(\infl|y)} > 0.$$
Suppose that the result of the lemma is not true, and for $\alpha$, $0 < \alpha < 0.5$, there exist two content producer $z_0,z' \in \comSet$, $z_0 \neq z'$, such that
$$\sum_{y \in \Cw{z'}} \delay{\mu(\infl|y)} < \alpha \epsilon_0$$
and
$$\sum_{y \in \Cw{z_0}} \delay{\mu(\infl|y)} < \alpha \epsilon_0.$$
As by assumption we have that
$$\sum_{y \in \comSet} \delay{\mu(\infl|y)} = \epsilon_0 > 0,$$
we obtain that 
$$ \delay{\mu(\infl|z_0)} > (1-\alpha)\epsilon_0$$
and
$$ \delay{\mu(\infl|z')} > (1-\alpha) \epsilon_0.$$
Using the fact that for $\alpha$ we have that
$$ 0 < \alpha < 0.5,$$
it follows that
$$\sum_{y \in \Cw{z'}} \delay{\mu(\infl|y)} > (1-\alpha) \epsilon_0 > \alpha \epsilon_0$$
and
$$\sum_{y \in \Cw{z_0}} \delay{\mu(\infl|y)} > (1-\alpha) \epsilon_0 > \alpha \epsilon_0.$$
This leads to a contradiction to our assumption that
$$\sum_{y \in \Cw{z'}} \delay{\mu(\infl|y)} < \alpha \epsilon_0$$
and
$$\sum_{y \in \Cw{z_0}} \delay{\mu(\infl|y)} < \alpha \epsilon_0.$$
Hence we have that there is a most one content producer $z_0$ such that for $\alpha$, $0 < \alpha < 0.5$, we have that
$$\sum_{y \in \Cw{z'}} \delay{\mu(\infl|y)} < \alpha \epsilon_0.$$
This establishes the result of the lemma.
\end{proof}

%\newpage

\begin{lemma}\label{lemma:proxy_producer_pos_rate}
For every $\mu_0 > 0$ there exists a constant $k_{\Infl}$ such that if $\coreBudget > N k_{\Infl}$, then the following is true.   
If $\allProxAllocationNE  = (\coreInteractVecOpt,\comAllocation^*,\contAllocation^*)$  is a  Nash equilibrium for the proxy game given by Definition~\ref{def:proxy_NE} such that 
$$\sum_{y \in \comSet} \mu^*(\infl|y) > 0,$$
then there exists a content producer $z_0 \in \comSet$ such that 
$$\coreInteractVecOpt(z) \geq \mu_0, \qquad z \in \Cw{z_0}.$$
\end{lemma}

\begin{proof}
From Lemma~\ref{lemma:pq_min}, we have that there exists a $p_{\min} > 0$ and $q_{\min} > 0$ such that
$$ p(x|y) \geq p_{\min}, \qquad y \in \Com, x \in \setT,$$
and
$$ p(x|y) \geq q_{\min}, \qquad y \in \Com, x \in \setT.$$

Given a target following rate $\mu_0$ and a constant $\alpha$, $0 < \alpha < 0.5$, let $k_{\Infl}$ then be such that
$$k_{\Infl} > \mu_0$$
and
$$\ddelay{k_{\Infl}} < \ddelay{\mu_0} p_{\min}  q_{\min} \alpha.$$
Such a $k_{\infl}$ exists as from Assumption~\ref{ass:delay} we have that
$$ \ddelay{\mu} > 0, \qquad \mu \in \setR_+,$$
and
$$\lim_{\mu \to \infty} \ddelay{\mu} = 0.$$
Furthermore, let $z_0 \in \comSet$ be such that
$$\sum_{y \in \Cw{z}} \delay{\mu^*(\infl|y)} > \alpha \sum_{y \in \comSet} \delay{\mu^*(\infl|y)}, \qquad z \in \Cw{z_0}.$$
Such a $z_0 \in \comSet$ exists by Lemma~\ref{lemma:proxy_alpha}.

Using these values of $\mu_0$ and $k_{\infl}$, as well as our choice of $z_0 \in \comSet$, we then show if $\allProxAllocationNE  = (\coreInteractVec^*,\comAllocation^*,\contAllocation^*)$ is a  Nash equilibrium of the proxy game given by Definition~\ref{def:proxy_NE} with a rate budget $\coreBudget > N k_{\Infl}$, then we have that
  $$ \coreInteractVecOpt(z) \geq \mu_0, \qquad z \in \Cw{z_0}.$$

We show this result by contradiction as follows. Suppose that  $\allProxAllocationNE  = (\coreInteractVec^*,\comAllocation^*,\contAllocation^*)$ is a  Nash equilibrium of the proxy game given by Definition~\ref{def:proxy_NE} with a rate budget $\coreBudget > N k_{\Infl}$, and there exists a content producer $z' \in \Cw{z_0}$ such  
$$ \coreInteractVecOpt(z_0) < \mu_0$$
and
$$ \coreInteractVecOpt(z') < \mu_0.$$
As by Property f) of Lemma~\ref{lemma:conditions_proxy_NE} we have that
$$\sum_{z \in \comSet} \coreInteractVecOpt(z) = \coreBudget = N k_{\Infl}$$
and by construction we have that
$$ \coreInteractVecOpt(z') < \mu_0 < k_{\Infl},$$
it follows that there exists a content producer $z'' \in \comSet$ such that
$$  \coreInteractVecOpt(z'') \geq k_{\Infl}.$$

We then proceed as follows. First note that for all content producers $z \in \Com$ we have 
$$p_{\min}  q_{\min}  \sum_{y \in \Cw{z}}  \delay{\mu(\infl|y)} B(z|y)  \leq  \sum_{y \in \Cw{z}} \delay{\mu(\infl|y)} B(z|y) \leq \sum_{y \in \comSet}  \delay{\mu(\infl|y)}.$$
Using  Lemma~\ref{lemma:proxy_alpha} and the fact that b

By Assumption~\ref{ass:delay} the function $\delay{\mu}$, $\mu \geq 0$, is concave, and we have
$$\mu_0 \geq \coreInteractVecOpt(z')$$
that
$$ \ddelay{\coreInteractVecOpt(z')} \geq  \ddelay{\mu_0},$$
Using Lemma~\ref{lemma:proxy_alpha} and our choice of $z_0$, we have for the content producer $z'$ as given above that
$$\sum_{y \in \Cw{z'}} \delay{\mu^*(\infl|y)} > \alpha \sum_{y \in \comSet} \delay{\mu^*(\infl|y)},$$
and
$$  \ddelay{\mu_0}  p_{\min}  q_{\min} \alpha \sum_{y \in \comSet}  \delay{\mu(\infl|y)}   B(z'|y) 
< \ddelay{\coreInteractVecOpt(z')} \sum_{y \in \Cw{z'}} \delay{\mu(\infl|y)} B(z'|y).$$
Similarly, as for the content consumer $z''$ as defined above we have 
$$  \coreInteractVecOpt(z'') \geq k_{\infl},$$
we obtain that
$$\ddelay{\coreInteractVecOpt(z'')} \sum_{y \in \Cw{z''}} \delay{\mu(\infl|y)}
 B(z'|y) \leq   \ddelay{k_{\infl}}\sum_{y \in \comSet}  \delay{\mu(\infl|y)}.$$

As we have chosen the constant $k_{\infl}$ such that
$$\ddelay{k_{\Infl}} < \ddelay{\mu_0} p_{\min}  q_{\min}  \alpha, $$
we have that
$$\ddelay{k_{\Infl}} \sum_{y \in \comSet}  \delay{\mu(\infl|y)} < \ddelay{\mu_0} p_{\min}  q_{\min}  \alpha \sum_{y \in \comSet}  \delay{\mu(\infl|y)} ,$$
and it follows from the above inequalities that
$$   \ddelay{\coreInteractVecOpt(z'')} \sum_{y \in \Cw{z''}} \delay{\mu(\infl|y)} B(z''|y) <  \ddelay{\coreInteractVecOpt(z')} \sum_{y \in \Cw{z'}}  \delay{\mu(\infl|y)} B(z'|y).$$
However, as by construction we have that
$$\coreInteractVecOpt(z'') \geq k_{\infl},$$
this violates Property g) of the Lemma~\ref{lemma:conditions_proxy_NE} which provides the necessary and sufficient conditions for a Nash equilibrium of the proxy game.
This contradicts our assumption that  $\allProxAllocationNE  = (\coreInteractVecOpt,\comAllocation^*,\contAllocation^*)$ is a Nash equilibrium of the proxy game as given by Definition~\ref{def:proxy_NE}. The result of the lemma then follows.
\end{proof}

%\newpage

\begin{lemma}\label{lemma:proxy_pos_utility_consumers}
For every $\mu_0 > 0$ there exists a $k_{\Infl}$ such that if $\coreBudget > N k_{\Infl}$ then the following is true.   
If $\allProxAllocationNE  = (\coreInteractVecOpt,\comAllocation^*,\contAllocation^*)$  is a  Nash equilibrium for the proxy game given by Definition~\ref{def:proxy_NE} such that 
$$\sum_{y \in \comSet} \mu^*(\infl|y) > 0,$$
then for every content consumer $y \in \comSet$ we have that
$$\sum_{z \in \Cw{y}} \delay{\coreInteractVecOpt(z)} \geq (N-2)\delay{\mu_0}.$$
\end{lemma}

The result of this lemma follows immediately from Lemma~\ref{lemma:proxy_producer_pos_rate}, and from Assumption~\ref{ass:delay} which states that the function $\delay(\mu)$, $\mu \geq 0$, is strictly increasing.

%\newpage

\subsection{Properties of Nash Equilibrium  $\allProxAllocationNE$ for $\sum_{z \in \Cw{y}}  \delay{\coreInteractVecOpt(z)} \geq N \alpha \kappa_0$}

We next characterize the properties of a Nash equilibrium $\allProxAllocationNE= (\coreInteractVecOpt,\comAllocation^*,\contAllocation^*)$ such that
$$\sum_{z \in \Cw{y}}  \delay{\coreInteractVecOpt(z)} \geq N \alpha \kappa_0.$$

\begin{lemma}\label{lemma:proxy_pos_rate_consumers_1}
Let $\leafBudget$ be a given rate budget.   
Then for every $\kappa_0 >0$  and $\alpha$, $0< \alpha < 0.5$, there exist constants $N_0$ and $\epsilon_0 >0$ such that if $N > N_0$, then the following is true.
If $\allProxAllocationNE  = (\coreInteractVecOpt,\comAllocation^*,\contAllocation^*)$  is a Nash equilibrium for the proxy game given by Definition~\ref{def:proxy_NE} such that 
$$\sum_{z \in \Cw{y}}  \delay{\coreInteractVecOpt(z)} \geq \alpha N \kappa_0, \qquad y \in \comSet,$$
then we have that
$$\mu(\infl|y) > \epsilon_0, \qquad y \in \comSet.$$
\end{lemma}

\begin{proof}
Let $\kappa_0>0$  and $\alpha$, $0< \alpha < 0.5$, be given constants.
From Lemma~\ref{lemma:pq_min}  we have that there exists a $p_{\min} > 0$ and $q_{\min} > 0$ such that
$$ p(x|y) \geq p_{\min}, \qquad  y \in \Com, x \in \setT$$
and
$$ p(x|y) \geq q_{\min}, \qquad y \in \Com, x \in \setT.$$
Note that we have that
$$\frac{p_{\min}q_{\min}\kappa_0}{B_0} > 0.$$

We then choose $N_0$ and $\epsilon_0$, $0 < \epsilon_0 < \leafBudget$, such that
$$\ddelay{\leafBudget - \epsilon_0} r_0 B_0 <   \ddelay{\epsilon_0} r_p p_{\min}q_{\min}  \alpha (N_0-2) \kappa_0.$$
Note that such constants exist.

Using these values for $N_0$ and $\epsilon_0$, we then show that if $N > N_0$ and $\allProxAllocationNE  = (\coreInteractVecOpt,\comAllocation^*,\contAllocation^*)$  is a Nash equilibrium for the proxy game given by Definition~\ref{def:proxy_NE}, such that for content consumer $y \in \comSet$ we have that
$$\sum_{z \in \Cw{y}}  \delay{\coreInteractVecOpt(z)} \geq \alpha N \kappa_0,$$
then we have that
$$\mu^*(\infl|y) > \epsilon_0.$$

We show this result by contradiction as follows. For $N > N_0$, suppose that  $\allProxAllocationNE  = (\coreInteractVec^*,\comAllocation^*,\contAllocation^*)$ is a  Nash equilibrium of the proxy game given by Definition~\ref{def:proxy_NE} such that for content consumer $y \in \comSet$  we have that
$$\sum_{z \in \Cw{y}}  \delay{\coreInteractVecOpt(z)} \geq \alpha N \kappa_0$$
and
$$\mu(\infl|y) \leq \epsilon_0.$$
By construction we have
$$0 < \epsilon_0 < \leafBudget,$$
and we obtain that 
$$\lambda^*(y) \geq \leafBudget - \epsilon_0 > 0.$$

We then proceed as follows.
Combining the fact that
$$\sum_{z \in \Cw{y}}  \delay{\coreInteractVecOpt(z)} \geq \alpha N \kappa_0$$
and
$$\mu(\infl|y) \leq \epsilon_0,$$
with Assumption~\ref{ass:delay} that states that the function $\delay{\mu}$, $\mu \geq 0$, is concave, we obtain that 
$$  \ddelay{\epsilon_0} p_{\min}q_{\min}  \alpha (N_0-2) \kappa_0 \leq  \ddelay{\mu^*(\infl|y)} \sum_{z \in \Cw{y}} \delay{\coreInteractVecOpt(z)} B(z|y).$$
Furthermore, as we have that
$$\lambda^*(y) \geq \leafBudget - \epsilon_0 >0,$$
and as by  Assumption~\ref{ass:delay} the function $\delay{\mu}$, $\mu \geq 0$, is concave, we obtain that
$$ \ddelay{\lambda^*(y)} B_0 \leq  \ddelay{\leafBudget - \epsilon_0} B_0.$$
By construction, we have that
$$\ddelay{\leafBudget - \epsilon_0} r_0 B_0 <   \ddelay{\epsilon_0} r_p p_{\min}q_{\min}  \alpha (N_0-2) \kappa_0$$
and we obtain from the above inequalities that
$$ \ddelay{\mu^*(\infl|y)} r_p \sum_{z \in \Cw{y}} \delay{\coreInteractVecOpt(z)}  B(z|y) >  \ddelay{\lambda^*(y)} r_0 B_0. $$
However, as by construction we have that
$$\lambda^*(y) \geq \leafBudget - \epsilon_0 > 0,$$
this violates  Property d) of Lemma~\ref{lemma:conditions_proxy_NE}, which provides necessary and sufficient conditions for a Nash equilibrium for the proxy game. This implies that the allocation   $\allProxAllocationNE  = (\coreInteractVecOpt,\comAllocation^*,\contAllocation^*)$ is not a Nash equilibrium for the proxy game given by Definition~\ref{def:proxy_NE}.

This contradicts our assumption that  $\allProxAllocationNE  = (\coreInteractVecOpt,\comAllocation^*,\contAllocation^*)$ is a Nash equilibrium for the proxy game given by Definition~\ref{def:proxy_NE}. The result of the lemma then follows.
\end{proof}

%\newpage

\subsection{Properties of Nash Equilibrium  $\allProxAllocationNE$}\label{app:prop_NE_proxy}
Finally, we derive the key properties of a  Nash equilibrium $\allProxAllocationNE  = (\coreInteractVecOpt,\comAllocation^*,\contAllocation^*)$  for the proxy game as given by Definition~\ref{def:proxy_NE}.

The first result provides conditions such that for a  Nash equilibrium $\allProxAllocationNE  = (\coreInteractVecOpt,\comAllocation^*,\contAllocation^*)$  of the proxy game we have 
  $$ \sum_{y \in \Com} \mu^*(\infl|y) > 0.$$

\begin{lemma}\label{lemma:proxy_pos_rate}
There exist constants $N_0$ and $k_{\Infl}$ such that if $N > N_0$ and $\coreBudget > N k_{\Infl}$, then the following is true. If $\allProxAllocationNE  = (\coreInteractVecOpt,\comAllocation^*,\contAllocation^*)$  is a Nash equilibrium for the game with perfect information as given by Definition~\ref{def:proxy_NE}, then  we have that 
  $$ \sum_{y \in \Com} \mu^*(\infl|y) > 0.$$
\end{lemma}

\begin{proof}
From Lemma~\ref{lemma:pq_min}  we have that there exists a $p_{\min} > 0$ and $q_{\min} > 0$ such that
$$ p(x|y) \geq p_{\min}, \qquad  y \in \Com, x \in \setT$$
and
$$ p(x|y) \geq q_{\min}, \qquad y \in \Com, x \in \setT.$$

We then choose  constants  $N_0$ and $k_{\Infl} > 0$ such that
$$\ddelay{0} r_0 B_0 <  \ddelay{0} \frac{(N_0-1) r_p p_{\min}q_{\min}\delay{k_{\Infl}}}{B_0}.$$
Note that such constants exist.

We then show that the following is true for these values for $N_0$ and $k_{\Infl}$. If for $N > N_0$ we have that $\allProxAllocationNE  = (\coreInteractVecOpt,\comAllocation^*,\contAllocation^*)$  is a Nash equilibrium with $\coreBudget > N k_{\Infl}$, then  we have 
  $$ \sum_{y \in \Com} \mu^*(\infl|y) > 0.$$

We prove this results by contradiction as follows. For $N > N_0$, suppose that  $\allProxAllocationNE  = (\coreInteractVec^*,\comAllocation^*,\contAllocation^*)$ is a  Nash equilibrium as given by Definition~\ref{def:proxy_NE} with $\coreBudget > N k_{\Infl}$, such  that 
$$ \sum_{y \in \Com} \mu^*(\infl|y) = 0.$$
Note that in this case we have that
$$ \lambda^*(y) = \leafBudget > 0.$$
  
We then proceed as follows.  As for the Nash equilibrium  $\allProxAllocationNE  = (\coreInteractVec^*,\comAllocation^*,\contAllocation^*)$  we have that 
$$ \sum_{y \in \Com} \mu^*(\infl|y) = 0,$$
we obtain from Assumption~\ref{ass:infl_alloc} that in this case we have
$$  \coreInteractVecOpt(z) = \coreBudget/N = k_{\Infl}, \qquad z \in \comSet.$$
It then  follows that for every content consumer $y \in \comSet$ we have
$$ \sum_{z \in \Cw{y}} \coreInteractVecOpt(z) = (N-1)\delay{k_{\Infl}}.$$

We then have that for $N > N_0$ that
$$ (N_0-1) \ddelay{0}  p_{\min}q_{\min}\delay{k_{\Infl}} \leq  \ddelay{0} \sum_{z \in \Cw{y}} \delay{\coreInteractVecOpt(z)} B(z|y).$$
Furthermore as by Assumption~\ref{ass:delay} the function $\delay{\mu}$, $\mu \geq 0$, is concave, we have that
$$  \ddelay{\leafBudget} B_0 \leq  \ddelay{0} B_0.$$

As by construction we have that
$$\ddelay{0} r_0 B_0 <  \ddelay{0} \frac{(N_0-1) r_p p_{\min}q_{\min}\delay{k_{\Infl}}}{B_0},$$
we obtain from the above inequalities that
$$ \ddelay{\leafBudget} r_0 B_0 < \ddelay{0} \sum_{z \in \Cw{y}} \delay{\coreInteractVecOpt(z)} B(z|y).$$
However, as by assumption we have that 
$$ \sum_{y \in \Com} \mu^*(\infl|y) = 0$$
and hence
$$ \lambda^*(y) = \leafBudget > 0,$$
this violates Property d) of the Lemma~\ref{lemma:conditions_proxy_NE} which provides the necessary and sufficient conditions for a Nash equilibrium of the proxy game.
This contradicts our assumption that  $\allProxAllocationNE  = (\coreInteractVecOpt,\comAllocation^*,\contAllocation^*)$ is a Nash equilibrium of the proxy game as given by Definition~\ref{def:proxy_NE}. The result of the lemma then follows.
\end{proof}

%\newpage

The next result states that  if the community size $N$ and the rate budget $\coreBudget$ of the influencer are large enough, then under a   Nash equilibrium of the proxy game we have that each  consumer follows the influencer with a positive rate.

\begin{prop}\label{prop:proxy_pos_rate_consumers}
Let $\leafBudget$ be a given rate budget.
Then there exist constants $N_0$, $k_{\Infl}$, and $\epsilon_0 > 0$, such that if $N > N_0$ and $\coreBudget > N k_{\Infl}$, the following is true.
If $\allProxAllocationNE  = (\coreInteractVecOpt,\comAllocation^*,\contAllocation^*)$  is a  Nash equilibrium for the proxy game given by Definition~\ref{def:proxy_NE}, then  we have that 
  %$$ \sum_{y \in \Com} \mu^*(\infl|y) > 0.$$
$$ \mu^*(\infl|y) > \epsilon_0, \qquad y \in \comSet.$$ 
\end{prop}

\begin{proof}
Given the rate budget $\leafBudget$, we choose the constants $N_0 > 2$, $k_{\Infl}$, and $\epsilon_0>0$, such that for  $N > N_0$ and $\coreBudget > N k_{\Infl}$ the following properties are true:
\begin{enumerate}
\item  If $\allProxAllocationNE  = (\coreInteractVecOpt,\comAllocation^*,\contAllocation^*)$  is a  Nash equilibrium for the proxy game given by Definition~\ref{def:proxy_NE}, such that 
$$\sum_{y \in \comSet} \mu^*(\infl|y) > 0,$$
then for every content consumer $y \in \comSet$ we have that
$$\sum_{z \in \Cw{y}} \delay{\coreInteractVecOpt(z)} \geq (N-2)\delay{\mu_0}.$$
\item   If $\allProxAllocationNE  = (\coreInteractVecOpt,\comAllocation^*,\contAllocation^*)$  is a Nash equilibrium for the proxy game given by Definition~\ref{def:proxy_NE}, such that for content consumer $y \in \comSet$ we have that
$$\sum_{z \in \Cw{y}}  \delay{\coreInteractVecOpt(z)} \geq \alpha N \kappa_0,$$
then we have that
$$\mu(\infl|y) > \epsilon_0.$$
\item    If $\allProxAllocationNE  = (\coreInteractVecOpt,\comAllocation^*,\contAllocation^*)$  is a Nash equilibrium for the proxy game given by Definition~\ref{def:proxy_NE}, then  we have that 
  $$ \sum_{y \in \Com} \mu^*(\infl|y) > 0.$$
\end{enumerate}
Such constants  $N_0 > 2$, $k_{\Infl}$ and $\epsilon_0$ exist by Lemma~\ref{lemma:proxy_pos_utility_consumers}, Lemma~\ref{lemma:proxy_pos_rate_consumers_1}, and  Lemma~\ref{lemma:proxy_pos_rate}.

Note that for these values of  $N_0 > 2$, $k_{\Infl}$ and $\mu_0 > 0$, we have from Property 2) that
$$\sum_{z \in \Cw{y}} \delay{\coreInteractVecOpt(z)} \geq \alpha N \kappa_0$$
with
$$\alpha = \frac{1}{3}$$
and
$$\kappa_0 = \delay{\mu_0} > 0.$$

The result of the lemma then follows immediately from Property 3).
\end{proof}

%\newpage
The next result states that  if the community size $N$ and the rate budget $\coreBudget$ of the influencer are large enough, then under a   Nash equilibrium of the proxy game we that the influener allocates a large rate to each producer $z \in \comSet$.

\begin{prop}\label{prop:proxy_pos_rate_producers}
Let $\leafBudget$ be a given rate budget, and let $\mu_0$ be a given rate. Then there exist constants $N_0$ and $k_{\Infl}$ such that if $N > N_0$ and $\coreBudget > N k_{\Infl}$, the following is true.
If $\allProxAllocationNE  = (\coreInteractVecOpt,\comAllocation^*,\contAllocation^*)$ is a Nash equilibrium of the proxy game given by Definition~\ref{def:proxy_NE}, then we have that 
  $$ \coreInteractVecOpt(z) \geq \mu_0, \qquad z \in \Com.$$
\end{prop}

\begin{proof}
From Lemma~\ref{lemma:pq_min}, we have that there exists a $p_{\min} > 0$ and $q_{\min} > 0$ such that
$$ p(x|y) \geq p_{\min}, \qquad y \in \Com, x \in \setT,$$
and
$$ p(x|y) \geq q_{\min}, \qquad y \in \Com, x \in \setT.$$

Given the rate budget $\leafBudget$, we choose the constants $N_0 > 2$, $k_{\Infl}>0$, and $\epsilon_0>0$, such that for  $N > N_0$ and $\coreBudget > N k_{\Infl}$ the following properties are true:
\begin{enumerate}
\item  If $\allProxAllocationNE  = (\coreInteractVecOpt,\comAllocation^*,\contAllocation^*)$  is a Nash equilibrium for the proxy game given by Definition~\ref{def:proxy_NE}, then  we have that
  $$ \mu^*(\infl|y) > \epsilon_0, \qquad y \in \comSet.$$
\item $\ddelay{k_{\Infl}} \delay{\leafBudget}< \ddelay{\mu_0} p_{\min}  q_{\min}  \delay{\epsilon_0} $.
\item $k_{\infl} > \mu_0$.
\end{enumerate}
Such constants exist by Proposition~\ref{prop:proxy_pos_rate_consumers} and by Assumption~\ref{ass:delay} which states that for the function $\delay{\mu}$, $\mu \geq 0$, we have that
$$ \ddelay{\mu} > 0, \qquad \mu \in \setR_+,$$
and
$$\lim_{\mu \to \infty} \ddelay{\mu} = 0.$$

Using these values of $\mu_0$, $k_{\infl}$, and $m$, we then show if $\allProxAllocationNE  = (\coreInteractVec,\comAllocation^*,\contAllocation^*)$ is an optimal Nash equilibrium of the proxy game given by Definition~\ref{def:proxy_NE} with a rate budget $\coreBudget > N k_{\Infl}$, then we have that 
  $$ \coreInteractVecOpt(z) \geq \mu_0, \qquad z \in \Com.$$

We show this result  by contradiction as follows. Suppose that  $\allProxAllocationNE  = (\coreInteractVec,\comAllocation^*,\contAllocation^*)$ is an optimal Nash equilibrium of the proxy game given by Definition~\ref{def:proxy_NE} with a rate budget $\coreBudget > N k_{\Infl}$, and there exists a content producer $z_0 \in \comSet$ such  
  $$ \coreInteractVecOpt(z_0) < \mu_0.$$
As by construction we have that
$$k_{\infl|} > \mu_0,$$
it follows from Property f) of Lemma~\ref{lemma:conditions_proxy_NE} that in this case there exists a content producer $z' \in \comSet$ such that
$$  \coreInteractVecOpt(z') \geq k_{\infl},$$

We then proceed as follows. First note that for every content producer $z \in \Com$ that
$$p_{\min}  q_{\min} (N-1) \delay{\epsilon_0} \leq  \sum_{y \in \Cw{z}} \delay{\mu_i} B(z|y)  \leq   (N-1) \delay{\leafBudget}.$$
We then have for the content producer $z_0$ as given above that
$$  \ddelay{\mu_0} (N-1) p_{\min}  q_{\min} \delay{\epsilon_0} \leq \ddelay{\coreInteractVecOpt(z_0)} \sum_{y \in \Cw{z_0}} \delay{\mu(\infl|y)} B(z_0|y).$$
Similarly, for the content consumer $z'$ as defined above, we have that
$$\ddelay{\coreInteractVecOpt(z')} \sum_{y \in \Cw{z'}} \delay{\mu(\infl|y)}  B(z_0|y) \leq   \ddelay{k_{\Infl}} (N-1)\delay{\leafBudget},$$
where we used the fact that by Assumption~\ref{ass:delay} the function $\delay{\mu}$ is concave. 

As we have chosen the constant $k_{\infl}$ such that
$$\ddelay{k_{\Infl}} \delay{\leafBudget}< \ddelay{\mu_0} p_{\min}  q_{\min}  \delay{\epsilon_0} $$
we have that
$$\ddelay{k_{\Infl}} (N-1) \delay{\leafBudget}< \ddelay{\mu_0} (N-1) p_{\min}  q_{\min}  \delay{\epsilon_0} ,$$
and it follows from the above two inequalities that
$$   \ddelay{\coreInteractVecOpt(z')} \sum_{y \in \Cw{z'}} \delay{\mu(\infl|y)}  B(z'|y) <  \ddelay{\coreInteractVecOpt(z_0)} \sum_{y \in \Cw{z_0}} \delay{\mu(\infl|y)}B(z_0|y) .$$
However, as by construction we have that
$$  \coreInteractVecOpt(z') \geq k_{\infl} > 0,$$
this violates Property g) of the Lemma~\ref{lemma:conditions_proxy_NE} which provides the necessary and sufficient conditions for a Nash equilibrium of the proxy game.
This contradicts our assumption that  $\allProxAllocationNE  = (\coreInteractVecOpt,\comAllocation^*,\contAllocation^*)$ is a Nash equilibrium of the proxy game as given by Definition~\ref{def:proxy_NE}. The result of the proposition then follows

\end{proof}

%\newpage
\section{Analysis of Nash Equilibrium $\allAllocationNE$ for Game with Perfect Information}\label{app:analysis_perfect}

In this section, we derive additional properties of a Nash equilibrium of the game with perfect information as given by Definition~\ref{def:stable_allocation} that we use to in Appendix~\ref{app:proof_NE_proxy_perfect} to prove Proposition~\ref{prop:NE_proxy_perfect}.

\subsection{Preliminary Result}
We first derive a result on the properties of an  allocation $\allAllocation = (\coreInteractVec,\comAllocation,\contAllocation)$ that we use in our analysis.

Recall from Section~\ref{sec:results} that $\possibleStrategy$ is the set of all admissible allocations  $\allAllocation = (\coreInteractVec,\comAllocation,\contAllocation)$, i.e. the set of all non-negative following rate allocations  $\coreInteractVec$ and $\comAllocation$ that satisfy the rate budget constraints $\coreBudget$ for influencer, and the rate budget constraint $\leafBudget$ for all content consumers $y \in \comSet$, respectively, and all the content allocations $X$ such that $x(z) \in \setT$, $z \in \comSet$.

We then have the following lemma.

\begin{lemma}\label{lemma:allocation_no_direct_rate}
  For a given rate budget $\leafBudget$, there constants $N_0$ and $\mu_0$ such that for $N> N_0$ the following is true.
If $\allAllocation  = (\coreInteractVec,\comAllocation,\contAllocation) \in \strategySpace$ is an admissible allocation such that
$$ \coreInteractVec(z) \geq \mu_0, \qquad z \in \comSet,$$
then we have for all content consumers $y \in \comSet$ that
$$ \ddelay{ \mu(z|y)} B(z|y) <  \ddelay{ \mu(\infl|y)} \sum_{z \in \Cw{y}} B(z|y) \delay{\coreInteractVec(z)}, \qquad z \in \Cw{y}.$$
\end{lemma}

\begin{proof}
From Lemma~\ref{lemma:pq_min}  we have that there exists a $p_{\min} > 0$ and $q_{\min} > 0$ such that
$$ p(x|y) \geq p_{\min}, \qquad  y \in \Com, x \in \setT$$
and
$$ p(x|y) \geq q_{\min}, \qquad y \in \Com, x \in \setT.$$

Given the rate budget $\leafBudget$, we choose the constants $N_0$ and $\mu_0$ such that
$$ \ddelay{0} < \ddelay{\leafBudget} (N_0-1)  p_{\min}  q_{\min} \delay{\mu_0}.$$
Note that such constants exist.

By Assumption~\ref{ass:delay} the function $\delay{\mu}$, $\mu \geq 0$, is concave, and we have that 
$$  \ddelay{ \mu^*(z|y)} B(z|y) \leq \ddelay{0}.$$
By Assumption~\ref{ass:delay} the function $\delay{\mu}$, $\mu \geq 0$, is concave and strictly increasing, and  we have that
$$\ddelay{\leafBudget} (N_0-1)  p_{\min}  q_{\min} \delay{\mu_0} \leq \ddelay{ \mu^*(\infl|y)} \sum_{z \in \Cw{y}} \delay{\coreInteractVecOpt(z)} B(z|y).$$
By construction we have that
$$ \ddelay{0} < \ddelay{\leafBudget} (N_0-1)  p_{\min}  q_{\min} \delay{\mu_0},$$
and it follows from the above inequalities that
$$ \ddelay{ \mu(z|y)} B(z|y) <  \ddelay{ \mu(\infl|y)} \sum_{z' \in \Cw{y}} \delay{\coreInteractVec(z')} B(z'|y), \qquad z \in \Cw{y}.$$
This establishes the result of the lemma.
\end{proof}

%\newpage

\subsection{Optimality Condition for  Nash Equilibrium  $\allAllocationNE$}
The next result provides necessary and sufficient conditions for a Nash equilibrium of the game with perfect information.

\begin{lemma}\label{lemma:conditions_perfect_NE}

  An $\allAllocationNE  = (\coreInteractVecOpt,\comAllocation^*,\contAllocation^*)$  is a  Nash equilibrium as given by Definition~\ref{def:stable_allocation}, if an only if the following conditions hold true,
\begin{enumerate}
\item[a)]  For all content producers $z \in \comSet$, the content allocation $x^*(z)$ is a solution for the maximization problem  $\optcontperiphery{\coreRateVec,\comAllocation}{y}$.

\item[b)] For all content consumers $y \in \comSet$ we have that
$$ \lambda^*(y) + \mu^*(\infl|y)  + \sum_{z \in \Cw{y}}  \mu^*(z|y)  = \leafBudget$$   
  
\item[c)] If for content consumer $y \in \comSet$ we have that $\lambda^*(y) > 0$, then we have
$$\ddelay{\lambda^*(y)} r_0 B_0 \geq \ddelay{\mu^*(z|y)} r_p B(z|y), \qquad z \in \Cw{y}$$
and
$$\ddelay{\lambda^*(y)} r_0 B_0 \geq  \ddelay{\mu^*(\infl|y)} r_p \sum_{z \in \Cw{y}} \delay{\coreInteractVecOpt(z)} B(z|y).$$

\item[d)] If  for content consumer $y \in \comSet$ we have that $\mu^*(\infl|y) > 0$, then we have
$$\ddelay{\mu^*(\infl|y)} \sum_{z' \in \Cw{y}} \delay{\coreInteractVecOpt(z')}  B(z'|y) \geq \ddelay{\mu^*(z|y)} B(z|y), \qquad z \in \Cw{y}$$
and
$$\ddelay{\mu^*(\infl|y)} r_p \sum_{z \in \Cw{y}} \delay{\coreInteractVecOpt(z)} B(z|y) \geq \ddelay{\lambda^*(y)} r_0 B_0.$$

\item[e)] If for content consumer $y \in \comSet$ and content producer $z \in \Cw{y}$ we have that $\mu^*(z|y) > 0$, then we have
$$\ddelay{\mu^*(z|y)} r_p B(z|y) \geq \ddelay{\lambda^*(y)} r_0 B_0,$$
  $$\ddelay{\mu^*(z|y)} B(z|y) \geq  \ddelay{\mu^*(\infl|y)} \sum_{z' \in \Cw{y}} \delay{\coreInteractVecOpt(z')} B(z'|y),$$
  and
  $$ \ddelay{\mu^*(z|y)} B(z|y) \geq \ddelay{\mu^*(z'|y)} B(z'|y), \qquad z' \in \Cw{y,z}.$$

\item[f)] We have that
  $$\sum_{z \in \comSet} \coreInteractVecOpt(z) =  \coreBudget.$$  
  
\item[g)] If for content producer $z \in \comSet$ we have that $\coreInteractVecOpt(z) > 0$, then we have for $z' \in \Cw{z}$ that 
$$ \ddelay{\coreInteractVecOpt(z)} \sum_{y \in \Cw{z}} \delay{\mu^*(\infl|y)}  B(z|y)\geq \ddelay{\coreInteractVecOpt(z')} \sum_{y \in \Cw{z'}} \delay{\mu^*(\infl|y)} B(z'|y).$$

\end{enumerate}

\end{lemma}

We can prove Lemma~\ref{lemma:conditions_perfect_NE} using the same line of argument that we used to prove Lemma~\ref{lemma:conditions_proxy_NE}, and we omit here a detailed proof.

%\newpage

\subsection{Properties of Nash Equilibrium  $\allAllocationNE$ for  $ \sum_{y \in \Com} \mu^*(\infl|y) > 0$}

We next characterize the properties of a Nash equilibrium $\allProxAllocation= (\coreInteractVecOpt,\comAllocation^*,\contAllocation^*)$ such that
\begin{equation}\label{eq:perfect_infl_pos_rate}
  \sum_{y \in \comSet} \mu^*(\infl|y) > 0.
\end{equation}
In Appendix~\ref{app:prop_NE_perfect} we provide conditions under which the inequality given by Eq.~\eqref{eq:perfect_infl_pos_rate} holds.

 The following lemmas, Lemma~\ref{lemma:perfect_infl_pos_utility}~-~\ref{lemma:perfect_pos_utility_consumers},  can be proved using the same argument as given in the proof for Lemma~\ref{lemma:proxy_infl_pos_utility}~-~\ref{lemma:proxy_pos_utility_consumers} that we obtained for the proxy game in Appendix~\ref{app:analysis_proxy}, and we omit here detailed proofs.

\begin{lemma}\label{lemma:perfect_infl_pos_utility}
If $\allAllocationNE  = (\coreInteractVecOpt,\comAllocation^*,\contAllocation^*)$  is a  Nash equilibrium for the game with perfect information as given by Definition~\ref{def:stable_allocation} such that 
$$\sum_{y \in \comSet} \mu(\infl|y) > 0,$$
then  we have that 
$$\sum_{y \in \comSet} \delay{\mu(\infl|y)} > 0.$$
\end{lemma}

\begin{lemma}
Let $\alpha$, $0 < \alpha < 0.5$, be a given constant, then the following is true.  
If $\allAllocationNE  = (\coreInteractVecOpt,\comAllocation^*,\contAllocation^*)$  is a  Nash equilibrium for the game with perfect information as given by Definition~\ref{def:stable_allocation} such that 
$$\sum_{y \in \comSet} \mu(\infl|y) > 0,$$
then there exists a content producer $z_0 \in \comSet$ such that
$$\sum_{y \in \Cw{z}} \delay{\mu^*(\infl|y)} > \alpha \sum_{y \in \comSet} \delay{\mu^*(\infl|y)}, \qquad z \in \Cw{z_0}.$$
\end{lemma}

\begin{lemma}\label{lemma:perfect_producer_pos_rate}
For every $\mu_0 > 0$ there exists a constant $k_{\Infl}$ such that if $\coreBudget > N k_{\Infl}$, then the following is true.   
If $\allAllocationNE  = (\coreInteractVecOpt,\comAllocation^*,\contAllocation^*)$  is a  Nash equilibrium for the game with perfect information as given by Definition~\ref{def:stable_allocation} such that 
$$\sum_{y \in \comSet} \mu(\infl|y) > 0,$$
then there exists a content producer $z_0 \in \comSet$ such that 
$$\coreInteractVecOpt(z) \geq \mu_0, \qquad z \in \Cw{z_0}.$$
\end{lemma}

\begin{lemma}\label{lemma:perfect_pos_utility_consumers}
For every $\mu_0 > 0$ there exists a $k_{\Infl}$ such that if $\coreBudget > N k_{\Infl}$ then the following is true.    
If $\allAllocationNE  = (\coreInteractVecOpt,\comAllocation^*,\contAllocation^*)$  is a  Nash equilibrium for the game with perfect information as given by Definition~\ref{def:stable_allocation} such that 
$$\sum_{y \in \comSet} \mu(\infl|y) > 0,$$
then for every content consumer $y \in \comSet$ we have that
$$\sum_{z \in \Cw{y}} \delay{\coreInteractVecOpt(z)} \geq (N-2)\delay{\mu_0}.$$
\end{lemma}

%\newpage

\subsection{Properties of Nash Equilibrium  $\allAllocationNE$ for  $\sum_{z \in \Cw{y}}  \delay{\coreInteractVecOpt(z)} \geq \alpha N\kappa_0$}

We next characterize the properties of a Nash equilibrium $\allAllocationNE= (\coreInteractVecOpt,\comAllocation^*,\contAllocation^*)$ such that
$$\sum_{z \in \Cw{y}}  \delay{\coreInteractVecOpt(z)} \geq N \alpha \kappa_0.$$

\begin{lemma}\label{lemma:perfect_pos_rate_consumers_1}
Let $\leafBudget$ be a given rate budget.   
Then for every $\kappa_0 >0$  and $\alpha$, $0< \alpha < 0.5$, there exist constants $N_0$ and $\epsilon_0 >0$ such that if $N > N_0$, then the following is true.
If $\allAllocationNE  = (\coreInteractVecOpt,\comAllocation^*,\contAllocation^*)$  is a Nash equilibrium for the game with perfect information as given by Definition~\ref{def:stable_allocation} such that
$$\sum_{z \in \Cw{y}}  \delay{\coreInteractVecOpt(z)} \geq \alpha N \kappa_0, \qquad y \in \comSet,$$
then we have that
$$\mu(\infl|y) > \epsilon_0, \qquad y \in \comSet.$$
\end{lemma}

\begin{proof}
Let $\kappa_0>0$  and $\alpha$, $0< \alpha < 0.5$, be given constants.  

From Lemma~\ref{lemma:pq_min}  we have that there exists a $p_{\min} > 0$ and $q_{\min} > 0$ such that
$$ p(x|y) \geq p_{\min}, \qquad  y \in \Com, x \in \setT$$
and
$$ p(x|y) \geq q_{\min}, \qquad y \in \Com, x \in \setT.$$
Note that we have that
$$\frac{p_{\min}q_{\min}\kappa_0}{B_0} > 0.$$

We then choose the integer $N_0$ and $\epsilon_0$ such that
$$ 0 <  \epsilon_0  < \leafBudget,$$
and
$$ \ddelay{0} \max\{r_p,r_0 B_0\} <   \ddelay{\epsilon_0} \alpha N_0 \kappa_0  r_p p_{\min}q_{\min}.$$
Note that such a $N_0$ and $\epsilon_0$ exist.

Using these values for $N_0$ and $\epsilon_0$, we then show that if $N > N_0$ and $\allAllocationNE  = (\coreInteractVecOpt,\comAllocation^*,\contAllocation^*)$  is a Nash equilibrium for the proxy game given by Definition~\ref{def:stable_allocation}, such that for content consumer $y \in \comSet$ we have that
$$\sum_{z \in \Cw{y}}  \delay{\coreInteractVecOpt(z)} \geq \alpha N \kappa_0,$$
then we have that
$$\mu^*(\infl|y) > \epsilon_0.$$

We show this result by contradiction as follows. For $N > N_0$, suppose that  $\allAllocationNE  = (\coreInteractVec^*,\comAllocation^*,\contAllocation^*)$ is a  Nash equilibrium of the proxy game given by Definition~\ref{def:stable_allocation} such that for content consumer $y_0 \in \comSet$ we have that
$$\sum_{z \in \Cw{y_0}}  \delay{\coreInteractVecOpt(z)} \geq \alpha N \kappa_0$$
and
$$\mu(\infl|y_0) \leq \epsilon_0.$$

Using Property b) of Lemma~\ref{lemma:conditions_perfect_NE}, we have in this case for content consumer $y_0 \in \comSet$ that either
\begin{equation}\label{eq:perfect_pos_rate_outside_y0}
  \lambda^*(y_0) > 0
\end{equation}  
or there exists a content producer $z \in \Cw{y}$ such that
\begin{equation}\label{eq:perfect_pos_direct_rate_y0}
 \mu^*(z|y_0) > 0.
\end{equation}

We then proceed as follows.
Combining the fact that
$$\sum_{z \in \Cw{y_0}}  \delay{\coreInteractVecOpt(z)} \geq \alpha N \kappa_0$$
and
$$\mu(\infl|y_0) \leq \epsilon_0,$$
with Assumption~\ref{ass:delay}, which states that the function $\delay{\mu}$, $\mu \geq 0$, is concave, we obtain that 
$$  \ddelay{\epsilon_0} p_{\min}q_{\min}  \alpha (N_0-2) \kappa_0 \leq  \ddelay{\mu^*(\infl|y_0)} \sum_{z \in \Cw{y_0}}  \delay{\coreInteractVecOpt(z)}B(z|y_0).$$
Furthermore, as by Assumption~\ref{ass:delay} the function $\delay{\mu}$, $\mu \geq 0$, is concave, we have that 
$$  \ddelay{\lambda^*(y_0)} B_0  \leq \ddelay{0} B_0$$
and
$$   \ddelay{\mu^*(z'|y_0)} B(z'|y) \leq \ddelay{0}.$$
By construction, we have that 
$$ 0 <  \epsilon_0  < \leafBudget,$$
and
$$ \ddelay{0} \max\{r_p,r_0 B_0\} <   \ddelay{\epsilon_0} \alpha N_0 \kappa_0  r_p p_{\min}q_{\min},$$
and we obtain from the above inequalities that
\begin{equation}\label{eq:fo_perfect_outside_y0}
  \ddelay{\lambda^*(y_0)} r_0 B_0  <  \ddelay{\mu^*(\infl|y_0)} r_p \sum_{z \in \Cw{y_0}} \delay{\coreInteractVecOpt(z)} B(z|y)
\end{equation}
and
\begin{equation}\label{eq:fo_perfect_direct_rate_y0}
  \ddelay{\mu^*(z'|y_0)} B(z'|y_0)  <  \ddelay{\mu^*(\infl|y_0)} \sum_{z \in \Cw{y_0}} \delay{\coreInteractVecOpt(z)} B(z|y_0).
\end{equation}  
However, by  Lemma~\ref{lemma:conditions_perfect_NE}, and Eq.~\eqref{eq:perfect_pos_rate_outside_y0} and  Eq.~\eqref{eq:perfect_pos_direct_rate_y0}, we either have that 
$$\lambda^*(y_0) > 0$$
or there exists a content producer $z \in \Cw{y_0}$ such that
$$\mu^*(z|y_0) > 0,$$
and the inequalities given by Eq.~\eqref{eq:fo_perfect_outside_y0} and  Eq.~\eqref{eq:fo_perfect_direct_rate_y0}  either violate Property c) or e) of Lemma~\ref{lemma:conditions_perfect_NE}.
This contradicts our assumption that  $\allProxAllocationNE  = (\coreInteractVecOpt,\comAllocation^*,\contAllocation^*)$ is a Nash equilibrium as given by Definition~\ref{def:stable_allocation}. The result of the lemma then follows.
\end{proof}

%\newpage

\subsection{Properties of Nash Equilibrium  $\allAllocationNE$}\label{app:prop_NE_perfect}
Finally, we derive the key properties of a  Nash equilibrium $\allAllocationNE  = (\coreInteractVecOpt,\comAllocation^*,\contAllocation^*)$  for the game with perfect information as given by Definition~\ref{def:stable_allocation}.

The first result provides conditions such that for a  Nash equilibrium $\allAllocationNE  = (\coreInteractVecOpt,\comAllocation^*,\contAllocation^*)$  of the game with perfect information  we have 
  $$ \sum_{y \in \Com} \mu^*(\infl|y) > 0.$$

\begin{lemma}
There exist constants $N_0$ and $k_{\Infl}$ such that if $N > N_0$ and $\coreBudget > N k_{\Infl}$, then the following is true.  
If  $\allAllocationNE = (\coreInteractVec,\comAllocation^*,\contAllocation^*)$  is a  equilibrium of the game with perfect information as given by Definition~\ref{def:stable_allocation}, then  we have that
$$ \sum_{y \in \Com} \mu^*(\infl|y) > 0.$$
\end{lemma} 

\begin{proof}

From Lemma~\ref{lemma:pq_min}  we have that there exists a $p_{\min} > 0$ and $q_{\min} > 0$ such that
$$ p(x|y) \geq p_{\min}, \qquad  y \in \Com, x \in \setT$$
and
$$ p(x|y) \geq q_{\min}, \qquad y \in \Com, x \in \setT.$$

We then choose  constants  $N_0$ and $k_{\Infl} > 0$ such that
$$\ddelay{0} \max\{1,B_0\} <  \ddelay{0} \frac{(N_0-1) p_{\min}q_{\min}\delay{k_{\Infl}}}{B_0}.$$
Note that such constants exists.

We then show that the following is true for these values for $N_0$ and $k_{\Infl}$. If for $N > N_0$ we have that $\allAllocationNE  = (\coreInteractVecOpt,\comAllocation^*,\contAllocation^*)$  is a Nash equilibrium with $\coreBudget > N k_{\Infl}$, then  we have 
  $$ \sum_{y \in \Com} \mu^*(\infl|y) > 0.$$

We prove this results by contradiction as follows. For $N > N_0$, suppose that  $\allProxAllocationNE  = (\coreInteractVec^*,\comAllocation^*,\contAllocation^*)$ is a  Nash equilibrium as given by Definition~\ref{def:stable_allocation} with $\coreBudget > N k_{\Infl}$, such that 
$$ \sum_{y \in \Com} \mu^*(\infl|y) = 0.$$

Using Property b) of Lemma~\ref{lemma:conditions_perfect_NE}, we have in this case for every content consumer $y \in \comSet$ that either
\begin{equation}\label{eq:perfect_pos_rate_outside}
  \lambda^*(y) > 0
\end{equation}  
or there exists a content producer $z \in \Cw{y}$ such that
\begin{equation}\label{eq:perfect_pos_direct_rate}
 \mu^*(z|y) > 0.
\end{equation}  

We then obtain the following results. 
As by Assumption~\ref{ass:delay} the function $\delay{\mu}$, $\mu \geq 0$, is concave, we have that 
$$  \ddelay{\lambda^*(y)} B_0 \leq  \ddelay{0} B_0, \qquad y \in \comSet,$$
and
$$  \ddelay{\mu^*(z|y)} B(z|y) \leq  \ddelay{0}, \qquad y \in \comSet, z \in \Cw{y}.$$
By Assumption~\ref{ass:infl_alloc}, we have for the  case where
$$ \sum_{y \in \Com} \mu^*(\infl|y) = 0$$
that
$$  \coreInteractVecOpt(z) = \coreBudget/N = k_{\Infl}, \qquad z \in \comSet.$$
It then follows that for $N > N_0$ we that
$$ (N_0-1) \ddelay{0}  p_{\min}q_{\min}\delay{k_{\Infl}} \leq \ddelay{0} \sum_{z \in \Cw{y}} \delay{\coreInteractVecOpt(z)} B(z|y).$$
As by construction we have that
$$\ddelay{0} \max\{r_p,r_0B_0\} <  \ddelay{0} \frac{(N_0-1) r_p p_{\min}q_{\min}\delay{k_{\Infl}}}{B_0},$$
we obtain from the above inequalities that
\begin{equation}\label{eq:fo_perfect_outside}
  \ddelay{\lambda^*(y)} r_0 B_0 <  \ddelay{0} \sum_{z \in \Cw{y}} \delay{\coreInteractVecOpt(z)} B(z|y), \qquad y \in \comSet,
\end{equation}  
  and
\begin{equation}\label{eq:fo_perfect_direct_rate}  
  \ddelay{\mu^*(z|y)} B(z|y) < \ddelay{0} \sum_{z \in \Cw{y}} \delay{\coreInteractVecOpt(z)}  B(z|y), \qquad y \in \comSet, z \in \Cw{y}.
\end{equation}  
However, by  Lemma~\ref{lemma:conditions_perfect_NE}, and Eq.~\eqref{eq:perfect_pos_rate_outside} and  Eq.~\eqref{eq:perfect_pos_direct_rate}, we either have that 
$$\lambda^*(y) > 0$$
or there exists a content producer $z \in \Cw{y}$ such that
$$\mu^*(z|y) > 0,$$
and the inequalities given by Eq.~\eqref{eq:fo_perfect_outside} and  Eq.~\eqref{eq:fo_perfect_direct_rate} either violate Property c) or e) of Lemma~\ref{lemma:conditions_perfect_NE}.
This contradicts our assumption that  $\allProxAllocationNE  = (\coreInteractVecOpt,\comAllocation^*,\contAllocation^*)$ is a Nash equilibrium as given by Definition~\ref{def:stable_allocation}. The result of the lemma then follows.
\end{proof}

%\newpage

The next result states that  if the community size $N$ and the rate budget $\coreBudget$ of the influencer are large enough, then under a   Nash equilibrium of the game with perfect information we have that each  consumer follows the influencer with a positive rate.

\begin{prop}\label{prop:perfect_pos_rate_consumers}
Let $\leafBudget$ be a given rate budget.
Then there exist constants $N_0$, $k_{\Infl}$, and $\epsilon_0 > 0$, such that if $N > N_0$ and $\coreBudget > N k_{\Infl}$, the following is true.  
 If $\allAllocationNE  = (\coreInteractVecOpt,\comAllocation^*,\contAllocation^*)$  is a  Nash equilibrium as given by Definition~\ref{def:stable_allocation}, then  we have  that
 $$ \mu^*(\infl|y)  \geq \epsilon_0, \qquad y \in \comSet.$$
%  $$ \sum_{y \in \Com} \mu^*(\infl|y) > 0.$$
\end{prop}
Proposition~\ref{prop:perfect_pos_rate_consumers} can be proved using the same argument as given in the proof for Proposition~\ref{prop:proxy_pos_rate_consumers}, and we omit here a detailed derivation.

The next result states that  if the community size $N$ and the rate budget $\coreBudget$ of the influencer are large enough, then under a   Nash equilibrium of the game with perfect information we that the influencer allocates a large rate to each producer $z \in \comSet$.

\begin{prop}\label{prop:perfect_pos_rate_producers}
Let $\leafBudget$ be a given rate budget, and let $\mu_0$ be a given rate. Then there exist constants $N_0$ and $k_{\Infl}$ such that if $N > N_0$ and $\coreBudget > N k_{\Infl}$, the following is true.
If $\allAllocationNE(\coreBudget)  = (\coreInteractVec^{(\coreBudget)},\comAllocation^*,\contAllocation^*)$ is a  Nash equilibrium of the game with perfect information as given by Definition~\ref{def:stable_allocation}, then we have that 
$$ \coreInteractVecOpt(z) \geq \mu_0, \qquad z \in \Com.$$
\end{prop}
Proposition~\ref{prop:perfect_pos_rate_producers} can be proved using the same argument as given in the proof for Proposition~\ref{prop:proxy_pos_rate_producers}, and we omit here a detailed derivation.

%\newpage

The next result states that if the community size $N$ and the rate budget $\coreBudget$ of the influencer are large, then the all consumers $y \in \comSet$ will get their content through the influencer.

\begin{lemma}\label{lemma:perfect_NE_no_direct_rate}
  For a given rate budget $\leafBudget$, there exist constants $N_0$ and $\mu_0$ such that for $N > N_0$ the following is true.
If $\allAllocationNE  = (\coreInteractVec,\comAllocation,\contAllocation)$ is a Nash equilibrium for the game with perfect information as given by Definition~\ref{def:stable_allocation}, such that
$$ \coreInteractVecOpt(z) \geq \mu_0, \qquad z \in \comSet,$$
then we have for all content consumers $y \in \comSet$ that
$$ \mu^*(z|y)  = 0, \qquad z \in \Cw{y}.$$
\end{lemma}

\begin{proof}
Let $\leafBudget$ be a given rate budget. 
From Lemma~\ref{lemma:allocation_no_direct_rate}, we have that there exist exist constants $N_0$ and $\mu_0$ such that for $N > N_0$ the following is true.
If $\allAllocationNE  = (\coreInteractVec,\comAllocation,\contAllocation)$ is a Nash equilibrium as given by Definition~\ref{def:stable_allocation} such that
$$ \coreInteractVecOpt(z) \geq \mu_0, \qquad z \in \comSet,$$
then we have for all content consumers $y \in \comSet$ that
$$ \ddelay{ \mu^*(z|y)} B(z|y) <  \ddelay{ \mu^*(\infl|y)} \sum_{z' \in \Cw{y}} \delay{\coreInteractVecOpt(z')} B(z'|y), \qquad z \in \Cw{y}.$$
Combining this result with Lemma~\ref{lemma:conditions_perfect_NE} which gives the necessary and sufficient conditions for a Nash equilibrium, we obtain that  there exist constants $N_0$ and $\mu_0$ such that for $N > N_0$ the following is true.
If $\allAllocationNE  = (\coreInteractVec,\comAllocation,\contAllocation)$ is a Nash equilibrium as given by Definition~\ref{def:stable_allocation} such that
$$ \coreInteractVecOpt(z) \geq \mu_0, \qquad z \in \comSet,$$
then we have for all content consumers $y \in \comSet$ that
$$ \mu^*(z|y)  = 0, \qquad z \in \Cw{y}.$$
This establishes the result of the lemma.
\end{proof}

%\newpage

\subsection{Proof of Proposition~\ref{prop:NE_proxy_perfect}}\label{app:proof_NE_proxy_perfect}

In this appendix we provide a proof for  Proposition~\ref{prop:NE_proxy_perfect}. Recall that  Proposition~\ref{prop:NE_proxy_perfect} states that
for a given rate budget $\leafBudget$, there exist constants $N_0$ and $\mu_0$ such that for $N > N_0$ and $\coreBudget > N k_{\Infl}$, then the following is true. 
The allocation $\allAllocationNE  = (\coreInteractVecOpt,\comAllocation^*,\contAllocation^*)$ is a Nash equilibrium of the game with perfect information as given by Definition~\ref{def:stable_allocation}, if and only if  $\allAllocationNE$ is a Nash equilibrium for the proxy game. 

We prove  Proposition~\ref{prop:NE_proxy_perfect} as follows.
Let $\leafBudget$ be a given rate budget. Then by Lemma~\ref{lemma:allocation_no_direct_rate} and Lemma~\ref{lemma:perfect_NE_no_direct_rate}  there exist constants $N_0$ and $\mu_0$ such that the following is true.
If $\allAllocationNE  = (\coreInteractVec,\comAllocation,\contAllocation)$ is a Nash equilibrium as given by Definition~\ref{def:stable_allocation} such that
$$ \coreInteractVecOpt(z) \geq \mu_0, \qquad z \in \comSet,$$
then we have for all content consumers $y \in \comSet$ that
$$ \ddelay{ \mu(z|y)} B(z|y) <  \ddelay{ \mu(\infl|y)} \sum_{z' \in \Cw{y}} \delay{\coreInteractVec(z')}  B(z'|y), \qquad z \in \Cw{y}$$
and
$$ \mu^*(z|y)  = 0, \qquad z \in \Cw{y}.$$
As a result, from Lemma~\ref{lemma:allocation_no_direct_rate}, Lemma~\ref{lemma:conditions_perfect_NE}, and Lemma~\ref{lemma:perfect_NE_no_direct_rate},  we have that there exist constants $N_0$ and $k_{\Infl}$ such that for $N > N_0$ and $\coreBudget > N k_{\Infl}$ the following is true.
An $\allAllocationNE  = (\coreInteractVecOpt,\comAllocation^*,\contAllocation^*)$  is a  Nash equilibrium as given by Definition~\ref{def:stable_allocation}, if an only if the following conditions hold true,
\begin{enumerate}
\item[a)]  For all content producers $z \in \comSet$, the content allocation $x^*(z)$ is a solution for the maximization problem  $\optcontperiphery{\coreRateVec,\comAllocation}{y}$.

\item[b)] For all content consumers $y \in \comSet$ we have that
  $$\sum_{z \in \Cw{y}} \mu^*(z|y) = 0.$$

\item[c)] For all content consumers $y \in \comSet$ we have that
$$ \lambda^*(y) + \mu^*(\infl|y)  + \sum_{z \in \Cw{y}}  \mu^*(z|y)  = \leafBudget$$   
  
\item[d)] If for content consumer $y \in \comSet$ we have that $\lambda^*(y) > 0$, then we have
$$\ddelay{\lambda^*(y)} r_0 B_0 \geq \ddelay{\mu^*(z|y)} r_p B(z|y), \qquad z \in \Cw{y}$$
and
$$\ddelay{\lambda^*(y)} r_0 B_0 \geq  \ddelay{\mu^*(\infl|y)} r_p \sum_{z \in \Cw{y}} \delay{\coreInteractVecOpt(z)} B(z|y).$$

\item[e)] If  for content consumer $y \in \comSet$ we have that $\mu^*(\infl|y) > 0$, then we have

$$\ddelay{\mu^*(\infl|y)}r_p  \sum_{z \in \Cw{y}} \delay{\coreInteractVecOpt(z)} B(z|y) \geq \ddelay{\lambda^*(y)} r_0 B_0.$$

\item[f)]  For all content consumers $y \in \comSet$ we have that
$$ \ddelay{0} B(z|y) < \ddelay{\mu^*(\infl|y)} \sum_{z' \in \Cw{y}} \delay{\coreInteractVecOpt(z')}B(z|y'), \qquad z \in \Cw{y}$$

\item[g)] We have that
  $$\sum_{z \in \comSet} \coreInteractVecOpt(z) =  \coreBudget.$$  
  
\item[h)] If for content producer $z \in \comSet$ we have that $\coreInteractVecOpt(z) > 0$, then we have for $z' \in \Cw{z}$ that 
$$ \ddelay{\coreInteractVecOpt(z)} \sum_{y \in \Cw{z}} \delay{\mu^*(\infl|y)}  B(z|y)\geq \ddelay{\coreInteractVecOpt(z')} \sum_{y \in \Cw{z'}} \delay{\mu^*(\infl|y)} B(z'|y).$$

\end{enumerate}

In addition, from Lemma~\ref{lemma:allocation_no_direct_rate} and Lemma~\ref{lemma:conditions_proxy_NE} we have that there exist constants $N_0$ and $k_{\Infl}$ such that for $N > N_0$ and $\coreBudget > N k_{\Infl}$ the following is true.
An $\allAllocationNE  = (\coreInteractVecOpt,\comAllocation^*,\contAllocation^*)$  is a  Nash equilibrium of the proxy game as given by Definition~\ref{def:proxy_NE}, if an only if the following conditions hold true,

\begin{enumerate}
\item[a')]   For all content producers $z \in \comSet$, the content allocation $x^*(z)$ is a solution for the maximization problem  $\optcontperiphery{\coreRateVec,\comAllocation}{y}$.

\item[b')] For all content consumers $y \in \comSet$ we have that
  $$\sum_{z \in \Cw{y}} \mu^*(z|y) = 0.$$

\item[c')] For all content consumers $y \in \comSet$ we have that
  $$ \lambda^*(y) + \mu^*(\infl|y)  = \leafBudget.$$

\item[d')] If for content consumer $y \in \comSet$ we have that $\lambda^*(y) > 0$, then we have
$$\ddelay{\lambda^*(y)} r_0 B_0 \geq  \ddelay{\mu^*(\infl|y)} r_p \sum_{z \in \Cw{y}} \delay{\coreInteractVecOpt(z)} B(z|y).$$

\item[e')] If  for content consumer $y \in \comSet$ we have that $\mu^*(\infl|y) > 0$, then we have

$$\ddelay{\mu^*(\infl|y)}r_p  \sum_{z \in \Cw{y}} \delay{\coreInteractVecOpt(z)} B(z|y) \geq \ddelay{\lambda^*(y)} r_0 B_0.$$

\item[f')]  For all content consumers $y \in \comSet$ we have that
$$ \ddelay{0} B(z|y) < \ddelay{\mu^*(\infl|y)} \sum_{z' \in \Cw{y}} \delay{\coreInteractVecOpt(z')}B(z|y'), \qquad z \in \Cw{y}$$    
 
\item[g')] We have that
  $$\sum_{z \in \comSet} \coreInteractVecOpt(z) =  \coreBudget.$$  
  
\item[h')] If for content producer $z \in \comSet$ we have that $\coreInteractVecOpt(z) > 0$, then we have for $z' \in \Cw{z}$ that
  $$ \ddelay{\coreInteractVecOpt(z)} \sum_{y \in \Cw{z}} \delay{\mu^*(\infl|y)} B(z|y)  \geq \ddelay{\coreInteractVecOpt(z')} \sum_{y \in \Cw{z'}} \delay{\mu^*(\infl|y)}B(z|y).$$
\end{enumerate}

Note that the Properties a)-h) are identical to the Properties a')-h'). It then follows that for a given rate budget $\leafBudget$, there exist constants $N_0$ and $\mu_0$ such that for $N > N_0$ and $\coreBudget > N k_{\Infl}$, then the following is true. 
The allocation $\allAllocationNE  = (\coreInteractVecOpt,\comAllocation^*,\contAllocation^*)$ is a Nash equilibrium of the game with perfect information as given by Definition~\ref{def:stable_allocation}, if and only if  $\allAllocationNE$ is a Nash equilibrium for the proxy game. 

%\newpage

\section{Analysis of Nash Equilibrium $\allAllocationInflOpt$ for Imperfect Information}\label{app:analysis_imperfect}

In this section, we derive additional properties of a Nash equilibrium of the game with imperfect information as given by Definition~\ref{def:att_NE} that we use to in Appendix~\ref{app:proof_NE_proxy_imperfect} to prove Proposition~\ref{prop:NE_proxy_imperfect}.

Most of the analysis follows the same argument as given in Appendix~\ref{app:analysis_perfect} for the analysis of a Nash equilibrium $\allAllocationNE$ of the game with perfect information. We repeat here only the key results and arguments; for all other results and arguments, we refer to the analysis in Appendix~\ref{app:analysis_perfect}.

\subsection{Optimality Condition for  Nash Equilibrium  $\allAllocationInflNE$}
Our first result provides necessary and sufficient conditions for a Nash equilibrium of the game with perfect information.

\begin{lemma}\label{lemma:conditions_imperfect_NE}

  An $\allAllocationInflNE  = (\coreInteractVecOpt,\comAllocation^*,\contAllocation^*)$  is a  Nash equilibrium as given by Definition~\ref{def:att_NE}, if an only if the following conditions hold true,
\begin{enumerate}
\item[a)]   For all content producers $z \in \comSet$, the content allocation $x^*(z)$ is a solution for the maximization problem  
$$ x^*(z) =	\argmax_{x(z) \in \setT} \delay{\coreInteract{z|\comAllocation^*,\contAllocation^*_{\{z,x(z)\}}}}.$$

\item[b)] For all content consumers $y \in \comSet$ we have that
$$ \lambda^*(y) + \mu^*(\infl|y)  + \sum_{z \in \Cw{y}}  \mu^*(z|y)  = \leafBudget$$   
  
\item[c)] If for content consumer $y \in \comSet$ we have that $\lambda^*(y) > 0$, then we have
$$\ddelay{\lambda^*(y)} r_0 B_0 \geq \ddelay{\mu^*(z|y)} r_p B(z|y), \qquad z \in \Cw{y}$$
and
$$\ddelay{\lambda^*(y)} r_0 B_0 \geq  \ddelay{\mu^*(\infl|y)} r_p \sum_{z \in \Cw{y}} \delay{\coreInteractVecOpt(z)} B(z|y).$$

\item[d)] If  for content consumer $y \in \comSet$ we have that $\mu^*(\infl|y) > 0$, then we have
$$\ddelay{\mu^*(\infl|y)} \sum_{z' \in \Cw{y}} \delay{\coreInteractVecOpt(z')}  B(z'|y) \geq \ddelay{\mu^*(z|y)} B(z|y), \qquad z \in \Cw{y}$$
and
$$\ddelay{\mu^*(\infl|y)} r_p \sum_{z \in \Cw{y}} \delay{\coreInteractVecOpt(z)} B(z|y) \geq \ddelay{\lambda^*(y)} r_0 B_0.$$

\item[e)] If for content consumer $y \in \comSet$ and content producer $z \in \Cw{y}$ we have that $\mu^*(z|y) > 0$, then we have
$$\ddelay{\mu^*(z|y)} r_pB(z|y) \geq \ddelay{\lambda^*(y)} r_0B_0,$$
  $$\ddelay{\mu^*(z|y)} B(z|y) \geq  \ddelay{\mu^*(\infl|y)} \sum_{z' \in \Cw{y}} \delay{\coreInteractVecOpt(z')} B(z'|y),$$
  and
  $$ \ddelay{\mu^*(z|y)} B(z|y) \geq \ddelay{\mu^*(z'|y)} B(z'|y), \qquad z \in \Cw{y,z'}.$$

\item[f)] We have that
  $$\sum_{z \in \comSet} \coreInteractVecOpt(z) =  \coreBudget.$$  
  
\item[g)] If for content producer $z \in \comSet$ we have that $\coreInteractVecOpt(z) > 0$, then we have for  $z' \in \Cw{z}$ that
$$ \ddelay{\coreInteractVecOpt(z)} \sum_{y \in \Cw{z}} \delay{\mu^*(\infl|y)}  B(z|y) \geq \ddelay{\coreInteractVecOpt(z')} \sum_{y \in \Cw{z'}} \delay{\mu^*(\infl|y)} B(z'|y).$$

\end{enumerate}

\end{lemma} 
Lemma~\ref{lemma:conditions_imperfect_NE} can then be proved using the same argument as given for Lemma~\ref{lemma:conditions_proxy_NE} which establishes the necessary and sufficient conditions for a Nash equilibrium of the proxy game, and we omit here a detailed derivation.

%\newpage
\subsection{Properties of Nash Equilibrium  $\allAllocationInflNE$}\label{app:prop_NE_imperfect}
Next, we derive the key properties of a  Nash equilibrium $\allAllocationInflNE  = (\coreInteractVecOpt,\comAllocation^*,\contAllocation^*)$  for the game with imperfect information as given by Definition~\ref{def:att_NE}.

The first result provides conditions such that for a  Nash equilibrium $\allAllocationNE  = (\coreInteractVecOpt,\comAllocation^*,\contAllocation^*)$  of the game with perfect information  we have 
  $$ \sum_{y \in \Com} \mu^*(\infl|y) > 0.$$
\begin{lemma}\label{lemma:imperfect_pos_rate_infl}
There exist constants $N_0$ and $k_{\Infl}$ such that if $N > N_0$ and $\coreBudget > N k_{\Infl}$, then the following is true.  
If  $\allAllocationInflNE = (\coreInteractVec,\comAllocation^*,\contAllocation^*)$  is a  equilibrium of the game with imperfect information as given by Definition~\ref{def:att_NE}, then  we have that
$$ \sum_{y \in \Com} \mu^*(\infl|y) > 0.$$
\end{lemma}
Lemma~\ref{lemma:imperfect_pos_rate_infl} can be proved using the same argument as given in the proof for Lemma~\ref{lemma:imperfect_pos_rate_infl}, and we omit here a detailed derivation.

The next result states that  if the community size $N$ and the rate budget $\coreBudget$ of the influencer are large enough, then under a   Nash equilibrium of the game with imperfect information we have that each  consumer follows the influencer with a positive rate.

\begin{prop}\label{prop:imperfect_pos_rate_consumers}
For a given rate budget $\leafBudget$, there exist constants $\epsilon_0 > 0$, $N_0$ and $\mu_0$ such that for $N > N_0$ and $\coreBudget > N k_{\Infl}$, the following is true.
 If $\allAllocationInflNE  = (\coreInteractVecOpt,\comAllocation^*,\contAllocation^*)$  is a  Nash equilibrium for the game with imperfect information as given by Definition~\ref{def:att_NE}, then  we have  that
 $$ \mu^*(\infl|y)  \geq \epsilon_0, \qquad y \in \comSet.$$
%  $$ \sum_{y \in \Com} \mu^*(\infl|y) > 0.$$
\end{prop}
Proposition~\ref{prop:imperfect_pos_rate_consumers} can be proved using the same argument as given in the proof for Proposition~\ref{prop:proxy_pos_rate_consumers}, and we omit here a detailed derivation.

In addition we have that  if the community size $N$ and the rate budget $\coreBudget$ of the influencer are large enough, then under a   Nash equilibrium of the game with perfect information we that the influencer allocates a large rate to each producer $z \in \comSet$.
\begin{prop}\label{prop:imperfect_pos_rate_producers}
Let $\leafBudget$ be a given rate budget, and let $\mu_0$ be a given rate. Then there exist constants $N_0$ and $k_{\Infl}$ such that if $N > N_0$ and $\coreBudget > N k_{\Infl}$, the following is true.
If $\allAllocationInflNE(\coreBudget)  = (\coreInteractVec^{(\coreBudget)},\comAllocation^*,\contAllocation^*)$ is a  Nash equilibrium of the game with imperfect information as given by Definition~\ref{def:att_NE}, then we have that 
$$ \coreInteractVecOpt(z) \geq \mu_0, \qquad z \in \Com.$$
\end{prop}
Proposition~\ref{prop:imperfect_pos_rate_producers} can be proved using the same argument as given in the proof for Proposition~\ref{prop:proxy_pos_rate_producers}, and we omit here a detailed derivation.

The next result states that if the community size $N$ and the rate budget $\coreBudget$ of the influencer are large, then the all consumers $y \in \comSet$ will get their content through the influencer. 
\begin{lemma}\label{lemma:imperfect_NE_no_direct_rate}
  For a given rate budget $\leafBudget$, there exist constants $N_0$ and $\mu_0$ such that for $N > N_0$ the following is true.
If $\allAllocationInflNE  = (\coreInteractVec,\comAllocation,\contAllocation)$ is a Nash equilibrium as given by Definition~\ref{def:att_NE} such that
$$ \coreInteractVecOpt(z) \geq \mu_0, \qquad z \in \comSet,$$
then we have for all content consumers $y \in \comSet$ that
$$ \mu^*(z|y)  = 0, \qquad z \in \Cw{y}.$$
\end{lemma}
Lemma~\ref{lemma:imperfect_NE_no_direct_rate} can be proved using the same argument as given in the proof for Lemma~\ref{lemma:perfect_NE_no_direct_rate}, and we omit here a detailed derivation.

%\newpage

\subsection{Proof of Proposition~\ref{prop:NE_proxy_imperfect}}\label{app:proof_NE_proxy_imperfect}

In this appendix we provide a proof for  Proposition~\ref{prop:NE_proxy_imperfect}. Recall that  Proposition~\ref{prop:NE_proxy_imperfect} states that
for a given rate budget $\leafBudget$, there exist constants $N_0$ and $\mu_0$ such that for $N > N_0$ and $\coreBudget > N k_{\Infl}$, then the following is true. 
The allocation $\allAllocationInflNE  = (\coreInteractVecOpt,\comAllocation^*,\contAllocation^*)$ is a Nash equilibrium of the game with imperfect information as given by Definition~\ref{def:att_NE}, if and only if  $\allAllocationInflNE$ is a Nash equilibrium for the proxy game.

We prove  Proposition~\ref{prop:NE_proxy_imperfect} as follows.
Using Proposition~\ref{prop:att_contProducer}, for a Nash equilibrium $\allAllocationInflNE  = (\coreInteractVecOpt,\comAllocation^*,\contAllocation^*)$ as given by Definition~\ref{def:att_NE} with
$$ \mu^*(z|y)  = 0, \qquad y \in \comSet, z \in \Cw{y},$$
as well as
$$\mu^*(\infl|y) > 0, \qquad y \in \comSet,$$
and
$$\coreInteractVecOpt(z) > 0, \qquad z \in \comSet,$$
we have for all content producers $z \in \comSet$ that
$$ \argmax_{x(z) \in \setT} \delay{\coreInteract{z|\comAllocation^*,\contAllocation^*_{\{z,x(z)\}}}}
= \argmax_{x(z) \in \setT} \prodOptUtilityInf{z}
= \argmax_{x(z) \in \setT}\prodOptUtility{z},$$
where
$$\prodOptUtilityInf{z}
=  r_p \sum_{y \in \Clw{z}}    \delayCoreUtility{z}{y}$$
and
$$\begin{aligned}
		\prodOptUtility{z}
		= \; & r_p  \sum_{y \in \Clw{z}} \delayCoreUtility{z}{y}  \\
		& + r_p \sum_{z \in \Clw{y}} \delayUtility{z}{y}.
	\end{aligned}$$
Therefore, in the case where for a Nash equilibrium $\allAllocationInflNE  = (\coreInteractVecOpt,\comAllocation^*,\contAllocation^*)$ as given by Definition~\ref{def:att_NE} with
$$ \mu^*(z|y)  = 0, \qquad y \in \comSet, z \in \Cw{y},$$
as well as
$$\mu^*(\infl|y) > 0,$$
and
$$\coreInteractVecOpt(z) > 0, \qquad z \in \comSet,$$
the necessary and sufficient conditions for a Nash equilibrium given in Lemma~\ref{lemma:conditions_proxy_NE} are the same as the ones given the proof of Proposition~\ref{prop:NE_proxy_perfect}. 
The result of Proposition~\ref{prop:NE_proxy_imperfect} can then be proven using the same argument as given in the proof for Proposition~\ref{prop:NE_proxy_perfect}, and we omit here a detailed derivation.

\end{document}